\input harvmac.tex
\input epsf.tex
\input amssym

\def\figin{\epsfcheck\figin}\def\figins{\epsfcheck\figins}
\def\epsfcheck{\ifx\epsfbox\UnDeFiNeD
\message{(NO epsf.tex, FIGURES WILL BE IGNORED)}
\gdef\figin##1{\vskip2in}\gdef\figins##1{\hskip.5in}
\else\message{(FIGURES WILL BE INCLUDED)}%
\gdef\figin##1{##1}\gdef\figins##1{##1}\fi}
\def\DefWarn#1{}
\def\figinsert{\goodbreak\midinsert}
\def\ifig#1#2#3{\DefWarn#1\xdef#1{fig.~\the\figno}
\writedef{#1\leftbracket fig.\noexpand~\the\figno} %
\figinsert\figin{\centerline{#3}}\medskip\centerline{\vbox{\baselineskip12pt
\advance\hsize by -1truein\noindent\footnotefont{\bf
Fig.~\the\figno:} #2}}
\bigskip\endinsert\global\advance\figno by1}




\def\unit{\relax{\rm 1\kern-.26em I}}
\def\nada{\relax{\rm 0\kern-.30em l}}
\def\tilde{\widetilde}

\def \la {\langle}
\def \ra {\rangle}
\def \pa {\partial}

\def \eps {\epsilon}


\noblackbox
\def\IL{\relax{\rm I\kern-.18em L}}
\def\IH{\relax{\rm I\kern-.18em H}}
\def\IR{\relax{\rm I\kern-.18em R}}
\def\IC{\relax\hbox{$\inbar\kern-.3em{\rm C}$}}
\def\IZ{\relax\ifmmode\mathchoice
{\hbox{\cmss Z\kern-.4em Z}}{\hbox{\cmss Z\kern-.4em Z}} {\lower.9pt\hbox{\cmsss Z\kern-.4em Z}}
{\lower1.2pt\hbox{\cmsss Z\kern-.4em Z}}\else{\cmss Z\kern-.4em Z}\fi}


\font\manual=manfnt \def\dbend{\lower3.5pt\hbox{\manual\char127}}

\lref\GaoGA{
  S.~Gao and R.~M.~Wald,
Class.\ Quant.\ Grav.\  {\bf 17}, 4999 (2000).
[gr-qc/0007021].
}

\lref\PenroseUD{
  R.~Penrose, R.~D.~Sorkin and E.~Woolgar,
[gr-qc/9301015].
}

\lref\AichelburgDH{
  P.~C.~Aichelburg and R.~U.~Sexl,
Gen.\ Rel.\ Grav.\  {\bf 2}, 303 (1971).
}

\lref\ArutyunovNW{
  G.~Arutyunov and S.~Frolov,
Phys.\ Rev.\ D {\bf 60}, 026004 (1999).
[hep-th/9901121].
}

\lref\HorowitzBV{
  G.~T.~Horowitz and A.~R.~Steif,
Phys.\ Rev.\ Lett.\  {\bf 64}, 260 (1990).
}

\lref\tHooftRB{
G.~'t Hooft,
Phys.\ Lett.\ B {\bf 198}, 61 (1987).
}

\lref\DrayHA{
  T.~Dray and G.~'t Hooft,
Nucl.\ Phys.\ B {\bf 253}, 173 (1985).
}

\lref\GallowayBR{
  G.~J.~Galloway, K.~Schleich, D.~Witt and E.~Woolgar,
Phys.\ Lett.\ B {\bf 505}, 255 (2001).
[hep-th/9912119].
}. 

\lref\GiddingsXS{
  S.~B.~Giddings,
[arXiv:1105.2036 [hep-th]].
}

\lref\ShapiroUW{
I.~I.~Shapiro,
Phys.\ Rev.\ Lett.\  {\bf 13}, 789 (1964).
}

\lref\DrummondPP{
  I.~T.~Drummond and S.~J.~Hathrell,
Phys.\ Rev.\ D {\bf 22}, 343 (1980).
}

\lref\HofmanAR{
  D.~M.~Hofman and J.~Maldacena,
JHEP {\bf 0805}, 012 (2008).
[arXiv:0803.1467 [hep-th]].
}

\lref\acausality{
  S.~Coleman,
  ``Acausality,''
In *Erice 1969, Ettore Majorana School On Subnuclear Phenomena*, New York 1970, 282-327.
}

\lref\WittenQJ{
  E.~Witten,
Adv.\ Theor.\ Math.\ Phys.\  {\bf 2}, 253 (1998).
[hep-th/9802150].
}

\lref\SkenderisDG{
  K.~Skenderis and B.~C.~van Rees,
JHEP {\bf 0905}, 085 (2009).
[arXiv:0812.2909 [hep-th]].
}

\lref\EncisoWU{
  A.~Enciso and N.~Kamran,
Phys.\ Rev.\ D {\bf 85}, 106016 (2012).
[arXiv:1203.2743 [math-ph]].
}

\lref\LiuBU{
  H.~Liu and A.~A.~Tseytlin,
Nucl.\ Phys.\ B {\bf 533}, 88 (1998).
[hep-th/9804083].
}

\lref\SatohBC{
  Y.~Satoh and J.~Troost,
JHEP {\bf 0301}, 027 (2003).
[hep-th/0212089].
}

\lref\Tolman{ 
R.C. Tolman, ``The theory of relativity of motion'', Berkely Univ. Press, 1917. See also S. Coleman, ``Acausality'',  Erice School 1969, New York (1970) page 282.
 }

\lref\RajuBY{
  S.~Raju,
Phys.\ Rev.\ Lett.\  {\bf 106}, 091601 (2011).
[arXiv:1011.0780 [hep-th]].
}

\lref\BanksDD{
  T.~Banks, M.~R.~Douglas, G.~T.~Horowitz and E.~J.~Martinec,
[hep-th/9808016].
}

\lref\WeinbergKQ{
  S.~Weinberg and E.~Witten,
Phys.\ Lett.\ B {\bf 96}, 59 (1980)..
}

\lref\KobesUA{
  R.~Kobes,
Phys.\ Rev.\ D {\bf 43}, 1269 (1991)..
}

\lref\BarnesJP{
  E.~Barnes, D.~Vaman, C.~Wu and P.~Arnold,
Phys.\ Rev.\ D {\bf 82}, 025019 (2010).
[arXiv:1004.1179 [hep-th]].
}

\lref\SachsZJ{
  I.~Sachs,
Fortsch.\ Phys.\  {\bf 52}, 667 (2004).
[hep-th/0312287].
}

\lref\PoissonNH{
  E.~Poisson, A.~Pound and I.~Vega,
Living Rev.\ Rel.\  {\bf 14}, 7 (2011).
[arXiv:1102.0529 [gr-qc]].
}

\lref\SmolicGZ{
  J.~Smolic and M.~Taylor,
[arXiv:1301.5205 [hep-th]].
}

\lref\CamanhoVW{
  X.~O.~Camanho and J.~D.~Edelstein,
JHEP {\bf 1004}, 007 (2010).
[arXiv:0911.3160 [hep-th]].
}

\lref\CaiDZ{
  R.~-G.~Cai,
Phys.\ Rev.\ D {\bf 65}, 084014 (2002).
[hep-th/0109133].
}

\lref\HorowitzGF{
  G.~T.~Horowitz and N.~Itzhaki,
JHEP {\bf 9902}, 010 (1999).
[hep-th/9901012].
}

\lref\HottaQY{
  M.~Hotta and M.~Tanaka,
Class.\ Quant.\ Grav.\  {\bf 10}, 307 (1993)..
}

\lref\CiafaloniESA{
  M.~Ciafaloni and D.~Colferai,
[arXiv:1406.6540 [hep-th]].
}

\lref\AdeULN{
  P.~A.~R.~Ade {\it et al.}  [Planck Collaboration],
[arXiv:1303.5082 [astro-ph.CO]].
}
 
\lref\AnninosUI{
  D.~Anninos, T.~Hartman and A.~Strominger,
[arXiv:1108.5735 [hep-th]].
}

\lref\BriganteGZ{
  M.~Brigante, H.~Liu, R.~C.~Myers, S.~Shenker and S.~Yaida,
Phys.\ Rev.\ Lett.\  {\bf 100}, 191601 (2008).
[arXiv:0802.3318 [hep-th]].
}

\lref\GrisaruVM{
  M.~T.~Grisaru, H.~N.~Pendleton and P.~van Nieuwenhuizen,
Phys.\ Rev.\ D {\bf 15}, 996 (1977)..
}

\lref\HofmanUG{
  D.~M.~Hofman,
Nucl.\ Phys.\ B {\bf 823}, 174 (2009).
[arXiv:0907.1625 [hep-th]].
}

\lref\FullerZZA{
  R.~W.~Fuller and J.~A.~Wheeler,
Phys.\ Rev.\  {\bf 128}, 919 (1962)..
}

\lref\HollowoodAS{
  T.~J.~Hollowood and G.~M.~Shore,
Int.\ J.\ Mod.\ Phys.\ D {\bf 21}, 1241003 (2012).
[arXiv:1205.3291 [hep-th]].
}

\lref\AdamsSV{
  A.~Adams, N.~Arkani-Hamed, S.~Dubovsky, A.~Nicolis and R.~Rattazzi,
JHEP {\bf 0610}, 014 (2006).
[hep-th/0602178].
}

\lref\KulaxiziJT{
  M.~Kulaxizi and A.~Parnachev,
Phys.\ Rev.\ Lett.\  {\bf 106}, 011601 (2011).
[arXiv:1007.0553 [hep-th]].
}

\lref\KomargodskiEK{
  Z.~Komargodski and A.~Zhiboedov,
[arXiv:1212.4103 [hep-th]].
}

\lref\GaoGA{
  S.~Gao and R.~M.~Wald,
Class.\ Quant.\ Grav.\  {\bf 17}, 4999 (2000).
[gr-qc/0007021].
}

\lref\AdamsSV{
  A.~Adams, N.~Arkani-Hamed, S.~Dubovsky, A.~Nicolis and R.~Rattazzi,
JHEP {\bf 0610}, 014 (2006).
[hep-th/0602178].
}

\lref\SchusterNH{
  P.~C.~Schuster and N.~Toro,
JHEP {\bf 0906}, 079 (2009).
[arXiv:0811.3207 [hep-th]].
}

\lref\PageXN{
  D.~N.~Page, S.~Surya and E.~Woolgar,
Phys.\ Rev.\ Lett.\  {\bf 89}, 121301 (2002).
[hep-th/0204198].
}

\lref\DrayHA{
  T.~Dray and G.~'t Hooft,
Nucl.\ Phys.\ B {\bf 253}, 173 (1985)..
}

\lref\BriganteNU{
  M.~Brigante, H.~Liu, R.~C.~Myers, S.~Shenker and S.~Yaida,
Phys.\ Rev.\ D {\bf 77}, 126006 (2008).
[arXiv:0712.0805 [hep-th]].
}

\lref\BriganteGZ{
  M.~Brigante, H.~Liu, R.~C.~Myers, S.~Shenker and S.~Yaida,
Phys.\ Rev.\ Lett.\  {\bf 100}, 191601 (2008).
[arXiv:0802.3318 [hep-th]].
}

\lref\PolchinskiRY{
  J.~Polchinski,
[hep-th/9901076].
}

\lref\DubovskyAC{
  S.~Dubovsky, A.~Nicolis, E.~Trincherini and G.~Villadoro,
Phys.\ Rev.\ D {\bf 77}, 084016 (2008).
[arXiv:0709.1483 [hep-th]].
}

\lref\McFaddenKK{
  P.~McFadden and K.~Skenderis,
JCAP {\bf 1106}, 030 (2011).
[arXiv:1104.3894 [hep-th]].
}

\lref\ElShowkDWA{
  S.~El-Showk, M.~F.~Paulos, D.~Poland, S.~Rychkov, D.~Simmons-Duffin and A.~Vichi,
[arXiv:1403.4545 [hep-th]].
}

\lref\HofmanAR{
  D.~M.~Hofman and J.~Maldacena,
JHEP {\bf 0805}, 012 (2008).
[arXiv:0803.1467 [hep-th]].
}

\lref\ChristodoulouUV{
  D.~Christodoulou and S.~Klainerman,
  ``The Global nonlinear stability of the Minkowski space,''
Princeton University Press, Princeton, 1993.
}

\lref\CaronHuotFEA{
  S.~Caron-Huot,
[arXiv:1309.6521 [hep-th]].
}

\lref\MetsaevYB{
  R.~R.~Metsaev and A.~A.~Tseytlin,
Phys.\ Lett.\ B {\bf 185}, 52 (1987)..
}

\lref\CaiSA{
  Y.~Cai and C.~A.~Nunez,
Nucl.\ Phys.\ B {\bf 287}, 279 (1987)..
}

\lref\SchusterNH{
  P.~C.~Schuster and N.~Toro,
JHEP {\bf 0906}, 079 (2009).
[arXiv:0811.3207 [hep-th]].
}

\lref\GoldB{
Goldberger, Marvin L., and Kenneth M. Watson, 
``Concerning the notion of   ``time interval'' in S-matrix theory,'' 
Physical Review 127.6 (1962): 2284-2286.
}

\lref\GoldBtwo{
Froissart, M., M. L. Goldberger, and K. M. Watson,
 ``Spatial separation of events in S-matrix theory,''
Physical Review 131.6 (1963): 2820.
}

\lref\RychkovSF{
  V.~S.~Rychkov,
Phys.\ Rev.\ D {\bf 70}, 044003 (2004).
[hep-ph/0401116].
}

\lref\EardleyRE{
  D.~M.~Eardley and S.~B.~Giddings,
Phys.\ Rev.\ D {\bf 66}, 044011 (2002).
[gr-qc/0201034].
}

\lref\GiddingsXY{
  S.~B.~Giddings and V.~S.~Rychkov,
Phys.\ Rev.\ D {\bf 70}, 104026 (2004).
[hep-th/0409131].
}

\lref\VeloUR{
  G.~Velo and D.~Zwanziger,
Phys.\ Rev.\  {\bf 188}, 2218 (1969)..
}

\lref\BuchelSK{
  A.~Buchel, J.~Escobedo, R.~C.~Myers, M.~F.~Paulos, A.~Sinha and M.~Smolkin,
JHEP {\bf 1003}, 111 (2010).
[arXiv:0911.4257 [hep-th]].
}

\lref\HeemskerkPN{
  I.~Heemskerk, J.~Penedones, J.~Polchinski and J.~Sully,
JHEP {\bf 0910}, 079 (2009).
[arXiv:0907.0151 [hep-th]].
}

\lref\FitzpatrickZM{
  A.~L.~Fitzpatrick, E.~Katz, D.~Poland and D.~Simmons-Duffin,
JHEP {\bf 1107}, 023 (2011).
[arXiv:1007.2412 [hep-th]].
}

\lref\AmatiWQ{
  D.~Amati, M.~Ciafaloni and G.~Veneziano,
Phys.\ Lett.\ B {\bf 197}, 81 (1987).
}

\lref\AmatiUF{
  D.~Amati, M.~Ciafaloni and G.~Veneziano,
Int.\ J.\ Mod.\ Phys.\ A {\bf 3}, 1615 (1988).
}

\lref\AmatiTN{
  D.~Amati, M.~Ciafaloni and G.~Veneziano,
Phys.\ Lett.\ B {\bf 216}, 41 (1989).
}

\lref\AmatiZB{
  D.~Amati, M.~Ciafaloni and G.~Veneziano,
Phys.\ Lett.\ B {\bf 289}, 87 (1992).
}

\lref\AmatiTB{
  D.~Amati, M.~Ciafaloni and G.~Veneziano,
Nucl.\ Phys.\ B {\bf 403}, 707 (1993).
}

\lref\HorowitzBV{
  G.~T.~Horowitz and A.~R.~Steif,
Phys.\ Rev.\ Lett.\  {\bf 64}, 260 (1990).
}

\lref\deVegaTS{
  H.~J.~de Vega and N.~G.~Sanchez,
Nucl.\ Phys.\ B {\bf 317}, 706 (1989).
}

\lref\CaronHuotFEA{
  S.~Caron-Huot,
[arXiv:1309.6521 [hep-th]].
}

\lref\GrossIV{
  D.~J.~Gross and E.~Witten,
Nucl.\ Phys.\ B {\bf 277}, 1 (1986).
}

\lref\MetsaevZX{
  R.~R.~Metsaev and A.~A.~Tseytlin,
Nucl.\ Phys.\ B {\bf 293}, 385 (1987)..
}

\lref\MetsaevYB{
  R.~R.~Metsaev and A.~A.~Tseytlin,
Phys.\ Lett.\ B {\bf 185}, 52 (1987).
}

\lref\MaloneyRR{
  A.~Maloney, E.~Silverstein and A.~Strominger,
[hep-th/0205316].
}

\lref\SusskindQZ{
  L.~Susskind,
Phys.\ Rev.\ D {\bf 1}, 1182 (1970)..
}

\lref\HofmanUG{
  D.~M.~Hofman,
Nucl.\ Phys.\ B {\bf 823}, 174 (2009).
[arXiv:0907.1625 [hep-th]].
}

\lref\ZoharToAppear{

L.~Di Pietro and Z.~Komargodski,
``Cardy Formulae for SUSY Theories in $d=4$ and $d=6$,''
(to appear soon).

}

\lref\CornalbaXK{
  L.~Cornalba, M.~S.~Costa, J.~Penedones and R.~Schiappa,
JHEP {\bf 0708}, 019 (2007).
[hep-th/0611122].
}

\lref\CornalbaXM{
  L.~Cornalba, M.~S.~Costa, J.~Penedones and R.~Schiappa,
Nucl.\ Phys.\ B {\bf 767}, 327 (2007).
[hep-th/0611123].
}

\lref\CornalbaZB{
  L.~Cornalba, M.~S.~Costa and J.~Penedones,
JHEP {\bf 0709}, 037 (2007).
[arXiv:0707.0120 [hep-th]].
}

\lref\BerendsAH{
  F.~A.~Berends and R.~Gastmans,
Annals Phys.\  {\bf 98}, 225 (1976)..
}

\lref\AldayMF{
  L.~F.~Alday and J.~M.~Maldacena,
JHEP {\bf 0711}, 019 (2007).
[arXiv:0708.0672 [hep-th]].
}
\lref\FitzpatrickYX{
  A.~L.~Fitzpatrick, J.~Kaplan, D.~Poland and D.~Simmons-Duffin,
JHEP {\bf 1312}, 004 (2013).
[arXiv:1212.3616 [hep-th]].
}

\lref\KomargodskiEK{
  Z.~Komargodski and A.~Zhiboedov,
JHEP {\bf 1311}, 140 (2013).
[arXiv:1212.4103 [hep-th]].
}

\lref\PenroseCausality{ R. Penrose, ``On Schwarzschild Causality- A problem for ``Lorentz Covariant'' General Relativity'', in ``Essays in General Relativity'', Edited by
F. Tipler, 1980, Elsevier.}

\lref\DrummondPP{
  I.~T.~Drummond and S.~J.~Hathrell,
Phys.\ Rev.\ D {\bf 22}, 343 (1980)..
}
\lref\HollowoodKT{
  T.~J.~Hollowood and G.~M.~Shore,
Phys.\ Lett.\ B {\bf 655}, 67 (2007).
[arXiv:0707.2302 [hep-th]].
}
\lref\DubovskyAC{
  S.~Dubovsky, A.~Nicolis, E.~Trincherini and G.~Villadoro,
Phys.\ Rev.\ D {\bf 77}, 084016 (2008).
[arXiv:0709.1483 [hep-th]].
}

\lref\StiebergerHQ{
  S.~Stieberger,
[arXiv:0907.2211 [hep-th]].
}

\lref\FullerZZA{
  R.~W.~Fuller and J.~A.~Wheeler,
Phys.\ Rev.\  {\bf 128}, 919 (1962)..
}
\lref\IsraelUR{
  W.~Israel,
Phys.\ Lett.\ A {\bf 57}, 107 (1976)..
}
\lref\MaldacenaKR{
  J.~M.~Maldacena,
JHEP {\bf 0304}, 021 (2003).
[hep-th/0106112].
}

\lref\MaldacenaXJA{
  J.~Maldacena and L.~Susskind,
Fortsch.\ Phys.\  {\bf 61}, 781 (2013).
[arXiv:1306.0533 [hep-th]].
}

\lref\KawaiXQ{
  H.~Kawai, D.~C.~Lewellen and S.~H.~H.~Tye,
Nucl.\ Phys.\ B {\bf 269}, 1 (1986)..
}

\lref\ArkaniHamedRS{
  N.~Arkani-Hamed, S.~Dimopoulos and G.~R.~Dvali,
Phys.\ Lett.\ B {\bf 429}, 263 (1998).
[hep-ph/9803315].
}

\lref\RandallEE{
  L.~Randall and R.~Sundrum,
Phys.\ Rev.\ Lett.\  {\bf 83}, 3370 (1999).
[hep-ph/9905221].
}

\lref\RandallVF{
  L.~Randall and R.~Sundrum,
Phys.\ Rev.\ Lett.\  {\bf 83}, 4690 (1999).
[hep-th/9906064].
}

\lref\MyersLVA{
  R.~C.~Myers, R.~Pourhasan and M.~Smolkin,
JHEP {\bf 1306}, 013 (2013).
[arXiv:1304.2030 [hep-th]].
}

\lref\EmparanCE{
  R.~Emparan,
Phys.\ Rev.\ D {\bf 64}, 024025 (2001).
[hep-th/0104009].
}

\lref\KabatTB{
  D.~N.~Kabat and M.~Ortiz,
Nucl.\ Phys.\ B {\bf 388}, 570 (1992).
[hep-th/9203082].
}

\lref\CostaMG{
  M.~S.~Costa, J.~Penedones, D.~Poland and S.~Rychkov,
JHEP {\bf 1111}, 071 (2011).
[arXiv:1107.3554 [hep-th]].
}

\lref\ZwiebachUQ{
  B.~Zwiebach,
Phys.\ Lett.\ B {\bf 156}, 315 (1985)..
}

\lref\DrummondPP{
  I.~T.~Drummond and S.~J.~Hathrell,
Phys.\ Rev.\ D {\bf 22}, 343 (1980)..
}

\lref\GiudiceCE{
  G.~F.~Giudice, R.~Rattazzi and J.~D.~Wells,
Nucl.\ Phys.\ B {\bf 630}, 293 (2002).
[hep-ph/0112161].
}
\lref\MaldacenaNZ{
  J.~M.~Maldacena and G.~L.~Pimentel,
JHEP {\bf 1109}, 045 (2011).
[arXiv:1104.2846 [hep-th]].
}

\lref\AharonovVU{
  Y.~Aharonov, A.~Komar and L.~Susskind,
Phys.\ Rev.\  {\bf 182}, 1400 (1969)..
}

\lref\McGadySGA{
  D.~A.~McGady and L.~Rodina,
[arXiv:1311.2938 [hep-th]].
}

\lref\BekaertVH{
  X.~Bekaert, S.~Cnockaert, C.~Iazeolla and M.~A.~Vasiliev,
[hep-th/0503128].
}

\lref\HeemskerkPN{
  I.~Heemskerk, J.~Penedones, J.~Polchinski and J.~Sully,
JHEP {\bf 0910}, 079 (2009).
[arXiv:0907.0151 [hep-th]].
}

\lref\WillVA{
  C.~M.~Will,
  ``The Confrontation between general relativity and experiment,''
Living Rev.\ Rel.\  {\bf 9}, 3 (2006).
[gr-qc/0510072].
}
\lref\WillKXA{
  C.~M.~Will,
  ``The Confrontation between General Relativity and Experiment,''
[arXiv:1403.7377 [gr-qc]].
}

\lref\MyersJV{
  R.~C.~Myers, M.~F.~Paulos and A.~Sinha,
JHEP {\bf 1008}, 035 (2010).
[arXiv:1004.2055 [hep-th]].
}

\lref\SenNFA{
  K.~Sen and A.~Sinha,
[arXiv:1405.7862 [hep-th]].
}

\lref\BenincasaXK{
  P.~Benincasa and F.~Cachazo,
[arXiv:0705.4305 [hep-th]].
}

\lref\WeinbergKQ{
  S.~Weinberg and E.~Witten,
Phys.\ Lett.\ B {\bf 96}, 59 (1980)..
}

\lref\SarkarXP{
  S.~Sarkar and A.~C.~Wall,
Phys.\ Rev.\ D {\bf 83}, 124048 (2011).
[arXiv:1011.4988 [gr-qc]].
}

\lref\MaldacenaVR{
  J.~M.~Maldacena,
JHEP {\bf 0305}, 013 (2003).
[astro-ph/0210603].
}

\lref\MaldacenaNZ{
  J.~M.~Maldacena and G.~L.~Pimentel,
JHEP {\bf 1109}, 045 (2011).
[arXiv:1104.2846 [hep-th]].
}

\lref\AdeXNA{
  P.~A.~R.~Ade {\it et al.}  [BICEP2 Collaboration],
[arXiv:1403.3985 [astro-ph.CO]].
}

\lref\MortonsonBJA{
  M.~J.~Mortonson and U.~Seljak,
[arXiv:1405.5857 [astro-ph.CO]].
}

\lref\FlaugerQRA{
  R.~Flauger, J.~C.~Hill and D.~N.~Spergel,
[arXiv:1405.7351 [astro-ph.CO]].
}

\lref\ShenkerPQA{
  S.~H.~Shenker and D.~Stanford,
JHEP {\bf 1403}, 067 (2014).
[arXiv:1306.0622 [hep-th]].
}

\lref\VasilievEV{
  M.~A.~Vasiliev,
Phys.\ Lett.\ B {\bf 567}, 139 (2003).
[hep-th/0304049].
}

\lref\BrittoFQ{
  R.~Britto, F.~Cachazo, B.~Feng and E.~Witten,
Phys.\ Rev.\ Lett.\  {\bf 94}, 181602 (2005).
[hep-th/0501052].
}

\lref\PageXN{
  D.~N.~Page, S.~Surya and E.~Woolgar,
Phys.\ Rev.\ Lett.\  {\bf 89}, 121301 (2002).
[hep-th/0204198].
}

\lref\DixonWI{
  L.~J.~Dixon,
In *Boulder 1995, QCD and beyond* 539-582.
[hep-ph/9601359].
}

\lref\MaldacenaRE{
  J.~M.~Maldacena,
Adv.\ Theor.\ Math.\ Phys.\  {\bf 2}, 231 (1998).
[hep-th/9711200].
}
\lref\WittenQJ{
  E.~Witten,
Adv.\ Theor.\ Math.\ Phys.\  {\bf 2}, 253 (1998).
[hep-th/9802150].
}
\lref\GubserBC{
  S.~S.~Gubser, I.~R.~Klebanov and A.~M.~Polyakov,
Phys.\ Lett.\ B {\bf 428}, 105 (1998).
[hep-th/9802109].
}

\lref\OsbornCR{
  H.~Osborn and A.~C.~Petkou,
Annals Phys.\  {\bf 231}, 311 (1994).
[hep-th/9307010].
}
\lref\ErdmengerYC{
  J.~Erdmenger and H.~Osborn,
Nucl.\ Phys.\ B {\bf 483}, 431 (1997).
[hep-th/9605009].
}

\lref\MaldacenaVR{
  J.~M.~Maldacena,
JHEP {\bf 0305}, 013 (2003).
[astro-ph/0210603].
}

\lref\StromingerPN{
  A.~Strominger,
JHEP {\bf 0110}, 034 (2001).
[hep-th/0106113].
}

\lref\WittenKN{
  E.~Witten,
[hep-th/0106109].
}

\lref\NappiNY{
  C.~R.~Nappi and L.~Witten,
Phys.\ Rev.\ D {\bf 40}, 1095 (1989)..
}

\lref\DuffEA{
  M.~J.~Duff, C.~N.~Pope and K.~S.~Stelle,
Phys.\ Lett.\ B {\bf 223}, 386 (1989)..
}

\lref\DidenkoDWA{
  V.~E.~Didenko and E.~D.~Skvortsov,
[arXiv:1401.2975 [hep-th]].
}

\lref\NachtmannMR{
  O.~Nachtmann,
Nucl.\ Phys.\ B {\bf 63}, 237 (1973)..
}



\Title{
\vbox{\baselineskip6pt
}}
{\vbox{\centerline{Causality Constraints on Corrections to}\vskip .25cm
 \centerline{the Graviton Three-Point Coupling     }
}}


\centerline{Xi\'an O. Camanho,$^1$ Jos\'e D. Edelstein,$^2$ Juan Maldacena$^3$ and Alexander Zhiboedov$^4$  }
\bigskip
\centerline{\it $^1$ Max-Planck-Institut f\"ur Gravitationsphysik, Albert-Einstein-Institut, 14476 Golm, Germany}
\centerline{\it $^2$ Department of Particle Physics and IGFAE, University of Santiago de Compostela, Spain}
\centerline{\it and Centro de Estudios Cient\'{\i}ficos CECs, Valdivia, Chile}
\centerline{\it $^3$ School of Natural Sciences, Institute for Advanced Study,Princeton, NJ, USA }
\centerline{\it $^4$ Department of Physics, Princeton University,  Princeton, NJ, USA}

\vskip .3in \noindent

We consider higher derivative corrections to the graviton three-point coupling within a weakly coupled theory of gravity. Lorentz 
 invariance allows further structures beyond the one present in the Einstein theory. We argue that these are constrained by causality. We devise a thought experiment involving a high energy scattering process which leads to causality violation if the graviton three-point vertex contains the additional structures. This violation cannot be fixed by  adding conventional particles with spins $J \leq 2$. But, it can be fixed by adding    an infinite  tower of extra massive  particles with higher spins, $J > 2$. In AdS theories this implies a constraint on  the conformal anomaly coefficients $\left|{a - c \over c} \right| \lesssim {1 \over \Delta_{gap}^2}$ in terms of $\Delta_{gap}$, the dimension of the lightest single particle operator with spin $J > 2$. For inflation, or de Sitter-like solutions, it  indicates  the existence of  massive higher spin  particles if the gravity wave non-gaussianity deviates significantly from the one computed in the Einstein theory.


 \Date{ }




 \listtoc\writetoc
\vskip .5in \noindent

\newsec{Introduction/Motivation } 

In this paper we consider weakly coupled gravity theories in the tree-level approximation. It  is well-known that at long distances such theories should reduce to the Einstein gravity theory. However,  at intermediate energies we can have higher derivative corrections. By intermediate energies we mean those that are low enough that the theory is still weakly coupled but high enough that we are sensitive to possible higher derivative corrections. An example of such a theory is weakly coupled string theory where the corrections appear at a length scale $\sqrt{\alpha'}$, which is much larger than the Planck length. The theory at energies comparable to $1/\sqrt{\alpha'}$ is still weakly coupled. In this case, the higher derivative corrections are accompanied by extra massive higher spin particles which appear at the same scale. For higher energies the description is via a string theory which departs significantly from ordinary local quantum field theory. It is reasonable to expect that this is a generic feature. Namely, that higher derivative corrections only arise due to the presence of extra states with masses comparable to the scales where the higher derivative corrections become important.

The objective of this paper is to sharpen this link for the simplest possible correction, that of the graviton three-point coupling. Due to the fact that the graviton has spin, the flat space on-shell three-point function is not uniquely specified. In general, it has three different possible structures. The most familiar is the one we get in the Einstein theory. The others can be viewed as arising from higher derivative terms in the gravitational action. The first new structure has two more derivatives. Relative to the size of the Einstein-Hilbert term, it scales like $\alpha p^2$, where $\alpha$ is a new quantity with dimensions of length squared which characterizes the relative importance of the new term. Here $p$ is the typical scale of the momenta, it is not a Mandelstam invariant, since they all vanish for on-shell three-point functions. We work in a regime where both of the three-point vertices are small, so that gravity is weakly coupled.

We find that new three-point vertices lead to a potential causality violation unless we get contributions from extra particles. This causality violation is occurring when the theory is still weakly coupled. It occurs in a high-energy, fixed impact parameter, scattering process at a center of mass (energy)$^2$,  $s$, which is large compared to $1/\alpha$ but still small enough for the coupling to be weak. In general relativity this scattering process leads to the well-known Shapiro time delay \ShapiroUW, which is one of the classical tests of general relativity \WillKXA. See also \tHooftRB. When the graviton three-point vertex is corrected, the new terms can lead to a time advance, depending on the spin of the scattered graviton. At short enough impact parameter this time advance can overwhelm Shapiro's time delay and lead to a causality problem. This troublesome feature arises at an impact parameter of order $b^2 \sim \alpha $. At tree-level, this problem can only  be fixed by  introducing {an infinite number of} new massive particles with spin\foot{In $D>4$ dimensions by spin $J>2$ particles we mean particles in the representations of little group $SO(D-1)$ with both of the following two properties  (see also appendix H): a) their maximal spin projection $J_{+-} \geq 2$; b) their representations are labeled by Young tableaux with three or more boxes.} $J > 2$ and $m^2$ comparable to $\alpha^{-1}$.  In other words, it cannot be fixed by adding particles with spins $J \leq 2$, or by considering the existence of extra dimensions. 

These causality constraints are similar in spirit to those considered in \refs{\AharonovVU,\AdamsSV} but they differ in two ways. First of all, here the problem will be shown to arise for small $t/s$, but large $s$. Second,  the fact that the graviton has spin is crucial. On the other hand, in both cases we have locally Lorentz-invariant Lagrangians that nevertheless can lead to causality violations in non-trivial backgrounds.

An example of a theory that is constrained by these considerations is given by the action 
\eqn\consth{
S = l_p^{2-D} \int d^D\!x \sqrt{g} \left[ R + \alpha  \left(  R^2_{\mu \nu \rho \sigma} - 4 R^2_{\mu \nu} + R^2 \right)  \right] ~,
}
where the second term is the Lanczos-Gauss-Bonnet term.\foot{It is the dimensional continuation of the four-dimensional Euler density. In four dimensions it is a topological term, while in higher dimensions it is not topological, in fact it contributes to the three-point coupling of the graviton.} The constant $\alpha$ has dimensions of length squared. 
For $ \alpha \gg l_p^2 $,  we will show that  the theory is not causal. Furthermore, there is no way to make it causal by adding 
local higher curvature terms. 
In fact, our discussion refers to on-shell data, namely the three-point function, which 
reflects the real physical information and does not depend on the particular way that we write the Lagrangian. In other words, the discussion is invariant under field redefinitions. 

Note that if we view gravity as a low-energy effective theory with a UV cutoff of order $M_{pl}$ and we add higher derivative terms with dimensionless 
coefficients which are of order one, then we have nothing to say. The remarks in this paper {\it only apply}  for theories where the coefficients of the
higher derivative terms are much larger.  The ``natural'' value for the coefficient $\alpha$ if we view \consth\ as an effective gravity theory is $\alpha \sim l_p^2$. 
We will only constrain larger values of $\alpha$. 
Of course, we are discussing this problem because it {\it is indeed  possible }  to have theories with $\alpha \gg l_{p}^2$, 
for example a weakly coupled string theory.

 Another example where this discussion is relevant is the following. Imagine that we consider a large $N$ gauge theory. Such a gauge
 theory is expected to have a weakly coupled string dual. This theory will have a weakly coupled graviton corresponding to the stress tensor operator
\refs{\MaldacenaRE,\WittenQJ,\GubserBC}. 
  However, we are not guaranteed that the dual will be an ordinary Einstein gravity theory. 
 It might even be a Vasiliev-like gravity theory \refs{\VasilievEV,\BekaertVH}. The field theory three-point functions of the stress tensor  
\refs{\OsbornCR,\ErdmengerYC} determine the three-point functions of the gravity theory  \refs{\ArutyunovNW,\HofmanAR}. 
 One can imagine a theory where the only light single trace operator is the stress tensor.  
 It is natural to expect that this theory will have an Einstein gravity dual. It would be nice to prove that. For scalar interactions,   
 \HeemskerkPN\ argued that the solutions of the crossing equation have solutions that correspond to local vertices in the bulk, see also 
  \FitzpatrickZM . However, one can worry 
 that the size of the vertices with higher derivatives might be comparable to the one in the Einstein theory.
  As a case in point, we can think about the graviton three-point coupling. 
 This is constrained to be a linear combination of three structures,  only one of which is the Einstein one. These extra structures necessarily lead to a causality problem unless we introduce new higher spin particles at the scale that appears in these
 new three-point functions. Thus we link the three-point function of the stress tensor to the operator spectrum of the theory. 
 These three-point functions were constrained by causality in \HofmanAR . Here we get stronger constraints because we are making further assumptions about the operator spectrum of the theory. 
   
 As another application we  consider the possibility that gravity waves during inflation were generated by a theory that indeed had 
 these higher derivative corrections with a size comparable to the Hubble scale. This is a possibility which is allowed by conformal invariance 
 and would be realized if the dual description to inflation (in the spirit of dS/CFT 
\refs{ \MaldacenaVR , \StromingerPN , \WittenKN }) was a weakly coupled theory or if inflation occurs in a  string theory where 
 the string scale is close to the Hubble scale. The theory is still weakly coupled, so that scalar and tensor fluctuations are small. In this case the 
 gravity wave non-gaussianities would be different from the ones in the Einstein theory 
 \refs{\MaldacenaNZ,\McFaddenKK}. The observation of such gravity wave signals, combined with the arguments in this paper, would imply the existence 
 of extra particles with spin $J >2$ during inflation. 
 
 Finally, as a further motivation we should mention the grand dream of deriving 
 the most general weakly coupled consistent theory of gravity. It is quite likely that the only 
 such theory is a string-like theory, broadly defined. We are certainly very far away from this dream, but
 hopefully our simple observation about three-point functions could be useful. In particular, this observation 
 highlights the importance of spin.  Spin is likely to grow in importance as we consider constraints on the four-point function, given what we got for the much simpler three-point function. 
 In  \refs{\BenincasaXK ,\SchusterNH,\McGadySGA} various interesting constraints were derived by using crossing symmetry and the correct factorization 
 on the pole singularities.  The constraints discussed in this paper are additional constraints, not covered by their analysis.

 This paper is organized as follows. 
 In section two we discuss the notion of asymptotic causality, both for asymptotically flat and asymptotically AdS spacetimes.
 We also discuss the propagation of particles and fields through a shock wave. The purpose of this discussion is to set the stage for 
 the more general argument in the following section so that it becomes intuitively clearer.
  In section three, we present the main thought experiment  which involves only on-shell amplitudes and does not refer explicitly to shock waves. 
 In section four we discuss the effects of adding extra massive particles. We show that particles with spins two or less cannot solve the problem. 
 We also discuss the appearance of these massive particles among the final states of the scattering process. In order to argue that the problem persists we 
 present an alternative presentation of the problem where we consider the analyticity properties of the $S$ matrix in impact parameter representation.
 In section five we discuss various aspects of the $AdS$ version of the thought experiment and its implications for properties of the dual theory. 
 In section 6 we briefly mention the implications of a possible time advance in the context of wormholes. In section 7 we discuss a cosmological application, 
 where we link the possible existence of new structures for the gravity wave non-gaussianity to the presence of higher spin massive particles during inflation.  
 

\newsec{Flat Space Causality and Shock Waves} 

In this section we consider the problem in asymptotically flat space. We start by discussing the causal structure
in asymptotically flat space. As a motivation for our later discussion we analyze the  
scattering of a probe graviton from the shock wave. We then  present the argument for causality  violation  in purely on-shell terms.

 \subsec{Statement of Flat Space Causality} 
 
When we consider a gravity theory in asymptotically flat space we expect to 
be able to define scattering amplitudes. In particular, this presupposes that one can fix
the asymptotic structure of the spacetime so that we can compare times between the past 
and the future asymptotic regions. In dimensions $D>4$ we expect to be able to do this. 
Let us give a simple argument. 
Imagine we have a series of observers that sit at  a large distance $L$ from each other, and also from the center. 
For simplicity imagine them at the vertices of a large spatial hypercube and moving along the time direction. 
We would like to argue that they can synchronize their clocks. If they were in flat space, they can just send signals to 
each other. On the other hand if there is an object of mass $m$ in the interior, then 
 the  metric components decay as $1/r^{D-3}$. This gives rise to a 
 redshift of order ${ G m \over L^{D-3} }$ at the position of the detectors. More importantly, as a signal travels from one detector
 to the other, staying at a distance of order $L$ from the center, it will get delayed by an amount 
$ \delta t = \delta L \sim { G m \over L^{D-4} } $ (see appendix A).  
 In $D\geq 5 $, we can make this as small as we want by moving out to large enough $L$. 
 For $D=4$ this delay does not go to zero and we have a problem. This problem has been discussed in 
 \PenroseCausality , and it seems closely related to the soft graviton issues that arise in perturbative attempts to define the 
 gravitational S-matrix. This issue is not present in $AdS_4$. 
  In order to deal with the $D=4$ case, Gao and Wald   \GaoGA\  have introduced another notion of
  causality saying that 
 {\it we cannot send signals faster than what is allowed by the asymptotic causal structure of the spacetime.}
In  general relativity, with the null energy condition, they argued that this notion of causality is respected, see
 \GaoGA\ for a precise statement of the theorem. 
 This holds in any number of dimensions. For $D>4$ this becomes the more naive notion of causality introduced above.
 Note that this is an asymptotic notion and we are not relying on the locally defined lightcone as in  \VeloUR .

 We expect that this notion of causality is actually a {\it requirement} for any theory of quantum gravity. 
 For asymptotically $AdS$ theories we expect that we should not be able to send signals through the bulk faster than through the boundary. 
 For theories of gravity that are dual to a quantum field theory in the boundary,  this is implied by   causality  in  the boundary theory. Also, since we expect that quantum gravity in asymptotically flat space is Lorentz invariant, then these time delays can lead to acausality or closed time-like curves (see appendix G).

 Another reason to require this notion is to ensure that Lorentzian wormholes, such as the one obtained form the eternal Schwarzschild 
 black hole and discussed in  \FullerZZA , 
 do not lead to causality violations in the ambient spacetime. This is required by the ER=EPR interpretation 
 of such geometries \refs{\IsraelUR,\MaldacenaKR,\MaldacenaXJA}.  
 
 In the rest of the paper we will {\it assume}  this notion of causality and derive constraints  on some 
 higher derivative corrections. 

 \subsec{Scattering Through a Plane Wave in General Relativity }
 
 A well-known general relativistic effect is that light going near a massive body (e.g. the Sun) would suffer a time 
 delay relative to the same propagation in flat space. This is known as the Shapiro time delay \ShapiroUW\ and constitutes one of the
 classical tests of general relativity (see appendix A).

We will recall here the derivation of this time delay in the shock wave approximation, which will be all we need for our 
purposes. 
First we review the basic properties of shock wave solutions in gravitational theories \AichelburgDH. 
This solution describes the  gravitational field of an ultrarelativistic particle in a generic theory of gravity \refs{ \DrayHA,\HorowitzBV}  and
is directly relevant for the high-energy scattering. 

\ifig\TimeDelay{ A particle creates a shock wave localized at $u=0$. A second probe particle propagates on the 
geometry and experiences a time delay, $\Delta v $. The two particles are separated along the transverse directions, which are suppressed in 
this diagram.} {\epsfxsize2.2in \epsfbox{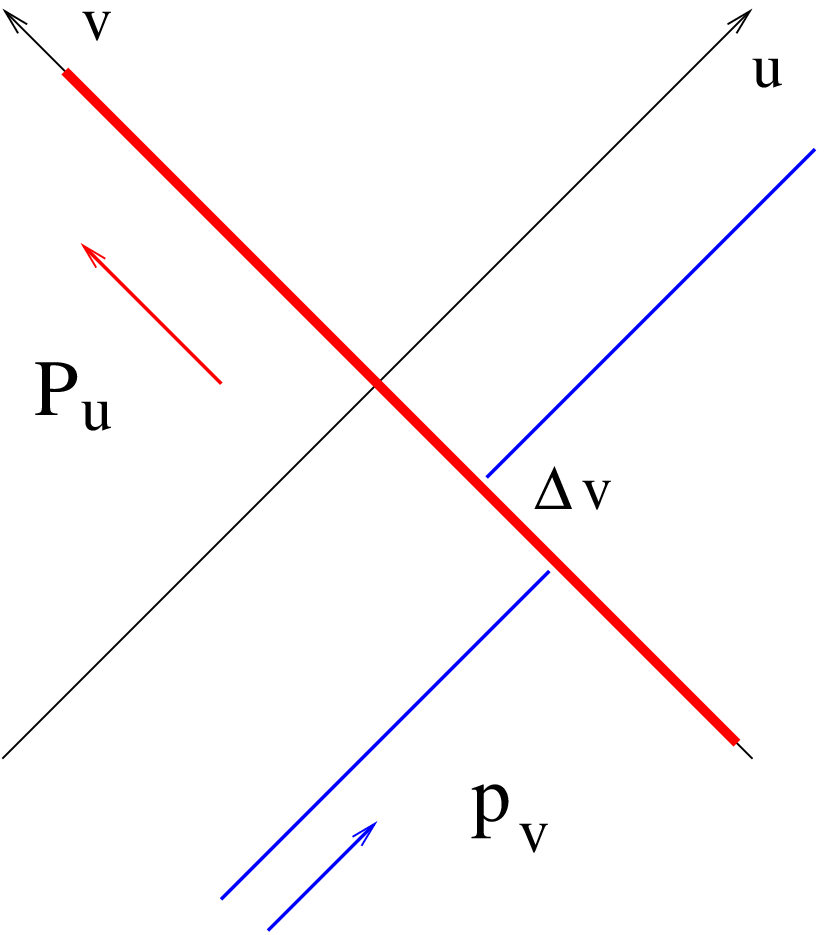} } 

A generic shock wave solution in flat space can be written in the following form
\eqn\shockwavesol{
d s^2 = - d u d v+ h(u,x_i)\, d u^2 + \sum_{i=1}^{D-2} ( d x_i )^2.
}
This geometry admits a covariantly constant null Killing vector $l^{\mu} \pa_{\mu} = \pa_{v}$. 
See \TimeDelay .

We will be interested in the case when this geometry is sourced by a particle that moves very fast in the
$v$ direction. Classically, we can model such particle via the stress tensor \DrayHA
\eqn\stresstensor{
T_{u u} = - P_{u} \delta(u) \delta^{D-2}(\vec x)
}
where $r = \sqrt{ \sum_{i=1}^{D-2} x_i^2}$ and $P_u < 0$ is the momentum of a particle. The Einstein equations then take the form
\eqn\translapl{
\pa^2_{\perp} h (u, x_i) = - 16 \pi G |P_u| \delta(u)  \delta^{D-2}(\vec x).
}
The solution is
\eqn\solution{
h (u, x_i) =  {4 \Gamma ({D - 4 \over 2}) \over \pi^{{D-4 \over 2}}} \delta(u) {G |P_u| \over r^{D-4} }.
}

Now we consider a probe particle that moves in the other light cone direction with momentum $p_v$ and is such that it crosses the 
shock with impact parameter $b$. In other words, the displacement in the transverse dimensions is $r = b$. 
The metric \shockwavesol\ is a bit peculiar because of the delta function $\delta(u)$ in $h (u, x_i)$. We can remove this delta function at the transverse position $b$ by defining a new coordinate 
\eqn\nerco{ 
v = v_{\rm new} +  {4 \Gamma ({D - 4 \over 2}) \over \pi^{{D-4 \over 2}}}  {G |P_u| \over b^{D-4} }   \theta(u)
}
This then cancels the $\delta(u)$ term in the metric at $r=b$. Thus, the geodesic that goes through this point is continuous in the $v_{\rm new}$ coordinates. 
This means that in the original coordinates it suffers a shift 
\eqn\shiftsu{ 
\Delta v = v_{\rm After~crossing} - v_{\rm Before~crossing} = {4 \Gamma ({D - 4 \over 2}) \over \pi^{{D-4 \over 2}}}  { G |P_u| \over b^{D-4} } > 0
}
This represents the Shapiro time delay, see   \TimeDelay . 
 
\ifig\shockwithoutdeflection{Left: we consider a particle propagating through the superposition of two left moving  shock waves localized at $u=0$. The particle trajectory is given by the arrows. 
 Right: in the transverse plane we separate shock waves by distance $2 b$ and send the particle between  them so that the net deflection angle  is zero. 
 The time delay is the sum of the time delays due to each shock wave.}
{\epsfxsize4.1in \epsfbox{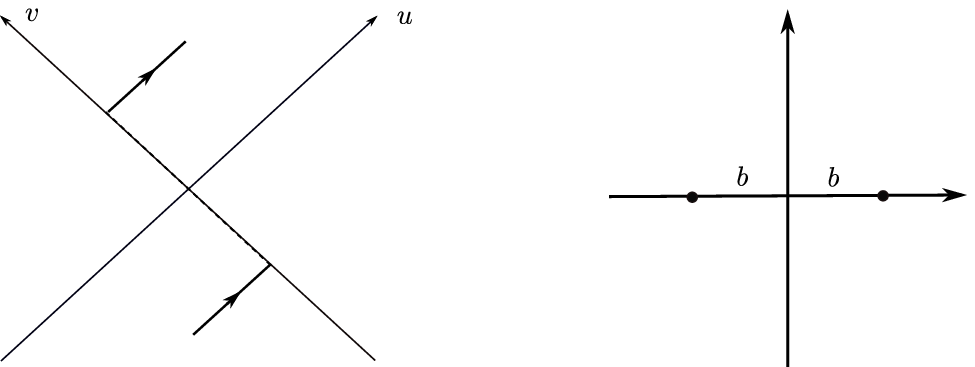} } 
 
In addition to this time delay, the trajectory is also subject to a deflection angle.
We might worry that the deflection angle would hamper our ability to see a possible time delay or time advance from far away. 
In fact, we could consider two shocks in succession, but separated in the transverse direction. Considering the probe particle coming at $r=0$, we set the shocks at $r =b$ opposite to each other as shown in   \shockwithoutdeflection. 
In this case, the probe particle does  not get a net deflection, but the time delays add. 
 
It is instructive to reproduce this formula for the time delay when we treat the probe as a quantum mechanical particle. 
 The wave equation for a scalar field  takes the form
\eqn\waveequation{\eqalign{
\nabla^2 \phi &=0 ~~~ \longrightarrow ~~~
\pa_u \pa_v \phi + h \pa^2_v \phi - { 1 \over 4 } \partial_i^2 \phi = 0,
}}
Let us now consider the change in the value of $\phi$ from $u=0^-$ to $u=0^+$. Since the variation of the $h$ term is much faster than the variation in 
the other variables, we neglect the $\partial_i$ derivatives and write
\eqn\soltic{ \eqalign{
\phi(u=0^+,v,x^i) &= e^{- \int_{0^-}^{0^+} du h  \partial_v } \phi(u=0^-,v,x^i) \cr
&= e^{- \Delta v\, \partial_v } \phi(u=0^-,v,x^i) = e^{- i \Delta v\, p_v } \phi(u=0^-,v,x^i)
}}
where $\Delta v$ is given in \shiftsu . 
Thus we see that we reproduce the answer we got through the geodesic analysis. Note that $p_v = -i \partial_v$ is the generator
of translations in $v$.  

In quantum field theory we would end the discussion here. Any time advance in $v$, relative to the background Minkowski metric would be a problem. 
In gravity, the situation is more subtle, in principle, we need to make observations from asymptotically far away. However, in that case, the $p_u$ energy also has 
a $p_v$ dependence\foot{See section 3 for the relevant kinematic configuration.} and it contributes positively to the time delay. Since $p_u = \vec q^2/4 p_v$, this is a small effect for large $p_v$. However, if we 
multiply by the total $u-$time elapsed, it can add to a very big time delay.   Here we will assume that we can sit far enough from the shock to be able to 
neglect the dynamics of spacetime, but close enough that we can neglect the $v-$time delay produced by the $p_u$ energy. This seems possible for small $G$.

\subsec{Connection with the Scattering Amplitude Computation}

Let us reproduce the computation above using scattering amplitudes  \tHooftRB . It is well-known that the shock wave computation can be reproduced using the so-called eikonal approximation \KabatTB. Consider the scattering amplitude for gravitating scalar particles. It is given by
\eqn\scatteringamplitude{
{\cal A}_{tree} (s,t) = - 8 \pi G {s^2 \over t}.
}
The eikonal approximation resums a particular set of diagrams (horizontal ladders) in the deflectionless limit when ${t \over s} \to 0$. 
Under favorable circumstances,\foot{This particular point will be discussed later in more detail.}  the amplitude   exponentiates in  
the impact parameter space (see e.g. \refs{\GiddingsXS,\CiafaloniESA} ) 
\eqn\eikonal{
i {\cal A}_{eik} = 2 s \int d^{D-2} \vec b \ e^{- i \vec q . \vec b} \left[ e^{i \delta (b,s)} - 1 \right]
}
where the phase is given by
\eqn\phaseeik{
\delta (b, s) = {1 \over 2 s} \int {d^{D-2}\vec q \over (2 \pi)^{D-2}} e^{i \vec q . \vec b} {\cal A}_{tree} (s, - {\vec q} ^{\, 2}) = {\Gamma ({D - 4 \over 2}) \over \pi^{{D-4 \over 2}}}  { G s \over b^{D-4} } .
}

This result matches the shock wave computation 
\eqn\deflectionless{
\delta(b, s) = - p_v \Delta v_{| r= b} ,
}
where we used $s = 4 P_u p_v$. As we will see below this picture can be naturally generalized to the scattering of particles with spin. 

\subsec{The Effect of Higher Derivative Interactions on Particles with Spin } 

If we had a photon propagating through the plane wave, it will have the same time delay we computed in \shiftsu . 
However, if the Lagrangian contains certain higher order interactions, such as
\eqn\conscte{ 
S = \int d^D x \sqrt{- g} { 1 \over 4  } \left[ F_{\mu \nu } F^{\mu \nu}  +  \hat \alpha_2  R^{\mu \nu}_{\ \ \sigma \delta } F_{\mu \nu} F^{\sigma \delta } \right] 
} 
then the second term gives rise to a time delay that depends on the polarization of the electromagnetic wave.
   The equations of motion take the form
\eqn\eomem{
\nabla^{\mu}F_{\mu\nu} - \hat \alpha_2 \, R_{\nu}^{\ \ \mu\alpha\beta}\nabla_\mu F_{\alpha\beta} = 0
}
In the shock wave background we have $R_{u i j u} = {1 \over 2} \pa_i \pa_j h$. For large $p_v$ \eomem\ is reduced to
\eqn\eomemb{
\pa_u F_{v i} + \left(\delta_{i j} h  + \hat \alpha_2 \pa_i \pa_j h  \right) \pa_v F_{v j}  = 0.
}
where we made use of the Bianchi identity in that limit, $\partial_u F_{vi}=\partial_v F_{vi}$; and the total time delay then has the form
\eqn\totaime{ \eqalign{
\Delta v & = \bigg[ 1  + \hat \alpha_2 { \eps^i \eps^j \pa_{b^i} \pa_{b^j} \over \eps^i \eps^i } \bigg] {4 \Gamma ({D - 4 \over 2}) \over \pi^{{D-4 \over 2}}}  { G |P_u| \over b^{D-4} } \cr
&= {4 \Gamma ( {D - 4 \over 2}) \over \pi^{{D-4 \over 2}}}  { G |P_u| \over b^{D-4} }
\left[ 1 + { \hat \alpha_2 (D-4) (D - 2) \over b^2 } \left( { (\eps.n)^2 \over \eps.\eps} - {1 \over D - 2} \right) \right]
} }
 where $\epsilon$ is the (real) transverse linear polarization direction of the electromagnetic wave. We also introduced $\vec n \equiv { \vec b \over b } $. 
The derivatives in \totaime\ come from the derivatives present in the Riemann tensor. The second term in \conscte\  can be viewed as a 
 spin dependent gravitational force.
 
 We see  that as  $b$ becomes small,    $b^2 <  \hat \alpha_2$,   the second term in  \totaime\ can overwhelm the first and, depending on the sign of $\hat\alpha_2$ and the polarization, may lead to time advance instead of time delay. 
 If the polarization is along $\vec b$, it has
 one sign, while if the polarization vector is orthogonal to $\vec b $,  $\vec \epsilon . \vec n =0$, then it has the other sign. If the polarization is a linear combination
 of these two,   we first decompose
 the wave in linear polarizations along and perpendicular to $\vec b$ and then exponentiate \totaime\ for each of these two cases separately, and then add the two 
 results. In other words, the expression for the time delay is now a matrix that can be diagonalized by choosing the polarizations to be along $\vec b$ or perpendicular
 to $\vec b $.     
 In conclusion, for either choice of the sign of $\hat \alpha_2$, there is a choice of polarization that can lead to time advance.

 Thus, if we require the theory to be causal, we see that $\hat \alpha_2$ should be set to zero. More precisely, it should be    small enough so that
 the computation we did above breaks down for some reason.
  An example of a theory where the  
 coupling  in \conscte\  arises at tree level is bosonic string theory \refs{\StiebergerHQ,\KawaiXQ}.  
  We will see later that in string theory the potential causality problem is fixed by the presence of extra massive states.

 As another example, let us consider the Gauss-Bonnet theory. This consists of the usual gravity action plus a specific $R^2$ interaction of the form  
\eqn\GBactionflat{
S = {1 \over 16 \pi G} \int d^D x \sqrt{- g} \left(R + \lambda_{GB} \left[ R_{\mu \nu \rho \sigma} R^{\mu \nu \rho \sigma} - 4 R_{\mu \nu} R^{\mu \nu} + R^2 \right] \right) .
}

The term in brackets is a topological invariant in $D=4$, but it is not topological for $D>4$. This theory has been extensively studied because it has the nice feature that the equations for small fluctuations around any background are second order \ZwiebachUQ .

As explained in \HorowitzBV\ the shock wave solution \shockwavesol\   is  also  an exact solution in the Gauss-Bonnet theory as well. 
We can consider propagation of a gravitational perturbation through 
the shock wave background. Before and after the shock the graviton moves as in flat space. All we need to know is what happens when it crosses the shock. 
 
We consider a high-energy graviton $\delta h_{i j}$ that propagates in the $v$ direction with momentum $p_v$ and traceless polarization in the transverse plane.
Near the shock we approximate the equations as 

\eqn\leadingGB{
\pa_u  \pa_v \delta h_{i j} + \left(\delta_{i k} + 4 \lambda_{GB} \pa_{i} \pa_{k} h \right) \pa_v^2 \delta h_{k j} = 0 
}

Using \leadingGB\ we can find the time delay which takes the following form 
\eqn\totaimeGB{ \eqalign{
\Delta v &= [ 1  + 4 \lambda_{GB} { \eps^{i k} \eps^{j k} \pa_{b^i} \pa_{b^j} \over \eps . \eps } ] {4 \Gamma ({D - 4 \over 2}) \over \pi^{{D-4 \over 2}}}  { G |P_u| \over b^{D-4} } \cr
&= {4 \Gamma ( {D - 4 \over 2}) \over \pi^{{D-4 \over 2}}}  { G |P_u| \over b^{D-4} }
\left[ 1 +  {4 \lambda_{GB} (D-4) (D - 2) \over b^2 } \left( { (\eps . n)^2 \over \eps.\eps} - {1 \over D - 2} \right) \right]
} }

Again, by choosing different polarizations we can get time advance for $b^2 \sim |\lambda_{GB} | $ for any sign of $\lambda_{GB}$. Notice that the formula for the time delay is  very similar
 to the ones discussed in the context of energy correlators in AdS/CFT with the parameters depending on the impact parameter of scattering $b$ (see e.g. \BuchelSK). We will see below that in the case of AdS the causality constraint interpolates between the usual energy correlator bounds and the flat space bounds obtained above. 

So we see that imposing positivity in all channels exclude $\lambda_{GB}$ completely unless new physics kicks in.  
In other words, the purely Gauss-Bonnet theory \GBactionflat\ is acausal.  As an aside, the Gauss-Bonnet theory was found in 
\SarkarXP\ to violate the second law of black hole thermodynamics in certain processes.\foot{More precisely, \SarkarXP\ considered the compactification of 
the Gauss-Bonnet theory to four dimensions on a circle. Then the Gauss-Bonnet term becomes topological. This leads to a constant contribution to the black hole
entropy. The sign of this constant contribution depends on the coefficient $\lambda$. When this contribution is negative, a small enough black hole can have negative
entropy. When the constant is positive,  a merger of two black holes can violate the second law. The constraints arise when the Schwarzschild radius $r_s^2 \sim \lambda_{GB}$.
 In this sense they are similar to the ones we have.  In both cases one needs $\lambda_{GB} \ll l_{Planck}$ in order to have a meaningful statement. 
\SarkarXP\  has the nice feature of also constraining purely topological terms in four dimensions. 
On the other hand, they might be modified by higher derivative corrections to the action, which could also modify black hole thermodynamics.  } 
 
We could also consider fermions coupled to gravity. At the level of the on-shell scattering amplitudes there is an additional spin-flipping structure that leads to time advance. This additional structure does not come from any known local Lagrangian and is known to be ruled out by considering consistency conditions imposed on the four-point amplitude \McGadySGA , at least, in four dimensions. If the same is true in all dimensions then the effects that we are discussing could not be observed for fermions. We leave exploration of this point for the future. Note that  having an explicit local Lagrangian which leads to second order equations of motions guarantees that all consistency conditions for the four-point amplitude
imposed in  \McGadySGA\ are obeyed. Thus, none of the problems discussed in \McGadySGA\ arise in the case of photons or gravitons with the modified
three-point functions that we are discussing.

 \newsec{General Constrains on the On-Shell Three-Point Functions} 
 
 The examples we have discussed so far have shown that there are causality issues with specific theories. 
 In this section we will isolate the crucial elements that produce the problem. It turns out that these causality problems 
 are produced purely and exclusively by the form of the {\it on-shell} three-point functions of the theory. Therefore they are 
 insensitive to higher order contact terms. First we will explain why the three-point functions determine the time delay. 
 Then we will recall some of the properties of three-point functions. 
 Finally, we will present a thought experiment that makes the causality violation more manifest.

 \subsec{The Phase Shift in Impact Parameter Representation From Three-Point Functions }

Let us consider the tree-level four-point amplitude ${\cal A}_4$. It depends on the kinematic invariants that we can produce with the four on-shell conserved momenta and polarization tensors of the external particles. Since it is a tree-level amplitude its only singularities are poles in the $s$, $t$ and $u$ Mandelstam variables. 
 We can now consider external momenta such that $s\gg t$, but $s$ small enough that the theory is still weakly coupled. 
 We can take the first incoming particle to have very large momentum along $p_u$ and the second with large momentum $p_v$. 

 \ifig\kinematics{Kinematics of the two-to-two scattering that we are interested in. Particle 1 has a very large momentum $p_u$ and particle two has a very large
 momentum $p_v$.}  {\epsfxsize2.4in \epsfbox{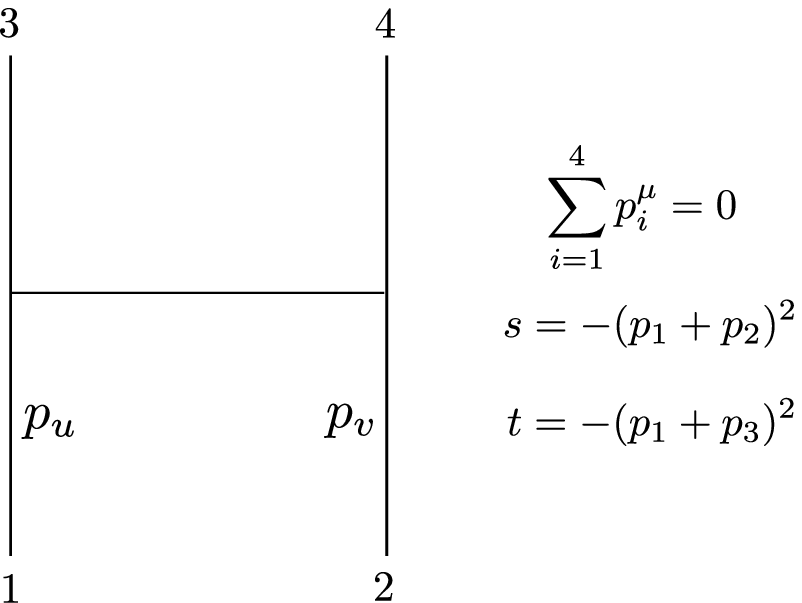} } 
 
 One can show that, in impact parameter space the amplitude is given by  (for review see appendix C)
 \eqn\phaseshift{
\delta (\vec b, s) = {1 \over 2 s} \int {d^{D-2}\vec q \over (2 \pi)^{D-2}} e^{i \vec q . \vec b} {\cal A}_4 ( \vec q) .
}
where ${\cal A}_4(\vec q) $ is really a short-hand for the four-point amplitude evaluated in the following momentum configuration 

\eqn\momconf{ \eqalign{ 
p_{1 \,\mu} = & \left(p_u, { q^2 \over 16 p_u} , {\vec q \over 2 } \right)~,~~~~~p_{2 \, \mu} =  \left({ q^2 \over 16 p_v} ,p_v,- { \vec q \over 2} \right)~,~~~~~~~
\cr
p_{3 \, \mu} = & -  \left( p_u,  { q^2 \over 16 p_u} , - { \vec q \over 2}\right) ~,~~~~~
p_{4 \, \mu} =   -  \left({q^2 \over 16 p_v} , p_v , { \vec q \over 2 }\right) ~,~~~~~~~
\cr
s \simeq & 4 p_u p_v ~,~~~~~~t \simeq   - ( \vec q)^2   
}}
where in all cases we indicated only the leading order term in the $t/s$ expansion, assuming $t/s \ll 1$. 

If we had scalar particles, the amplitude would depend only on the Mandelstam invariants and we can 
write   ${\cal A}_4( \vec q) = {\cal A}_{4}(s,t=- \vec q^{\, 2 } ) $. However, in the case of particles with spin, the amplitude depends also on 
the polarization vectors contracted with the various momenta. 

Let us assume that $\vec b$ in \phaseshift\ is chosen along $\vec b = (b, 0,0,\cdots)$, $b = |\vec b|$. Then let us consider the integral over the first 
component of   
$\vec q$, call it $q_1$.  Due to the exponential factor in \phaseshift\ this integrand is suppressed if we give $q_1$ a positive imaginary part. 
Here we assumed that the amplitude does not grow exponentially. This is true if we consider particles with polynomial interactions. 

Setting $q_1 = i \kappa+$real, we see that the exponential in \phaseshift\ is suppressed in this region as $e^{ - \kappa b}$, $\kappa>0$. Thus we would 
get a vanishing result (in the large $\kappa$ limit) unless we cross poles under this contour shift.  In fact, we do cross poles. For example the pole at $t=m^2$ 
coming from the exchange of a particle of mass $m$ in the $t$-channel leads to a pole at
\eqn\condp{ 
\kappa^2_* = m^2 + (\vec q_{\rm rest} )^2 
}
where $\vec q_{\rm rest}$ are the rest of the components of $\vec q$ except the first one. These are still real. 
The residue of the pole is given by a product of on-shell three-point functions. These three-point functions are non-vanishing because the intermediate momentum has one imaginary and one real component. Thus, we can think of the whole on-shell three-point function as being in two time directions. 
Note that this is a particular computation where the on-shell three-point function is relevant in ordinary signature. 
Notice that somewhat abstract notion of complex factorization in mixed signature \refs{\BrittoFQ,\SchusterNH} spacetimes has a direct physical meaning in the context of computing scattering amplitudes in the impact parameter representation.

The pole \condp\  gives a contribution to the amplitude  of the form 
$e^{ - \kappa_* b}$. For  a massive particle this gives something going like $e^{ - m b}$ for large $m b$, a Yukawa-like potential.\foot{
More precisely, after we integrate over the rest of the components of $\vec q$  we get the following expression $(2\pi)^{2-D\over 2} \left( { m \over b} \right)^{D-4\over 2 } K_{D-4 \over 2 } ( m b)$.}
For a massless particle, the integral over the rest of the components of $\vec q$ produces an inverse power of $b$, $1/b^{D-4}$.  
In addition, factors of $\vec q$ which are contracted with the  polarization tensors give derivatives with respect to $\vec b$. In a theory which only contains a massless graviton we just get the 
massless graviton pole. 

It should be noted that in impact parameter representation, for non-zero $b$,  we only get a contribution from the diagram that contains 
an on-shell particle in the $t$-channel.  In particular, $s$-channel exchanges do not contribute. The reason is that the two incoming particles have to actually overlap 
in order to give rise to the intermediate particle in the $s$-channel. Similarly, a four point contact interaction does not contribute for the same reasons. 
In both of these remarks we used that we are looking at the tree level diagrams to leading order in the weak coupling expansion. At higher orders there can be
other contributions. However, since we are at weak coupling, we can ignore them.  In these paragraph, we have used that the interactions are local. Any non-local 
effect has to come from the propagation of a physical particle in the $t$-channel.\foot{ If our theory contains extended objects, such as strings, we can 
actually create particles in the $s$-channel by going to small enough (but non-zero) impact parameter.  We discuss this below. }

\ifig\OnShell{  At fixed impact parameter we find a contribution to the amplitude given by the on-shell propagator in the intermediate channel, labeled by the letter $I$.
 The contribution involves 
a product of two on-shell intermediate amplitudes, ${\cal A}^{13 I}$ and ${\cal A}^{I 24 } $, which are circled in the figure. Particles 1 and 3 carry a large momentum $p_u$ 
and 2 and 4 have a large $p_v$. We get a factor of $s^J$ from contracting $p_u$ on the left side with $p_v$ on the right side through the sum over 
polarizations of the intermediate particle.   }  {\epsfxsize2.0in \epsfbox{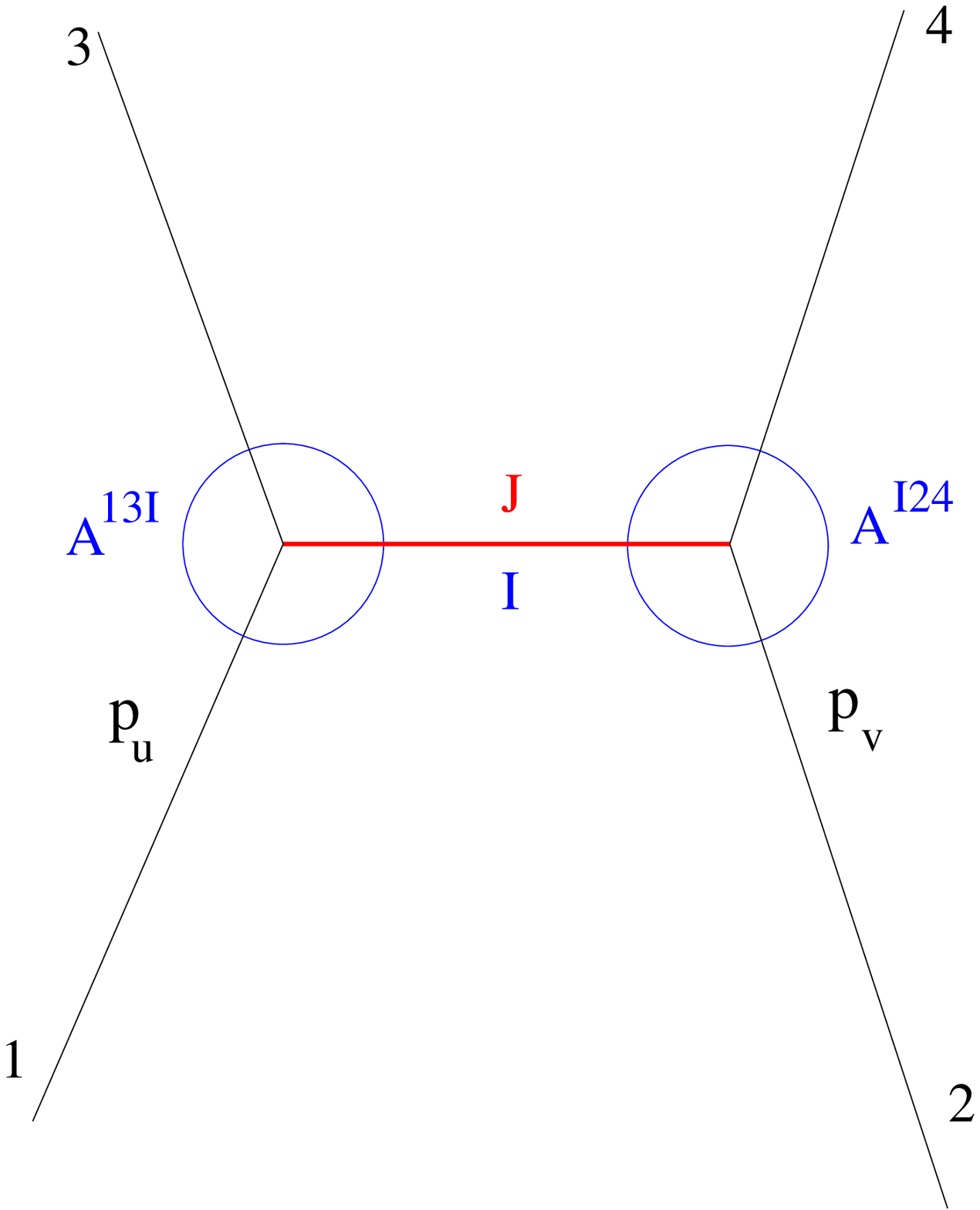} } 

We are interested in the  part of the amplitude that increases fastest with the Mandelstam variable $s$.   Since there are no kinematic invariants in the 
on-shell three-point amplitudes, a factor of $s$ can only come
from factors of $p_u$ or $p_v$  that are in each of the  three-point function contracted, via the sum of the polarizations of the intermediate particle, see figure
\OnShell  .  
With a particle of spin $J$, the maximum power we can get in the amplitude is ${\cal A}_4 \sim s^J $. This leads 
to a factor  of $s^{J-1}$ in the phase shift \phaseshift. In fact, this sum over polarizations over the intermediate state receives a large contribution only from 
one specific polarization, when the intermediate state has all $u$ indices. In other words, the polarization tensor in the intermediate state appearing in the 
left three-point function has the form $\epsilon^{I \, u \cdots u } $ and the one in the right three-point function is $\epsilon^{I \, v\cdots v} $. These are 
contracted with the corresponding factors of the large momentum to give $ (p_u p_v)^J \sim s^J$. 

The polarization tensors of the external particles can be written as products of vectors $\epsilon_{\mu \nu}  = \eps_\mu \eps_\nu $ where for particles one
and three we can write  
\eqn\canrwi{ 
\eps^{ 1\, \mu } = (-{ \vec q . \vec e_1 \over 2 p_u},0 , \vec e_1 ) ~,~~~~~\eps^{3\, \mu } =  ( { \vec q . \vec e_3 \over 2 p_u},0 , \vec e_3 )
}
where $\vec e_{1,3}$ are vectors in the purely transverse directions. 
The on-shell three-point functions will contain factors of the form 
\eqn\factfo{
 \eps^1 . \eps^3 =  \vec e_1 . \vec e_3 ~,~~~~ \eps^1. p^3 = \vec q . \vec e_1 ~,~~~~\eps^3 . p^1 = \vec q . \vec e_3
}
The conclusion is that we can think of the polarizations of the external particles as contained effectively in the transverse space. In addition, when 
we contract them
with the external states we get factors of $\vec q$. These translate into derivatives with respect to $\partial_{\vec b }$.

The final result from a massless pole is
 \eqn\phaseshifta{\eqalign{
\delta (\vec b, s) &= {  {\cal A}^{13I}_3 ( - i \pa_{\vec b}) {\cal A}^{I24}_3 ( - i \pa_{\vec b}) \over 2 s}  \int {d^{D-2}\vec q \over (2 \pi)^{D-2}} { e^{i \vec q . \vec b} \over \vec q^2 } \cr
&= {\Gamma ({D - 4 \over 2}) \over 4 \pi^{{D - 2 \over 2}}}{  {\cal A}^{13I}_3 ( - i \pa_{\vec b}) {\cal A}^{I24}_3 ( - i \pa_{\vec b}) \over 2 s} {1 \over |\vec b|^{D-4}} .
}}
where the three-point functions are evaluated on three momenta \momconf . As explained above, in each three-point function there is only one relevant polarization
for the intermediate state that produces the factor of $s^J$. 
The polarizations of the external particles can be viewed as living purely in the transverse space, after we use \canrwi\ and \factfo . 
  
In this way we can compute  $\delta (\vec b , s)$ using on-shell methods. For more details  see appendix B. 
 This  gives the phase shift to leading order in perturbation theory. The answers agree  with 
  with the infinitesimal form of the phase shift computed using the shock wave computation. In fact, the shock wave itself can be viewed as the on-shell graviton intermediate state we discussed above. The fact that the metric contains only the $h_{uu}$ component is related to the fact that only one polarization of the
  intermediate state produces the most factors of $s$.

 \subsec{The Possible Forms of Three-Point Functions in Various Theories} 
 
Given that the answer for the time delay depends on the on-shell three-point functions, it is useful to recall their possible structures. In four dimensions we can use the usual helicity basis. The Einstein-Hilbert gravity action gives rise to the $++-$ and $--+$ three-point functions. We can  also have  $+++$ and $---$ structures which could come from (Riemann)$^3$ terms. There are two combinations, one of them being parity violating. 

In higher dimensions, $D>4$, we have three possible structures for the graviton three-point functions, all parity preserving. They have the schematic structure, writing $\epsilon_{\mu \nu} = \eps_\mu \eps_\nu$,  
 \eqn\schem{  \eqalign{
  {\cal A}_{R} = &  ( \eps_1 . \eps_2 \eps_3 . p_1 + \eps_1 . \eps_3 \eps_2 . p_3 + \eps_2 . \eps_3 \eps_1 . p_2)^2 \cr
 {\cal A}_{R^2} = &  ( \eps_1 . \eps_2 \eps_3 . p_1 + \eps_1 . \eps_3 \eps_2 . p_3 + \eps_2 . \eps_3 \eps_1 . p_2) \eps_1.p_2 \eps_2 . p_3 \eps_3 . p_1 \cr
  {\cal A}_{R^3} = & (\eps_1.p_2 \eps_2 . p_3 \eps_3 . p_1)^2 .
}}
The first is the usual one coming from Einstein gravity, while the second can arise from the Lanczos-Gauss-Bonnet term \GBactionflat. The third one can arise from a  (Riemann)$^3$ term.\foot{(Riemann)$^3  = R_{\mu \nu \sigma \delta} R^{ \sigma \delta \rho \gamma} R^{~~\mu \nu }_{\rho \gamma }$. In four dimensions we can replace one of the curvatures by $R_{\mu\nu \sigma \delta} \to \tilde R_{\mu\nu \sigma \delta } = \epsilon_{\mu \nu \rho \gamma} R^{\rho\gamma}_{~~\sigma \delta } $ to obtain the parity violating term. This parity violating term gives rise to the three-point vertex 
 $\epsilon_{\mu\nu\delta \sigma} p_1^\delta p_2^\sigma \epsilon_{1,\mu\mu'} \epsilon_{2,\nu \nu'} p_2^{\mu'} p_1^{\nu'} \epsilon_{3,\gamma \rho} p_1^\rho p_2^\gamma $, which, despite appearances, is properly symmetric under exchange of any of the three particles. When this term is present the coefficient of the 
 $+++$ amplitude is complex and the coefficient of the $---$ amplitude is the complex conjugate. }
They can be viewed as products of the two possible structures that we can have for on-shell spin one particles. 
 In a given theory, the total three-point function is given by a linear combination of  these three answers 
 \eqn\totalr{ 
 {\cal A}_{ggg} = \sqrt{32 \pi G} \left[ {\cal A}_{R} + \alpha_2 \, 
 {\cal A}_{R^2} + \alpha_4 {\cal A}_{R^3}  \right]
  }
  where $\alpha_2$ and $\alpha_4$ are two parameters with dimension of  (length)$^2$ and (length)$^4$ respectively.
   Notice that we have an overall coupling, given by $G$, which we take
  to be parametrically smaller than the other two parameters. 

  In the high energy limit the three-point functions appearing in \phaseshifta\ simplify further and become 
  \eqn\schembet{ \eqalign{ 
  {\cal A}^{13 I}_{R} = &  2 p_u^2 (\vec e_1. \vec e_3)^2
    \cr
    {\cal A}^{13 I}_{R^2} = & 2 p_u^2  (\vec e_1. \vec e_3) ( \vec q . \vec e_1) (\vec q . \vec e_3) 
    \cr
     {\cal A}^{13 I}_{R^3} = & 2 p_u^2   ( \vec q . \vec e_1)^2 (\vec q . \vec e_3) ^2
    }}
  for the three terms in \schem . We used \canrwi . 
  The amplitude ${\cal A}^{I24}$ has a similar expression with $p_u \to p_v$ and $1,3 \to 2,4$.  The reader can express these in terms of transverse 
  traceless two index tensors by using the replacement rule $  e_i e_j \to e_{ij} $ in \schembet , where $i,j$ are indices in the $D-2$ transverse
  directions. 
    
  Note that the third structure in \schem\ or   \schembet\ is not allowed in a supersymmetric theory. This can be seen as follows. In $D=4$ this structure 
  gives rise  to $+++$ and $---$ amplitudes. However,  by the methods described in   \GrisaruVM\ it is possible to show that supersymmetry implies that this 
  structure should be set to zero. This is actually true in all dimensions. The reason is that we can start with a theory in $D$ dimensions and consider external graviton 
  three-point functions with four-dimensional kinematics. If we had a non-zero contribution from the third structure in $D$ dimensions, then it would lead to 
  a contribution to the $+++$ or $---$ amplitudes in four-dimensional kinematics. Since the arguments in \GrisaruVM\ are purely kinematical, based on the 
  symmetries of the theory, then they also force the amplitudes to vanish. In conclusion, supersymmetry implies that $\alpha_4 =0$. 
  In the heterotic string we have a non-zero $\alpha_2$. And this is also true for compactifications of the string to $D$ dimensions. Thus $\alpha_2$ is compatible 
  with half maximal supersymmetry.  With maximal supersymmetry, e.g. ${\cal N} =8$ in $D=4$, we should also have $\alpha_2 =0$. This can be understood 
  from the fact that the four point amplitude for the full supergravity multiplet is determined up to a unique function. Thus, there is no freedom to introduce the 
  polarization dependent terms that would arise if we had the freedom to switch on $\alpha_2$. Of course, in standard maximal supergravity
   $\alpha_2$ is not present, therefore
  the only value consistent with maximal supersymmetry is $\alpha_2 =0$. As an aside, notice that this is also related to the fact that in four-dimensional 
${\cal N}=4$ superconformal theories, the three-point functions of the stress tensor are completely fixed by supersymmetry in terms of the two-point functions.

Between a graviton and two photons the number of possible three-point functions is two
\eqn\twostruct{\eqalign{
{\cal A}_{F^2}^{13I} &=\eps^I_{\mu \nu} \left[p_1^{\mu} p_3^{\nu} (\eps_1 . \eps_3) -  \eps_1^{\mu} p_3^{\nu} (\eps_3 . p_1) - \eps_3^{\mu} p_1^{\nu} (\eps_1 . p_3) \right], \cr
{\cal A}_{R F F}^{13I} &=\eps^I_{\mu \nu}  p_1^{\mu} p_3^{\nu} (\eps_1 . p_3)  (\eps_3 . p_1) ,
}}
so that ${\cal A}_{g \gamma \gamma} = \sqrt{32 \pi G} \left[ {\cal A}_{F^2} + \hat \alpha_2 \, 
 {\cal A}_{R F F} \right]$.
 One arising from the usual electrodynamics and the other from the second term in \conscte . Again this second structure is forbidden in a supersymmetric theory. 
  As we did in \schembet , in the high energy limit we can write them as 
 \eqn\writ{\eqalign{ 
  {\cal A}^{13 I}_{F^2} = &2  p_u^2 (\vec e_1. \vec e_3) ,
    \cr
 {\cal A}^{13 I}_{R FF} = &2 p_u^2    ( \vec q . \vec e_1) (\vec q . \vec e_3) .
 }}
 
 We emphasize  that we work here with the on-shell three-point functions, independently of the precise way we write the 
 Lagrangian. This discussion depends only on on-shell three-point functions and not on other contact terms. Any contact four point interaction does not
 give rise (at tree level)  to the long range force at a non-zero value of  the impact parameter. 
 
 \subsec{Problems with Higher Derivative Corrections to the Three-Point Functions} 
 
We discussed above how the three-point functions give rise to the leading order expression for the phase shift $\delta(\vec b,s) = s F(\vec b)$. 
 If this result were exponentiated, as $e^{ i \delta } $, then we could get a time advance  problem similar to what we found for the shock waves. 
 Here we would like to explain how to get a time advance problem without using the particular non-linear structure of shock waves. The goal is to 
 present the problem in a way that depends only on very general principles. 
 
 \ifig\shockcomb{We imagine a particle going through a set of successive scattering events. The intrinsic quantum uncertainty in $v$ is $\Delta_q v$. 
 We have drawn a situation where there is a final   time advance after going through all the shocks that is larger than the quantum uncertainty. 
  In this figure we have neglected the delay of the $u$-localized 
 particles.  }  {\epsfxsize2.9in \epsfbox{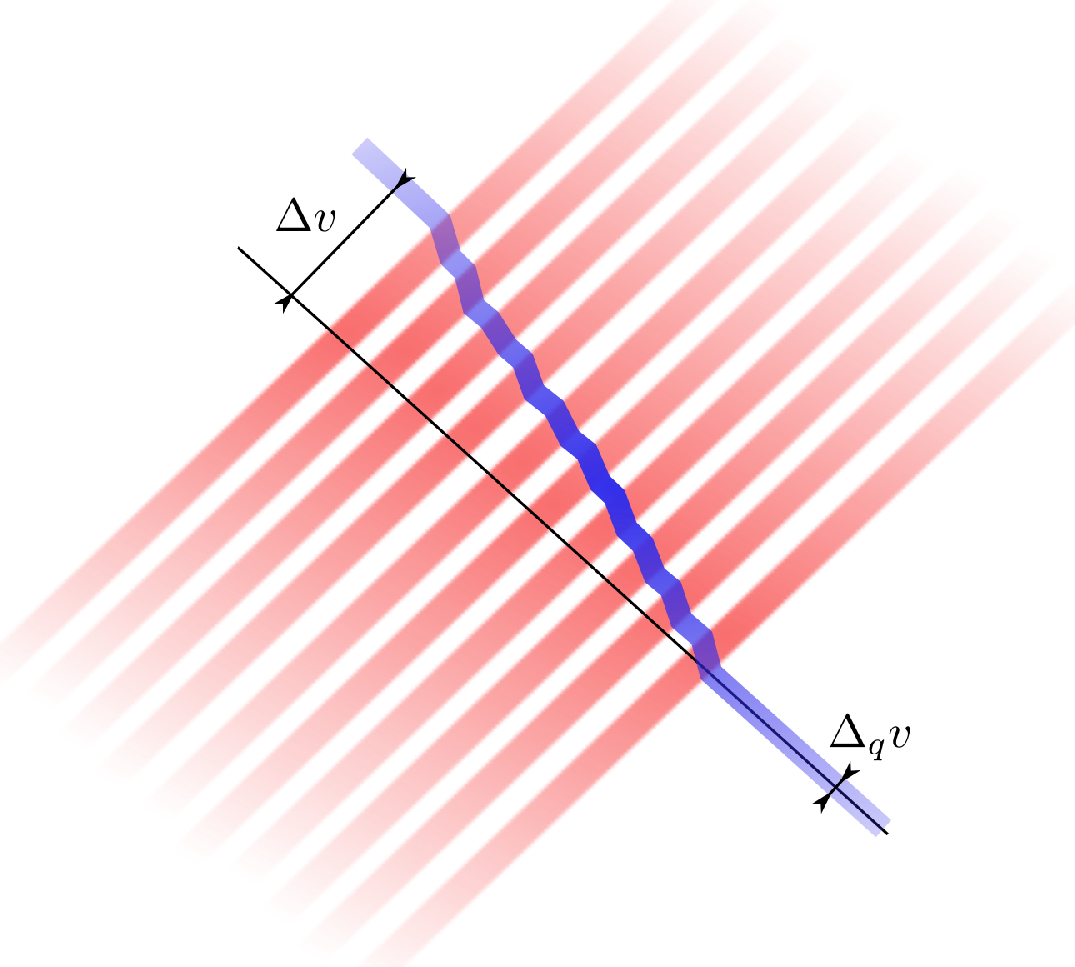} } 

 First note that in order for time delay to be a problem we would like to find that the time delay $\Delta v = \partial_{p_{2, v } } \delta$ is 
 larger than the quantum mechanical uncertainty that is implicit in the definition of the wave packet for a particle of momentum $p_{2,v}$. 
 This uncertainty is of the order of $\Delta_{q} v \sim 1/p_{2,v}$. Thus the figure of merit is 
 \eqn\figmet{ 
 { \Delta v \over \Delta_q v } = p_{2, v}  \partial_{p_{2,v} } \delta = \delta 
 }
 here we used that $\delta$ is linear in $s$, and therefore linear in $p_{2,v}$ ($s \propto p_{1,u} p_{2,v}$).
 Thus in order to see a problem, we expect that $\delta $ should be greater than one. On the other hand, the validity of perturbation theory 
 suggests that $\delta$ should be much less than one. 
 
 In order to amplify the effect, we imagine that particle number  two undergoes $N$ successive instances of particle number one, 
 see \shockcomb . Through each instance it gets a 
 small phase shift which leads to a  factor in the out state of the form $(1+ i\delta)$ with small $\delta$. 
 If we repeat this $N$ times we get
 \eqn\getcn{ 
 ( 1 +i \delta )^N \sim e^{ i N \delta }  ~~~~~~~\delta \ll 1 ~,~~~~N \gg 1 .
 }

 This is the total phase shift, and as explained in \figmet, we want $N \delta$ to be of order one. This can be achieved by taking $N \sim 1/\delta $. 
 In addition, we would like to make sure that the approximations that we used remain valid. In particular, we have said that particle 2 remains localized at 
 some distance $b$ through the whole process. We can choose light-cone coordinates for the evolution of particle two so that $u$ is time, then we have a non-relativistic
 problem with mass $m \sim p_v$. The spreading of the wavefunction during the time $U$ that the whole process takes is $\Delta b \sim \sqrt{ U/p_v}$. 
 The time $U$ that the process takes is determined as follows. We want to separate the $N$ instances of the scattering process from each other so that
 we can view them as independent and we can use \getcn . 
 The best we can localize each of the $N$ particles of type 1 is by an amount $\Delta_q u \sim 1/p_u$. Thus $U = N/p_u$ and then 
 $\Delta b \sim \sqrt{ N \over s } $. We want $\Delta b \ll b$. This translates into the condition 
 \eqn\condit{ 
  N \ll s b^2  ~~~~ \longrightarrow ~~~~  { 1 \over s b^2 } \ll \delta 
  }
where we used that $N \sim 1/\delta$ in order to have a problem. We see that the simultaneous validity of \condit\ and \getcn\ can be achieved if 
$s b^2 \gg 1$ and if $s$ can also be increased so as to achieve 
\eqn\exponcond{ 
{ 1 \over s b^2 } \ll \delta \ll 1 
}
which can be done if $\delta $ grows with $s$. 
An additional issue is that we wanted to neglect the deflection angle. Then we can replace each of the 1 particles by a pair of particles localized in the transverse 
dimension at a distance $b$ with respect to particle number 2 on opposite sides of its trajectory,
as shown in \shockwithoutdeflection .

  For spin zero particles, notice that if we had a $g \phi^3$ vertex, then the scalar exchange leads to 
 $\delta \sim  { 1 \over s } { g^2 \over b^{D-4}}  $. 
  In this case we cannot obey the conditions \exponcond \ to 
 obtain a possible causality problem. In fact, \condit\ implies that $g^2/b^{D-6} >1$, which means that we are at strong coupling. 
 Thus we see that a crucial feature that we used is that the phase shift increases as a function of $s$. In the case that we exchange a spin one field, $\delta$ is independent of $s$ 
 and there is also no causality issue. The reason is simple, the spin two field is effectively changing the metric and causal structure 
 while the spin zero or one fields are not.

The conclusion is that we have justified the exponentiation \getcn\ and thus the derivation of the time delay problem in a way that does not depend on the 
details of the shock wave solution. Let us emphasize that the preceding shock wave discussion was motivational, but the thought experiment that we have set up 
in this section does not require us to rely on the non-linear structure of the shock wave. It was all derived from the on-shell three-point functions plus certain 
assumptions about the locality of the theory that allowed us to view each shock as an independent event. 

In fact, there are some cases where    the shock wave computation does not give the right answer. For example, consider the pure gravity case, where 
$\delta = G s/b^{D-4}$. If the energy is large enough to form a black hole, then the time delay computed from the shock wave is not the 
correct description for the physics. We expect to form a black hole when the center-of-mass energy is such that the associated Schwarzschild radius 
$r_s^{D-3} = \sqrt{s} G$ is larger than $b$  \refs{\EardleyRE,\RychkovSF,\GiddingsXY}.
 It is easy to check that this never happens if \exponcond\ is obeyed. It is not obeyed  
even  if we consider the energy of the $N$ particles that we used for the argument  (with $N\sim 1/\delta$).   

There is still one more complication that we need to deal with in order to make the argument clearer. 
In the shock wave discussion of section two, the spin of the particle creating the shock did not matter. 
However, with the modified three-point functions, the spins of the scattered particles can change. The full 
interaction is a spin dependent force which acts of both spins. It acts on both the spin of the left and the right moving particles (particle one and two, see \kinematics ). 
I will be necessary for us to    be able to fix the polarizations of particles one and three, see \kinematics . 
This can be achieved by replacing particle one, by a coherent state of particles. In this case, due to the usual Bose enhancement factor, particle 
three will have a larger probability of remaining with the same polarization. In this set up we can set the spins, or polarization vectors, 
of particles one and three to be the same. Or, more precisely, $\epsilon_3 = \epsilon_1^* $. Since we are at weak coupling we can consider a coherent 
state with a mean occupation number which is large enough for us to be able to neglect the spin flips but small enough that the total scattering amplitude is 
still small. In other words, the use of coherent states allows us to effectively select the final state for particle 3. 
More explicitly, say that we form a coherent state for the oscillator mode created by $a^\dagger$, $ e^{ \lambda a^\dagger} |0 \rangle$. 
We could then have terms in the interaction Hamiltonian that  leave this  oscillator the same  $ H_{1,int} = h_1 a^\dagger a $ or that mix it with a second
oscillator $H_{2,int} = h_2 b^\dagger a + h.c. $. Here $h_1$ and $h_2$ can act on other degrees of freedom. 
Then, for a coherent state, the matrix elements where there is no change are enhanced, due to the usual Bose enhancement factors, relative to the ones where
there is a change
\eqn\changefli{ \eqalign{ 
\langle 0 | e^{ \bar \lambda a } \,   H_{1,int}  \, e^{ \lambda a^\dagger} | 0 \rangle &\propto  |\lambda|^2 \langle h_1 \rangle 
\cr
\langle 0 | e^{ \bar \lambda a } b \,   H_{2,int}  \, e^{ \lambda a^\dagger} | 0 \rangle & \propto \lambda  \langle h_2 \rangle     \, 
}}
We see that for large $\lambda$ the first term is enhanced. We have not indicated explicitly the initial and final states on which $h_1$ and $h_2$ act, since they 
involve other oscillators. 
Since we are at weak coupling, we can choose $\lambda$ large enough so that the first term dominates
relative to the second term, while still the whole process is in the weakly coupled approximation, or the total effect of the interaction Hamiltonian is small. 
Of course an alternative way to say this is that we are creating a classical background with the particles of type 1 in \kinematics\ the terms that have a non-zero 
expectation value in this classical background dominate over the others. We want a classical background with small enough amplitude that we can still trust the 
leading order perturbative expansion of the interaction Hamiltonian. 

The use of coherent states also allows us to select final states for particle 3 in \kinematics\ with a a small momentum $p_v$. Namely,  we form a coherent state
out of a superposition of particles with large momenta $p_{1,\mu}$. We need a superposition since we need to localize this particles within the transverse plane to 
a location smaller than $b$. Thus we have some dispersion in the transverse momenta $\vec q$. With a large $p_u$ component of the momentum, we then 
get a small momentum along the $p_{v} = { {\vec q}^{\,2} \over 4 p_u }$ direction. Since we have a coherent state, the particle 3 is also taken out of this superposition and
has the same range of values for the momentum. Therefore the total momentum transfer in the $t$-channel along the 
$v$direction is very small. Then the kinematics chosen in \momconf\ 
is representative for the process in question. This still allows a possibly large amount of momentum transfer along the $u$ direction.
 We will discuss this in subsection 4.3 . 

Another minor point,  is that the phase shift $\delta$ represents a time delay for both particles, it affects both particle 1 and 
particle 2. So far we have been focusing only on the effects on particle 2. These coherent states also allow us to effectively select a final state for particles 3, so that
we can focus more clearly on the time delay with which particle four emerges, see \kinematics . 

\subsec{Scattering of Gravitons in $D>4$ Dimensions} 

It is also easy to argue that $\alpha_2$ and $\alpha_4$ structures lead to causality problem in $D>4$. To show this we consider the probe graviton
that scatters off the coherent state. The phase shift takes in this case the following form (see appendix B for details)
\eqn\phaseshiftgenD{\eqalign{
\delta &\sim  Gs ( e_1^{i j} e_3^{i j} + \alpha_2 e_1^{i j} e_3^{i k} \pa_{b^j} \pa_{b^k} + \alpha_4 e_1^{i j} e_3^{k l} \pa_{b^i} \pa_{b^j} \pa_{b^k} \pa_{b^l}) \times \cr
& (e_2^{i j} e_4^{i j} + \alpha_2 e_2^{i j} e_4^{i k} \pa_{b^j} \pa_{b^k} + \alpha_4 e_2^{i j} e_4^{k l} \pa_{b^i} \pa_{b^j} \pa_{b^k} \pa_{b^l}) {1 \over |\vec b|^{D-4}}.
}}

In order to find problems we will be choosing various polarizations for the particles. For example, let us firt consider particle 1  with polarization
 $e^1_{xx} =-e^1_{yy} $ and the other components equal to zero. Here $ x,y$ represent to two directions in the 
 transverse plane.  We call this the $\oplus$ polarization. 
 We also choose $e^3 = e^1$. We enforce this by 
 sending   a coherent state with this polarization. 
 Now for particle 2 we can choose the same polarization or the crossed polarization, called $\otimes$, 
  given by $e^2_{xy} = e^2_{yx} =1/\sqrt{2}$ and all 
 other components equal to zero. 
 Then we find the following. If $\vec b $ is along the $\hat x$ direction, then these two different polarizations for particle two do not mix as they go through the shock.
 They diagonalize the phase shift matrix.
  For small enough $b$ the $\alpha_4^2$ terms dominate since they are the most singular in the small $b$ expansion. These terms have the form 
 \eqn\delayboth{ \eqalign{ 
 \delta_{\oplus \oplus} \, \sim \,  & G s      \alpha_4^2 O_{\oplus}  O_{\oplus}  \,  { 1 \over b^{D-4} } = { G s  \alpha_4^2 \over b^{ D+4} } ( {\rm positive} )
 \cr 
 \delta_{\oplus \otimes } \, \sim \,    &   G s       \alpha_4^2 O_{\oplus} O_{\otimes} \,  { 1 \over b^{D-4} } =  { G s  \alpha_4^2 \over b^{ D+4} } ( {\rm negative } )
 \cr
 O_{\oplus}  = & ( \partial_{b_x}^2- \partial_{b_y}^2 )^2  ~,~~~~~O_{\otimes} = 4 \partial_{b_x}^2 \partial_{b_y}^2 
}}
where we take the derivatives first and then set $\vec b = ( b_x , 0,\cdots ,0)$. Where the terms in parenthesis are polynomials in $D$ which are positive or negative
definite.\foot{ For $D>4$, (positive)$ = (D-4) (D^2-4) D (336 + 128 D + 20 D^2 + 4 D^3 + D^4 )$ and (negative)$= - 4 (D-4)(D^2-4) D (D+3)(20+6 D + D^2)$. For
$D=4$ the derivatives act on $ - \log b$ and we get (positive) = -(negative) =80640.}
  Notice that the positivity of the first case is due to the following argument. 
 If the polarizations of particle 2 and 4 are the same as those of 1 and 3, then the configuration is constrained by 
 unitarity along the $t$-channel.\foot{$t$-channel unitarity is the  following statement. When the polarizations of 2 and 4 are related by conjugation and reflection
 along the $\vec b$ axis to the polarizations of 1 and 3 respectively, then the residue of the $t$-channel pole is positive.}  Therefore in this case we should get 
    a strictly  positive answer for the time delay for the contribution of any particle with a non-zero coupling. Since we obtained a negative time delay for the
    second case in \delayboth , we conclude that 
 $\alpha_4$ should be set to zero unless new particles are present. 

Once $\alpha_4$ has been set to zero, we can discuss $\alpha_2$. In 
that case we can still choose the $\otimes_{xy}$ polarization for particles 1 and 3 and the $\otimes_{yz}$ polarizations for particles 2 and four. We then focus on the
terms proportional to $\alpha_2^2 $ since they are the dominant terms at small $b$ (once we have set $\alpha_4 =0$).
We then get 
\eqn\resnow{ \eqalign{ 
\delta_{\otimes_{xy} , \otimes_{yz} } \, \sim \, &  G s \alpha_2^2 \hat O_{xy} \hat O_{yz } { 1 \over b^{D - 4   } }= - {Gs \alpha_2^2 \over b^{D} }  2 (D-4)(D-3)(D-2)
\cr
\hat O_{xy}  = & -\partial_{b_x}^2 - \partial_{b_y}^2 ~,~~~~~~\hat O_{yz}  =   -\partial_{b_y}^2 - \partial_{b_z}^2
}}
 
 Then we conclude that $\alpha_2$ should also be set to zero unless new massive particles appear.

\subsec{Scattering of Gravitons in Four Dimensions} 

Let us now discuss in more detail the four-dimensional case, $D=4$, which is a bit special.   
 First, need to take into account that the Einstein term produces a $\log(L/b)$ time 
delay, where $L$ is an IR cutoff. Second we need to take into account the parity violating structure. 
The logarithm can be taken into account by modifying the causality criterion in the form suggested by Gao and Wald, who define it by comparing to the 
behavior of the same metric far away. In this way the $\log L $ term is eliminated and it is easy for a power law behavior produced by $\alpha_4$, which goes as 
$1/b^4$ to overwhelm the logarithm. Also we will later repeat the computations for $AdS_4$ space and we will see that $L \to R_{AdS_4}$. In conclusion, this is 
not a real issue.

Note that in four dimensions the $\alpha_2$ structure is identically zero. This
 is related to the fact that the Gauss-Bonnet term becomes topological in four dimensions. Thus, we have only the Einstein term structure and the $\alpha_4$ one.
With four-dimensional kinematics, we can consider the situation with coherent states of particles of type 1. Let us choose the spin of these particles to be plus. 
Due to the coherent state considerations, the spin of particle 3 also needs to be plus (in the outgoing notation, or minus in the incoming notation). In other words the amplitude does not have a spin flip. In fact, without a spin flip the    $\alpha_4$ structure   does  not contribute in four dimensions.
 Thus in the vertex involving particles one, three, and the intermediate one, the only structure that contributes is the Einstein one. This contribution is effectively the same as the spin zero one. Then we can run an argument similar to the one above.

 Let us now discuss the parity violating structure, together with the parity preserving one. 
 By considering the coherent state for particle one (see \kinematics ),
  with definite helicity (positive or negative) we get that only the Einstein structure contributes to the 
 ${\cal A}^{13I}$ three-point amplitude (see \OnShell ).  The we get the following matrix form for the phase shift for particle two
 \eqn\phseficpa{ \eqalign{ 
 \delta = &  { 2 G s } \left(   {\bf 1 } ~+ ~  \gamma \partial_{\beta}^4 |- \rangle \langle +|   ~+ ~\gamma^* \partial_{\beta^*}^4  |+ \rangle \langle -|  \right)(- \log |\beta|  )
\cr
 =&  { 2 G s } \left(   - {\bf 1 } \log |\beta|  ~+ ~ {3  \gamma  \over \beta^4}  |- \rangle \langle +|   ~+~{ 3  \gamma^* \over {\beta^*}^4}   |+ \rangle \langle -|  \right) \,,
}}
where we introduced a complex variable $\beta = b_1 + i b_2$ in the two dimensional transverse space. We also used $\pa_{\beta}= {1 \over 2} \left( \pa_{b^1} - i \pa_{b^2} \right)$ and $e_{\pm} \propto e_{x} \mp i e_{y}$.
Here $\gamma$ is the coefficient of the $+++$ amplitude and $\gamma^*$ the coefficient of the $---$ one. 
Notice that these physically imply that the particle two undergoes a spin flip. We see that $\delta$ is a two by two matrix in the space of helicities. 
In \phseficpa , ${\bf 1}$ represents the identity matrix in this two dimensional space. 
The matrix  in \phseficpa\  can be diagonalized by choosing the polarization directions 
\eqn\polardi{ 
p_{1,2} \propto  |+ \rangle \pm \sqrt{ \gamma \,  {\beta^*}^4 \over \gamma^* \, \beta^4} |- \rangle 
}
Then we find that 
\eqn\phaseonetwo{ 
\delta_{1,2} = 2  G s \left( -\log |\beta|  \pm   3 { |\gamma| \over |\beta|^4 } \right) \,.
}
Thus we see that we have a causality problem for small enough $b = |\beta|$.

 \newsec{Fixing the Causality Problem by Adding Massive Particles}

 Let us discuss how to evade the causality problem that we found above.  
 This problem can be evaded by adding new particles at the scale $\alpha_2$ or $\alpha_4$. 
 We will discuss the case of a weakly coupled theory where the problem should be  fixed at tree level. 
 This is indeed what happens in string theory,  see appendix E. For a case involving loops see appendix C. 
 Let us first consider the corrections to the time delay
due to new particles being exchanged in the $t$-channel.
The new particles have to lead to a phase shift growing like $s$, or a higher power of $s$. 
{\it Thus, we should add massive particles with spin $J \geq 2$. }

  We will now argue that massive spin two particles  do not help and that we need particles of higher spin. 
In particular, this will then rule out  a solution  involving Kaluza-Klein gravity, which would be a special example of the 
addition of massive spin two particles.  
For this reason we will analyze it in detail.

  \subsec{Massive Spin Two Particles Do Not Fix the Problem in $D=4$}

Let us first discuss the four-dimensional case. 
Since the external states are massless spin two particles, the on-shell three-point vertices involve two massless particles and a massive spin two particle.

\ifig\WeinbergWitten{ Consider the coupling of a massive spin two 
particle to two massless gravitons. Let us choose the kinematic configuration so that the massive 
particle decays into two massless gravitons along the $\hat z$ axis. The $+-$ helicity configuration is impossible since the angular momentum
 along the $z$ axis would be $+4$. The $++$ configuration is allowed. 
 }  {\epsfxsize2.5in \epsfbox{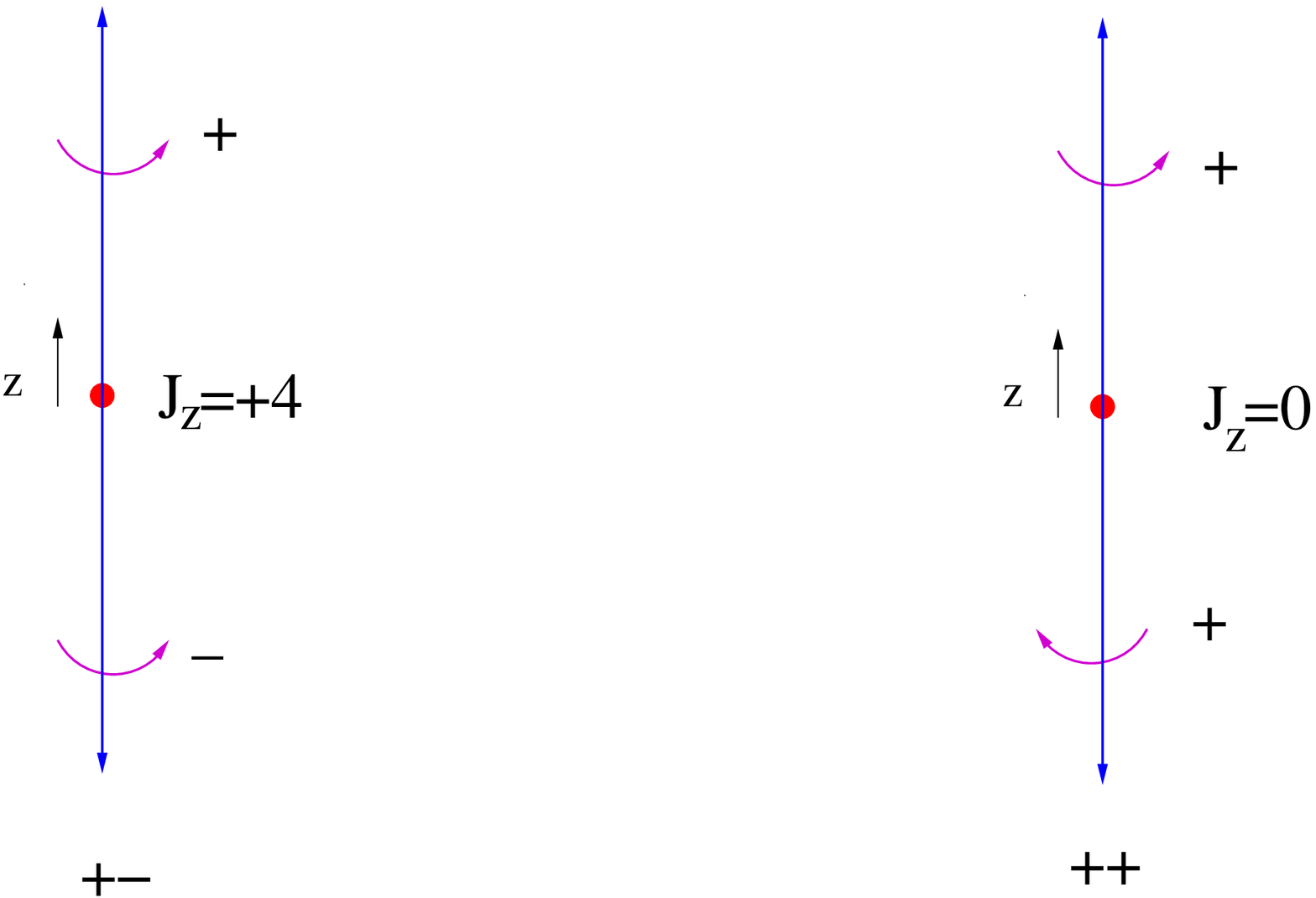} } 

In four dimensions, we can label the massless particles by their helicities. An important result is that, in all incoming notation, the only non-zero amplitudes 
involve $++$ or $--$ helicities for the massless particles. In particular the $+-$ combinations are zero. 
The argument is essentially the same as in the Weinberg Witten theorem \WeinbergKQ , or the statement that gravity does not have a local stress tensor operator.\foot{ 
The matrix elements of the stress tensor operator between two on-shell graviton states is like the coupling to a massive spin two particle, where the square of the
momentum of the stress tensor,  $q^2$ corresponds to the mass of the massive spin two particle. } 
 Imagine that we have the massive spin two particle in its rest frame. We let it decay into two massless spin two particles. Let us suppose
that the two decay products move in opposite directions along the $\hat z$ axis, see \WeinbergWitten . In the $+-$ configuration the total sum of the spins of the decay products 
along the $\hat z$ axis is $+4$ or $-4$. However, the initial massive particle had spin at most $\pm 2$. Therefore a $+-$ configuration is impossible. 
With a $++$ or $--$ configuration there is not problem because the sum of the spins is zero.   One can also write down explicitly the corresponding 
three-point amplitude 
\eqn\amplie{ \eqalign{ 
 &\tilde \alpha_4 \epsilon^I_{\mu \nu} p_1^\nu p_3^\mu \left[ ( \epsilon_1 .p_3) (\epsilon_3 . p_1) - (\epsilon_1 .\epsilon_3) (p_1.p_3) \right]^2 
 \cr
&{ \to 2 \tilde  \alpha_4 p_u^2 \left[ (\vec  e_1 . \vec q) (\vec e_3 . \vec q)  + { m_I^2 \over 2 } (\vec e_1 .\vec e_3)   \right]^2}
 }}
 where $\epsilon^{1}_{\mu \nu} = \epsilon^1_\mu \epsilon^1_\nu$, and  we used that the   component of $\epsilon^{I \mu \nu}$ that 
 contributes the largest factor of $s$ in the sum over intermediate states  is 
 $\epsilon^{I uu} = 2 $. 
  We have denoted the coupling by $\tilde \alpha_4$ since it   reduces to the $\alpha_4$ structure in the massless limit. 
 Here 1 and 3 are the massless particles. Of course, $p_1.p_3$ is given by the 
 mass of the massive particle. We see that this result is invariant under $\epsilon^1 \to \epsilon^1 + p^1 $, and so on. In the second line of \amplie\ we have 
 written the three-point amplitude including the leading terms in the high energy limit. This is written in terms of the purely transverse polarization vectors (or tensors)
 introduced in \canrwi .  In $D=4$ there is also a parity violating 
 structure which we will not need to write explicitly. 

 If we now consider particles 1 and 3 in \kinematics\  to be associated to a coherent state with definite spin, then we have no spin flip allowed and this 
 coherent state does not couple to the massive spin two particles. Therefore in four dimensions the massive spin two particles cannot solve the problem, they 
 simply do not couple to the type of source that we are considering. Note that it is important that the massless intermediate gravitons are still coupling to the 
 1-3 coherent state through the Einstein three-point function, and as discussed in section 3.5, it leads to a causality problem for particle two in \kinematics . 
 
 We can further show that the massive spin two particle with a coupling \amplie\ by itself also leads to a causality problem and should therefore not 
 be present. In fact, it will be useful for our later argument to understand this in more detail.
  For simplicity let us set to zero the parity violating massive structure.  
   For the coherent state that involves particles one and three in \kinematics\ we 
    choose the $\oplus$ polarization with  $e^1_{xx} =-e^1_{yy} $ and the other components equal to zero, and the same for $e^3$.  
 Here $ x,y$ represent the two directions in the 
 transverse plane. Now for particle 2 we can choose the same polarization or the crossed polarization, called $\otimes$, 
  given by $e^2_{xy} = e^2_{yx} =1/\sqrt{2}$ and all 
 other components equal to zero.  
 Then we find the following. If $\vec b $ is along the $\hat x$ direction, then these two different polarizations for particle two do not mix as they go through the shock.
 They diagonalize the phase shift matrix. If the polarizations of particle 2 and 4 are the conjugate to   those of 1 and 3, and reflected along $\vec b$, 
 then the configuration is constrained by 
 unitarity along the $t$-channel to give a strictly  positive answer for the   contribution  to the time delay of any particle with a non-zero coupling. 
 On the other hand, if we average over all polarizations for particle 2, it is possible to see that the terms involving $\alpha_4$ or $\tilde \alpha_4$ (the massive
 particle contributions) all vanish. Thus, the contribution from the crossed polarization has to have the opposite sign. In other words, unitarity fixes a plus sign for
 the time delay for one polarization and this implies a negative sign for the other.
  Indeed, it is possible to see this 
 explicitly by computing the massive particle contribution to  both answers, which are 
 \eqn\delayboth{ \eqalign{ 
 \delta_{\oplus ,\oplus} =& 4 G s (  \sum_m  \tilde \alpha_4^2 O_m O_m ) K_0(mb) 
 \cr 
 \delta_{\oplus, \otimes } =  & - 4 G s ( \sum_m \tilde \alpha_4^2 O_m O_m) K_0(mb) 
 \cr
 O_m  \equiv & \partial_{b_x}^2 \partial_{b_y}^2 - { m^4 \over 8} 
 }}
 where the first subindex of $\delta $ is the polarization of particles 1 and 3 and the second that of particles 2 and 4 in \kinematics . 
 By acting with this operator explicitly one can see that it gives  a negative answer in the second case. This is independent of the sign of $\tilde \alpha_4$. 
 In fact, it is also negative for the contribution of the massless case when we have the $\alpha_4$ structure on both sides. 
 The full phase shift also has the general relativity contribution. Once we have a single massive particle, it is possible to go to a small enough $b$ so that
 we overwhelm the positive contributions from the General Relativity vertices. 
 
 This shows that in a theory with up to spin two particles we cannot solve the causality problem that arises when  $\alpha_4$ is nonzero. In addition, we see any massive 
 spin two particles, even if present, they should have $\tilde \alpha_4 =0$ in order not to cause further causality problems.

  \subsec{Massive Spin Two Particles Do Not Fix the Problem in $D>4$ }

 Now we now move on to a higher dimensional gravity theory,   $D>4$. 
 The three-point amplitudes for two gravitons and a massive spin two particle  now have two possible structures, first the one in \amplie ,
  which can be multiplied by a coefficient 
  which we will still call $\tilde \alpha_4$. And a second one of the form 
  \eqn\apliothes{\eqalign{ 
  & \tilde \alpha_2   \eps^I_{\mu \nu}  \left[ \eps_1^{\mu} p_3^{\nu} (\eps_3 . p_1) + \eps_3^{\mu} p_1^{\nu} (\eps_1 . p_3) -
   p_1^{\mu} p_3^{\nu} (\eps_1 . \eps_3) - \eps_1^{\mu} \eps_3^{\nu} (p_1 . p_3)\right] \times 
   \cr
   &~~~~~~~~~~~~~~~~~~~~~~~~~~~~~\left[ ( \epsilon_1 .p_3) (\epsilon_3 . p_1) - (\epsilon_1 .\epsilon_3) (p_1.p_3) \right]
  \cr
& \to {  2 \tilde \alpha_2 p_u^2 \left[ e^1_{k i } q^i e^3_{k j} q^j + {m^2 \over 2} e^1_{ij} e^3_{ij} \right] }
  }} 
  where again $\epsilon_{\mu \nu}^I$ is the intermediate state polarization vector and we used that we only care about its $\epsilon^{I uu} =2$ component. 
  We have introduced a new coefficient $\tilde \alpha_2$. 
  In the second line we have indicated the form that it takes in the high energy 
  limit. In the last line the polarization tensors are purely in the transverse directions and $q$ is the momentum transfer.

We can first consider a setup with four-dimensional kinematics. Namely, we can consider particles 1 and 3 to be associated to a coherent state which 
 is uniformly distributed along $D-4$ of the original dimensions.  In this case the problem is essentially four-dimensional and the three-point amplitudes involving 
 $\alpha_2$ and $\tilde \alpha_2$ (both massless and massive) do not contribute. If we want to avoid causality problems, and without spin $>2$ particles, 
 we conclude that both $\alpha_4$ and $\tilde \alpha_4$ should be zero.  The argument is the same as the one we presented in the four-dimensional discussion. 
 Note that since we are getting to four dimensions by effectively 
  dimensionally reducing the higher dimensional theory, then the parity violating four-dimensional structure does
 not arise. 
  
We would now like to rule out the possibility of having contributions with non-zero $\alpha_2$. For this we will assume that $\alpha_4$ and $\tilde \alpha_4$ are 
zero, as shown by the previous argument. 

Let us first consider the case $D=5$. Now we have three transverse directions, let us call them $x,~y,~z$. 
We choose the polarizations of $1$ and $3$ to be of the $\oplus_{xy}$ type. Inserting this into \apliothes\ we see that this produces a factor of 
$ O_{xy} = m^2 - \partial_{b_x}^2 - \partial_{b_y}^2$ acting on the massive propagator, which is simply ${ 1 \over b} e^{ - m b } $. 
For the particles 2 and 4 we choose the polarization $\otimes_{yz}$, which also produces a similar operator $O_{yz}$.
The final result has the form 
\eqn\contsi{ \eqalign{ 
\delta_{\oplus_{xy} ,\otimes_{yz} } ~&\sim~   G  s     \left( \sum_m (\tilde \alpha_2)^2 O_{xy} O_{yz}  { 1\over b }e^{ -m b }\right)
\cr
&=  - Gs \sum_m ( \tilde \alpha_2)^2  { (b^3 m^3+5 b^2 m^2+12 b m+12  )\over b^5}  e^{ - m b}
  }}
  where we have set $\vec b = (b ,0,0)$ after taking the derivatives. 
  We see that for any sign of $\tilde \alpha_2$ this produces a negative result. Furthermore, in the massless case, $m=0$, this also gives the part of the graviton 
  contribution proportional to $\alpha_2^2$. The graviton also contains other contributions involving the ordinary Einstein piece on both three-point functions, as well 
  as  a mixed term. These contributions behave like $1/b$ and $\alpha_2/b^3$ respectively. Thus, for small $b$, the graviton contribution involving $\alpha_2^2$ dominates, 
  since it  
  goes as $\alpha_2^2 b^{-5}$. 
 Therefore, we conclude that if $\alpha_2$ is non-zero, then by going to small enough $b$ we get a causality problem. Furthermore, this problem cannot 
 be fixed by adding massive spin two particles. 
 
 For $D>5$ one can run a similar argument. But of course, we could also set up the problem with five-dimensional kinematics. In other words, we choose a coherent 
 state spread over $D-5$ of the dimensions and we get the same as what we discussed above. 
 
 The final conclusion is that if we have extra structures in the graviton three-point function (if $\alpha_2 $ or $\alpha_4$ are nonzero), 
  they lead to  a causality problem {\it which cannot be fixed by 
 adding massive particles with spins } $J \leq 2$. 
 
 With a risk of being repetitive, let us summarize the argument that rules out massive spin two particles. First we go to four-dimensional kinematics where
 the massless or massive couplings proportional to $\alpha_2$ or $\tilde \alpha_2$ do not contribute and we rule out 
 both $\alpha_4$ and the similar coupling $\tilde \alpha_4$ to massive intermediate gravitons. Then we go to five (or higher) dimensional kinematics and we 
 rule out both $\alpha_2$ and $\tilde \alpha_2$. 
 In particular, notice that, even in the case of ordinary Einstein gravity, with $\alpha_2 = \alpha_4=0$, we have ruled out tree level 
 couplings to massive gravitons (or massive spin two particles).

 \subsec{Exciting the Graviton Into New Particles }

\ifig\Xproduction{  When the particles scatter, the graviton can become another massive particle, here labeled by $X$.  }
 {\epsfxsize1.1in \epsfbox{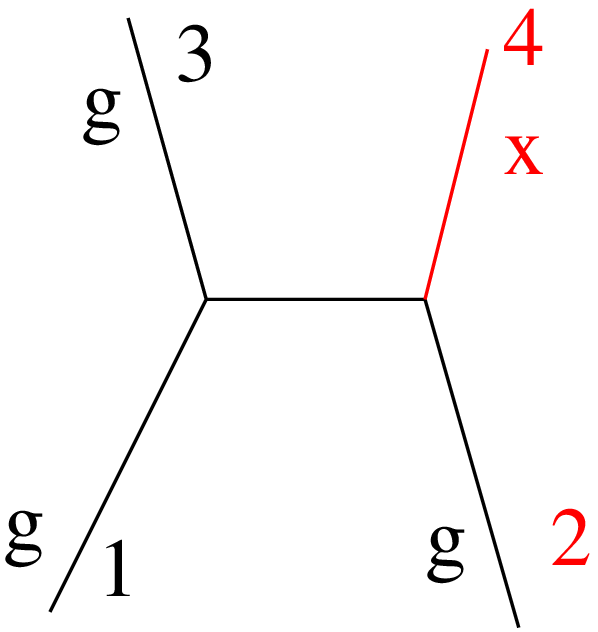} } 

In the above discussion we
 ignored the possibility of exciting a graviton when it passes through the shock and transforming it into a new state.\foot{In string theory these are  called tidal
excitations of the string.}  In this section we discuss this possibility and conclude that it cannot fix the problem. 

Since the available energy is large, compared to $1/b$, it is possible to turn the incoming graviton into an outgoing massive particle, let us call it $X$. 
If we use coherent states for particles 1 and 3 in \kinematics , then we suppress the processes where particle 3 becomes a new massive particle and we 
only  have to worry about the possibility of particle 2 turning into this new massive particle. 
      This can happen even if the mass of the new particle, call it $X$, is much larger than $b^{-1}$, but smaller than 
$\sqrt{s}$. The reason is that there can be some $p_u$ energy transfer from particles 1 and 3 to particle 4 in \kinematics . 

Let us view the process of the graviton  2 passing through the shock as a signal transmission problem. Focusing on the $v$ dependence of the signal we can say 
that the out-signal, given an in-signal must be causal. Namely, if the in-signal vanishes for $v<0$, then the out-signal must vanish for $v<0$. 
In Fourier space these signals  are related by $f_{out}(\omega) = S(\omega) f_{in}(\omega) $, with $\omega = - p_v$. Here
we are using that particles 1 and 3 carry a negligible amount of $p_v$, since we are using $v-$translation invariance. 
Of course, if we have physical particles, we cannot
localize them sharply in $v$ because they only have positive frequencies. In order to obtain a sharp causality bound we need to invoke the vanishing of the 
field commutators, $ [ \phi_{out}(v) , \partial_v \phi_{in} (v') ] =0$ for $v< v'$ (we put the derivative to remove possible zero mode issues).\foot{In a theory 
of gravity we do not have local field operators. However, we can imagine defining such operators in the asymptotic past and future. More precisely, in order to run
into the causality problems we need to put them far enough to the past and future that we can neglect the change in the spacetime metric but close enough so
that the quantum mechanically dictated momentum $p_u$ of the signal particles does not wash out the possible time advance.  Here we  will assume that it 
is possible to do this.   }
As reviewed in appendix D , causality implies that $S(\omega)$ is analytic in the upper half plane. In addition, the fact that we can produce other particles can only 
make the strength of the graviton signal in the future smaller. This, in turn, implies that the S matrix element for graviton going into graviton, call it $S_{gg}(\omega)$
should be be smaller than one in the upper half plane, 
\eqn\boundssgg{
|S_{gg}(\omega)|\leq 1  ~,~~~~~~{\rm for}~~~~~ {\rm Im}(\omega) > 0
} 
See appendix D for further discussion. 

In situations where we have some time advance for the graviton, we are getting an infinitesimal matrix element of the form 
\eqn\infint{ 
 S_{gg} = 1 -  i \Delta v   p_v = 1 + i \Delta v \omega 
}
with $\Delta v < 0$. Then if we set $\omega \to  i \gamma$, with $\gamma > 0$ we get
 $S_{gg} = 1 - \gamma \Delta v $ which is bigger than one 
in the upper half plane.  Note that we do not need to go to very large values of $\gamma$ to obtain a violation, we only need $p_v$ or $\gamma$, to be large
enough so that this impact parameter description is good enough. 

In this presentation of the argument, it is clear that adding extra particles as possible extra final states does not help. We need to modify  $S_{gg}$. 
In other words, the 
transformation of the graviton to $X$ is irrelevant for this argument because we are considering the graviton-graviton $S$ matrix element. The 
transformation to $X$ and then back to the graviton can contribute to this matrix element at higher orders in $G$. But,   this 
cannot fix the problem we run into \infint, which is of first order in $G$.

 \subsec{Massive Higher Spin Particles Can Solve the Problem}

   Now we consider the exchange of a massive spin $J>2$ particle in the $t$-channel. 
   Its contribution to the phase shift will rise with energy like $ G s^{J-1}$ and at high energies it will dominate 
   over the graviton contribution. This can happen even in the regime that the theory is weakly coupled. 
   If we have a single contribution of this type, we also run into a problem. The problem is the following. We can think of the propagation of the 
   particle number $2$ as signal transmission problem where time is $v$. In other words, we start with a signal $f_{in}(v) $ which vanishes for $v<0$, then
   the out-signal $f_{out}(v) $ should be zero for $v< 0 $. In addition we want that the total norm of the out wave should not grow. From these two conditions 
   we can deduce that the $S(\omega)$ matrix as a function of the ``energy'',   $\omega = -p_v$, should be analytic in the upper half complex $\omega$-plane
   and, in addition, it should be bounded $|S(\omega)|\leq 1 $ in the upper half plane. See appendix D for a review of these properties. 
   However, a particle of spin $J>2$ leads to a contribution $S \propto 1 + i G s^{J-1}   + \cdots $ which  becomes bigger than one 
    in some regions of the upper half complex  $s$ plane. Notice that the problem arises already at weak coupling, for a small value for $Gs^{J-1}$.

   Thus, a {\it finite}  number of higher spin particles does not fix the problem. In fact, it generates problems of its own. 
   On the other hand, an {\it infinite} number of particles with higher spin can solve the problem. An example is string theory. 
   This problem has been discussed extensively  in the classic papers by Amati, Ciafaloni and Veneziano 
   \refs{\AmatiWQ,\AmatiUF,\AmatiTN,\AmatiZB,\AmatiTB} (see also \SusskindQZ ). 
   In fact,   the amplitude Reggeizes  
    \eqn\detlstring{ 
    \delta \propto {  G s \over t }    s^{   { \alpha' \over 2 }  t } e^{ - i \pi  { \alpha'  \over 4 } t }  
    }
    for large $s$   small $t$ ($s \alpha'\gg1 $, $t \alpha' \ll 1 $). 
   This expression has a cut in the $s$-channel, due to the creation of physical   states along the $s$-channel. These are simply the massive closed string that 
   are present along the $s$-channel.  
   For spacelike $t$, $t< 0$, we see that this effectively leads to a phase shift that decreases faster than $s$ 
   at large $s$. 
   Taking \detlstring\ and transforming to the impact parameter representation we find that for   $b^2 \ll \alpha' \log s $ we get a behavior \AmatiTN\ 
   \eqn\strinphs{ 
   \delta  \sim  {\rm Pol } 
   { G s \over  ( \log (\alpha' s/4)  )^{ D-4 \over 2} } \left( { 1  \over D-4} - { b^2 \over 2 (D-2) \alpha' \log(\alpha' s/4)}  + \cdots   + i { \pi 
   \over 4    \log (\alpha' s/4 ) } + \cdots \right)
   }
   where Pol $=1 + \alpha_2 \epsilon . \partial_b \epsilon . \partial_b + \cdots $ is the part coming from the polarization tensors, which includes the new structures 
   in the three-point functions.\foot{For  type II superstrings Pol$=1$.}
    This is indeed compatible with causality. We also get a large imaginary part that 
   is reflecting the fact that we are creating strings  along the $s$-channel. 
   Notice that we had argued before that in a local theory we expect that by going to impact parameter space we can suppress tree level $s$-channel processes. 
   This is not true in string theory, which contains extended objects. Furthermore, since their size increases with mass logarithmically, we see that at high energies
   their effects appear at   $b^2 \sim \alpha' \log (s \alpha')$ rather than the more naive expectation of $b^2 \sim \alpha'$.  This justifies the small $t$ expansion used in \detlstring . 
   Further aspects of the string case are discussed in appendix E.

    The conclusion is that an infinite number of higher spin particles can solve the problem. We need a tower of particles 
     with increasing spins and intricate relations  between them so that the expansion can be resummed into an amplitude that does not have a problem.   
     Besides string theory, we do not know if there are other ways of doing this.

  \subsec{Compositeness and the Extra Structures for Graviton Scattering }

 In this subsection, with the string theory case as an inspiration, we make some remarks about the extra structures for the graviton scattering.   
 
 Imagine that the graviton has a composite structure.\foot{Note that we cannot make the graviton as 
 a zero energy bound state in a local relativistic quantum field theory that 
 contains a stress tensor operator \WeinbergKQ  .}
  Let us 
  imagine that the graviton is given by a pair of particles which are in a bound state given by a wavefunction 
$\psi(r)$ where $r$ is the relative distance.  We further assume that these particles scatter through the shock via the usual general relativity three-point functions. 
However, since the two particles feel slightly different forces we find that the total scattering amplitude will have the form
\eqn\totals{ 
 \delta_0 (b_0) + { \partial_i \partial_j } \delta_0(b)   \langle \psi_{\epsilon} | r_i r_j | \psi_{\epsilon } \rangle  + \cdots  
 }
 where $\delta_0$ is the general relativistic expression \phaseeik , and the subindex $\epsilon$ indicates the spin of the graviton.
  Notice that since the Laplacian $\nabla^2_b \delta_0 =0$, the only terms that can contribute to the 
 second part are those which are not rotationally invariant in the relative coordinates.  These are possible because the graviton spin or polarization $\epsilon $.  
 Notice that this is a simple argument for the presence of extra structures in the graviton three-point function. Even though we motivated this with a graviton 
 composed with two particles, the same final formula works if the graviton is made out of many more elementary constituents as it happens in string theory, 
 when it is a string.  
 In any case, the size of the new structure, $\alpha_2$, 
  due to compositeness, is of   order $\alpha_2  \sim r_s^2$ where $r_s$ is the typical 
 size of the graviton. Given that the bound state has this typical size, then we also expect that it can be excited to other states with masses $m^2 \sim 1/r_s^2$. 
   Indeed, by imposing the causality constraint, we found that there should be new particles with masses of this order of magnitude.

\newsec{Anti-de Sitter Discussion} 

 The case of asymptotically AdS space is very similar to the asymptotically flat space one. The causal 
structure is defined by the causal structure of the boundary. We then require that signals that go through
the bulk cannot go faster than signals that remain on the  boundary.
As argued in \refs{\PageXN, \GaoGA}, general relativity with the null energy condition implies that this is obeyed. 

In terms of the dual CFT, this is just the statement that CFT observers cannot exchange information 
faster than light. Equivalently boundary CFT operators commute outside the boundary light-cone. 

\subsec{Motivation: The Emergence of Bulk Locality Should Happen in the Classical Theory}

In this subsection we discuss in more detail the AdS considerations that motivated the present paper. 

We expect that the dual of a large $N$ gauge theory should be a weakly coupled string  theory with 
coupling $g_{s} \sim 1/N$. This should be true both at weak and strong 't Hooft coupling. 
As we increase  the 't Hooft coupling we are supposed to interpolate between a Vasiliev-like theory  
and an ordinary Einstein-like theory at strong 't Hooft coupling. This whole interpolation happens within
the classical string theory in the bulk. Of course, in the ordinary Einstein description we see a local 
theory in the bulk. Thus the emergence of bulk locality is something that should be contained within 
{\it classical} string theory.  
It is for this reason that it is interesting to understand the constraints of tree level interactions of 
gravitons and the link between the masses of the higher spin particles and the size of the corrections
to Einstein gravity. Here we attempted to link them via the causality considerations for the simplest 
gravitational interactions. 
Given the interest of the AdS case, we will discuss in more detail some of its features.

\subsec{Statement of AdS Causality } 

In asymptotically AdS gravitational theories the causal structure given by the Minkowski light-cone on the boundary of AdS. This allows us to
formulate the causality criterion in a very simple manner. None of the subtleties that existed in flat space regarding the definition of causality appear for gravitational theories in AdS. The basic condition we would like to impose is that 
\eqn\cleanestway{
 \la \Psi | [T_{\mu \nu}(y), T_{\rho \sigma}(0)] | \Psi \ra = 0,~~~ y^2 > 0,
}
where $\Psi$ is some nontrivial state in the theory. We will be interested in the commutator computed on the shock wave background
\refs{\CornalbaXK,\BriganteGZ,\HottaQY,\HorowitzGF,\HofmanUG}. The Aichelburg-Sexl shockwave in AdS can be created by inserting a pair of operators creating a 
coherent state bulk  wavefunctions that localize the bulk stress tensor on the light ray \refs{\CornalbaXK ,\CornalbaXM ,\CornalbaZB}. 

As before, instead of computing the commutator, we study propagation of an energetic graviton through the bulk and impose positivity of the time delay.
 The whole discussion is similar to the flat space one. The only difference is that the $t$-channel propagator is now in $AdS$. Instead to computing it directly, 
 we consider plane wave solutions which encode intermediate state gravitons with the properties we need. 
Plane  wave solutions in $AdS_{D}$ have the form
\eqn\planwav{ 
ds^2 =  {  -d u d v + h(u,y_i,z)\, d u^2 + \sum_{i=1}^{D-3} d y_i^2 + d z^2 \over z^2 } 
}
Here the function $h$ is only constrained by the Laplace  equation in the transverse hyperbolic space 
\eqn\laplac{ 
 {  z^{ D-2} } \partial_z ( z^{-( D-2)} \partial_z h) + \partial_{y_i}^2 h      = 0 
}
which can be equivalently written as
\eqn\laplacb{
\nabla^2_{D-2} f - (D-2) f = 0
}
where $f = {h \over z }$.
In appendix F we give an argument that this is a solution, to all orders in the derivative expansion. 

We can also   a plane wave with a delta function source so that instead of \laplac\ we write
\eqn\laplsou{ 
 {  z^{D-2} } \partial_z ( z^{-(D-2)} \partial_z h) + \partial_{y_i}^2 h   =- 16 \pi G |P_u| \delta(u)  z_0^{D-2}  \delta^{D-3}(\vec y - \vec y_0) \delta(z-z_0).
}
where the RHS corresponds to an insertion of a delta-function source in the hyperbolic space in \laplacb . The Green's function in the hyperbolic space is well-known, so that we get
\eqn\exactsolution{\eqalign{
h &= {z \varpi (\rho) \over 1 - \rho^2 }\delta(u) \cr
&= 16 \pi G |P_u| \delta(u)  {z (4 \pi)^{{2-D \over 2}} \Gamma ({D \over 2}) \over (D-1)(D-2)} \left( \rho^2 \over 1- \rho^2 \right)^{2-D}\!\!\!\!\! \ _2 F_1 \left(D-2,{D \over 2}, D ; - {1 - \rho^2 \over \rho^2}\right) \,,
}}  
where $\vec y_0$ and $z_0$ are the coordinates of the source in the bulk, and
\eqn\exprho{
\rho = \sqrt{{(z-z_0)^2 + |\vec y - \vec y_0|^2  \over (z+z_0)^2 + |\vec y - \vec y_0|^2}} \,.
}
It can be checked that $P_u$ is also the total momentum from the boundary point of view 
by  integrating the 
boundary  stress tensor (read off from the small $z$ expansion of this metric) on the boundary. 
  For example, in $AdS_4$ and $AdS_5$  \exactsolution\ takes the following form
\eqn\fourd{\eqalign{
\varpi_4 (\rho) /( G |P_u|) &= - 8 \left(1 - \rho^2 + (1 + \rho^2 ) \log \rho  \right), \cr
\varpi_5 (\rho) /( G |P_u|) &= 2  {(1- \rho)^4 \over \rho }.
}}
More generally, we have 
\eqn\limitsshock{\eqalign{
\lim_{\rho \to 0}  \varpi (\rho) &\sim  {1 \over \rho^{D-4}}, \cr
\lim_{\rho \to 1}  \varpi (\rho)  &\sim (1 - \rho)^{D-1}.
}}

Let us understand better the symmetries of the problem. The shock wave is localized around $u=0$ and is probed by a particle which is localized in $v$. The role of the transverse plane in flat space is played here by the transverse $H_{D-2}$. It is convenient to think of the probe crossing this hyperbolic space at the center.\foot{
 From the CFT point of view we can create such a probe by acting with the operator of given energy and zero momentum in the 
 AdS Poincare coordinates for which $u=0$  is the future null infinity as explained in \HofmanAR . In \HofmanAR\ terms we are working
  here in the $y$-coordinates, while the operator with given momentum is inserted in the $x$-coordinates. In pure AdS case isometries of $H_{D-2}$ at fixed $u=0$ correspond to the usual Lorentz symmetry group in the $x$-coordinates of \HofmanAR. }

The shock centered at $(\vec y_0 , z_0)$ has a number  of Killing vectors that depend on $f(u)$. For arbitrary $f(u)$ the background has 
an obvious $SO(D-3)$ symmetry that rotates $\vec y$ and a translation in $v$. For $f(u)=\delta(u)$ the geometry has extra Killing vectors
 which enhance the rotational symmetry to $SO(D-2)$, as we had in flat space. In the original coordinates \planwav , some of the extra symmetries involve special 
 conformal generators. 
 Since we are working in the high energy limit, we effectively have a delta function in $u$, so that we also expect to have this extra $SO(D-2)$ symmetry. 
 See \HorowitzGF\ for a coordinate system that makes this manifest. 

General properties of the AdS shock wave and its different limits are considered in appendix F. 
In particular,  when the center of the shock goes to the boundary $z_0 \to 0$ the problem becomes very similar to the one arising in the 
 computation of  energy correlators \HofmanAR, whereas in the limit $z_0 \to \infty$ it reduces to the setup used in  \HofmanUG\ to study causality. Our formulas will reduce to the ones considered before in those limits. 

\subsec{The Effect of Higher Derivative Interactions on Particles with Spin}

Now we would like to consider different type of probes and compute the time delay for them. We start with a simple example
of a scalar probe and then move to the case of particles of spin one and two.

If we consider a minimally coupled scalar in the shock wave background its equation of motion takes the form
\eqn\scalarbackgr{
\nabla^2\phi=0 .
}
In our setup we are interested in corrections to this equation which are second order in derivatives.\foot{Since we need derivatives to bring down large factors of momentum. }
Considering terms yielding two derivative equations of motion for $\phi$ we have to consider terms like
\eqn\addscalar{
{\cal H}^{\mu\nu} \nabla_\mu\nabla_\nu\phi
}
where the tensor ${\cal H}$ is made from the background metric, Riemann and covariant derivatives. However one can check that there is no two index symmetric
tensor  that is not vanishing on-shell \HorowitzGF . Of course, this statement is equivalent to the uniqueness of the scalar-scalar-graviton three-point vertex. 
In the high energy limit we get similar to flat space
\eqn\scalareom{
\partial_u\partial_v \phi + f(u) h(z, \vec y) \partial_v^2 \phi = 0
}
which produces the  time delay
\eqn\phaseshiftscalar{
\Delta v = {\varpi(\rho) \over 1 - \rho^2 }
}
reproducing the   flat space computation for small $\rho$. 
We assumed that the perturbation crosses the shock at $z=1$ and $\vec y =  0$.

For the gauge boson we imagine at the level of two derivative the following equation
\eqn\actionAa{
\nabla^{\mu}F_{\mu\nu} + {\cal H}_\nu^{~\mu\alpha\beta}\nabla_\mu F_{\alpha\beta}
}
where ${\cal H}$ is built out of the Riemann tensor and its covariant derivatives. Using the properties of the background discussed above (we defer the details to appendix F) one can check that the only term that we can have is a correction analogous to that appearing in the case of flat space
\eqn\actionA{
\nabla^{\mu}F_{\mu\nu} - \hat \alpha_2\, \check{R}_{\nu}^{\ \ \mu\alpha\beta}\nabla_\mu F_{\alpha\beta} = 0.
}
where $\check R$ reduces to the Weyl tensor on-shell (see appendix F for details).

If we compute the time delay using the same action \actionA, considering that each mode corresponds to a different constant polarization, $\epsilon_i$, we get
\eqn\timedelaygauge{
\Delta v ={ \varpi(\rho) \over 1 - \rho^2 } \left(1 -\hat \alpha_2\left(1- \rho^2\right)^2
{\varpi'(\rho)-\rho\varpi''(\rho) \over 4 \rho\varpi(\rho)}\left({ \eps.n^2 \over \eps. \eps }-{1 \over D-2}\right)\right).
}

The final result is very similar to the one obtained in flat space. The only different is that the polarization dependent delay is slightly more complicated.
The flat space result \totaime\  is reproduced by considering $\rho \to 0$ limit, whereas the energy correlator constrained is recovered in the limit $\rho \to 1$.

Similarly, in the case of gravity we are interested in the most general form of the second order equations. We choose to parameterize the equations of motions for perturbations as follows
\eqn\gravitonpert{
\delta R_{\mu\nu} + \alpha_2 \; \check{R}_{(\mu}^{\ \ \rho\alpha\beta}\delta R_{\nu)\rho\alpha\beta} + {\alpha_4 \over 2} \; [\nabla_{(\mu}\nabla_{\nu)}\check{R}^{\alpha\beta\rho\sigma}]\delta R_{\alpha\beta\rho\sigma}=0
}
where the parameters $\alpha_i$ are in units of the $AdS$ radius $R_{AdS}$ that we set to one. In this case, even though there are several possible ways to contract the indices in each of the above terms, we may concentrate on the contributions to the transverse equations of motion that are the ones yielding the time delay. For that purpose  \gravitonpert\ is the most general parametrization.\foot{There is also the possibility of considering more than two derivatives acting on the perturbation as $\nabla^n \delta R$. These contributions in general change the number of degrees of freedom of the theory and will not be considered here.}


The time delay for these equations of motion is then given by
\eqn\timedelay{\eqalign{
\Delta v &= {\varpi(\rho) \over 1 - \rho^2 } \left(   1 + t_2(\rho) \left({ (\eps . n)^2 \over \eps . \eps } - {1 \over D-2} \right) + t_4 (\rho)  \left({ (\eps.n)^4 \over  (\eps . \eps)^2} - {2 \over D (D-2)} \right)\right), \cr
t_2(\rho) &=\left( 1-\rho^2 \right)^2 {\varpi'-\rho\varpi'' \over 4\rho\,\varpi(\rho)} \left(- \alpha_2  +  \alpha_4  { D (1 + \rho^2)^2 - 2 \rho^2 \over \rho^2} \right), \cr
t_4(\rho) &= - D \alpha_4  \left( 1-\rho^2 \right)^2 {\varpi'-\rho\varpi'' \over 4\rho\,\varpi(\rho)} { D (1 + \rho^2)^2 + 2 (1 + \rho^4) \over 4 \rho^2} .
}}
where $\vec n$ is a vector pointing from the center of the shock to the probe particle, and $\epsilon $ is the polarization of the probe particle. 
These time delays can become negative for small enough $\rho$ if $\alpha_2$ or $\alpha_4 $ are non-zero. 

These results can be also reproduced using slightly different method of evaluating the on-shell action in an explicit gravitational theories in the shock wave background like Lovelock or quasi-topological theories (see e.g. \refs{\MyersJV,\SenNFA}).

In the limit $\rho \to 0$ these constraints reproduce the flat space analysis of section B.1 with $\rho = {b \over 2}$. In the $\rho \to 1$ limit the above result reproduces constraints discussed in the past.  If we take the limit $\rho \to 1$ by taking the shock center to the boundary $z_0 \to 0$ we recover energy correlator computation \HofmanAR. If we on the other hand consider $\rho \sim 1$ by taking the shock center to the horizon $z_0 \to \infty$ we recover the shock wave discussed by Hofman \HofmanUG.

\subsec{Implications for $a,c$ in Theories with Large Operator Dimensions}

Imagine an abstract CFT$_4$ with large $N \gg 1$ and large gap $\Delta_{gap} \gg 1$, where $\Delta_{gap}$ is the dimension of the lightest higher spin 
single trace operator.  This theory is described by a gravitational theory in the bulk 
 with potentially some higher derivative corrections. 
 String theory inspired intuition suggests that higher derivative corrections should be suppressed by the ${1 \over \Delta_{gap}}$ factor.
  Our argument  shows  that this is indeed the case  for the simplest  corrections, which are the ones 
    affecting the three-point function of stress tensor. 
    Indeed, as we showed in \timedelay , we run into a potential problems with causality at impact parameters $\rho_c \sim (\alpha_2)^{1/2}, (\alpha_4)^{1/4}$. 
Since this can only be fixed by higher spin particles, we conclude that $\Delta_{gap}$ has to be small enough so that it can start correcting the amplitude before 
we run into this problem.

For $\Delta_{gap} \gg 1$ then the relevant impact parameters are such that we can approximate the formulas by the flat space limit. 
  Thus, we get the bound of the type
\eqn\bound{\eqalign{
(\alpha_2 )^{1/2} \lesssim {1 \over \Delta_{gap}},~~~ (\alpha_4)^{1/4} \lesssim {1 \over \Delta_{gap}},
}}
where $\lesssim$ stands for some numerical coefficient that we cannot fix using our simple analysis.

In the case of ${\cal N} = 1$ superconformal theories, $\alpha_4=0$ by supersymmetry and $\alpha_2\propto { a-c \over c } $. Then we get 
\eqn\boundac{
\left|{a - c \over c} \right| \lesssim {1 \over \Delta_{gap}^2} .
}

 In the case of ${\cal N } = 4$ SYM \bound\ is satisfied trivially since $\Delta_{gap} \sim \lambda^{1/4}$ and $a=c$.

It will be very interesting to find an independent field theoretic argument that leads to the bounds of the type \bound\ or \boundac . It would also be nice to find the precise numerical coefficients in \bound\ and \boundac . In the supersymmetric case, $a-c$ was argued to control asymptotic density of BPS operators in \ZoharToAppear . It would be very interesting to understand if there is any relationship between their work and our analysis.

\subsec{Implications for Dimensions of Double Trace Operators } 

This computation 
of the time delay can be viewed as  a special 
 four-point correlation function with  particular wave functions for external operators.
  We can use the OPE to expand the four-point function computation 
  in terms of the time evolution eigenstates (or, equivalently, local operators). 
  One can ask then the following question: what is the relation between the time delay and the OPE data?

\ifig\DoubleTraceTrajectory{ We consider two energetic particles in AdS that oscillate back and forth with energy $E$ and angular momentum $J$. This models the high twist, high spin double trace operator in the dual CFT. }  {\epsfxsize2.2in \epsfbox{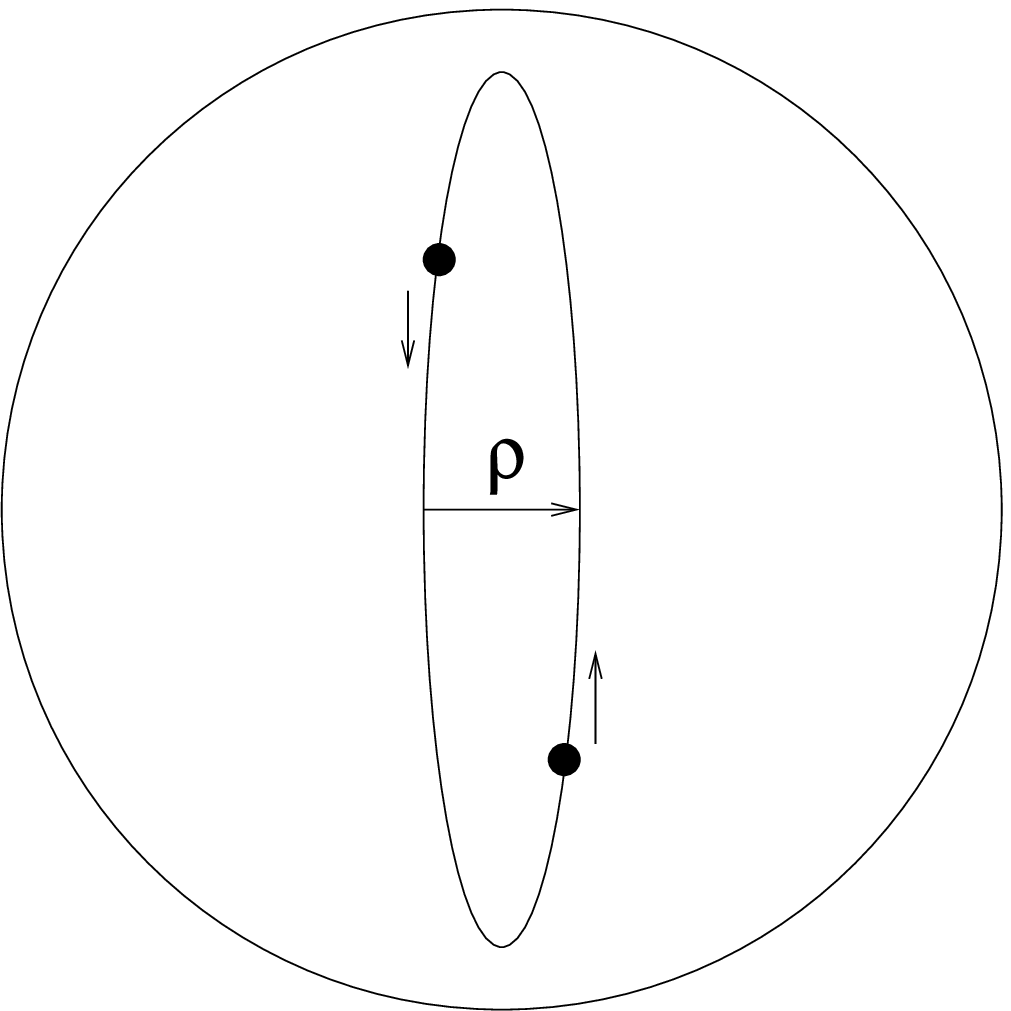} } 

This question was addressed in a nice series of papers \refs{\CornalbaXK,\CornalbaXM,\CornalbaZB} where the 
 time delay was shown to be equal to anomalous dimension of double trace operators of 
 the type ${\cal O} (\pa^2)^n \pa_{\mu_1} ... \pa_{\mu_j} {\cal O}$ for large $n$ and large $j$ 
\eqn\relationads{\eqalign{
\delta (s, \rho) &= - \pi \gamma(n,j), \cr
\rho &= {j \over j + 2 n}, \cr
s &= 4 n (j+n).
}}
where the relations between parameters on both sides of the equation are reviewed below. Intuitively, \relationads\ follows from the fact that the phase shift of the correlator $e^{i \delta}$ is given by $e^{- i (\Delta_* - 2 \Delta_{{\cal O}}) \Delta \tau}$ where $\Delta_*$ is the dimension of the operator that dominates the OPE, and $\Delta \tau$ is the global AdS time that passed from the beginning to the end of the process. The time delay that we computed corresponds to particles starting at the boundary, getting close to each other at the center and then  reaching the boundary again, see   \DoubleTraceTrajectory .
 It takes $\Delta \tau = \pi$ for this process to occur. From this fact \relationads\ follows. Even though \relationads\ was derived in general relativity in the limit when $\delta \ll 1$ it follows simply from the AdS graviton diagram exchanged. 
 In the impact parameter representation, only the on-shell $t$-channel exchange diagram constributes. 
 This diagram is fixed in terms of three-point function $\la {\cal O} {\cal O} T_{\mu \nu} \ra$ in a generic gravitational 
 theory with generic three-point couplings similar to our flat space analysis.  

Let us briefly review  the results in \refs{\CornalbaXK,\CornalbaXM,\CornalbaZB}. 
The basic idea is the following: the state created by ${\cal O} (\pa^2)^n \pa_{\mu_1} ... \pa_{\mu_j} {\cal O}$ for large $n$ and large $j$ can be thought of as two highly energetic particles that follow null geodesics in AdS, see   \DoubleTraceTrajectory . 

For two geodesics that are characterized by total energy $E$ and spin $J$ the minimal separation is achieved in global coordinates at
\eqn\relour{
\rho = {J \over E} \simeq {j \over j + 2 n}.
}
where we matched energy and spin of the pair of particles to the ones of the double trace operator $J = j$, $E = 2 \Delta + 2 n + j$ and used that $n,s \gg 1$, with 
$\Delta$ also of order one.

Thus, we see that probing distances much smaller than AdS radius ($\rho \ll 1$) corresponds to considering operators with $n \gg j$. On the other hand $n \ll j$ corresponds to scattering at very large impact parameters. The Mandelstam variable $s$ is given by 
\eqn\relationtoours{
s = E^2 - J^2 \simeq 4 n (j+n)
}
and the relation to the anomalous dimensions is that
\eqn\anomalousdim{
\pi \gamma(n,j)=p_v \Delta v =- G s {\varpi (\rho) \over 1 - \rho^2} .
}
Let us consider different limits of this formula. First, consider very large impact parameters $\rho \sim 1$ or $j \gg n$. In this limit 
we get
\eqn\limitlargespin{
\gamma(n,j) \sim-  G {n^{D-1} \over j^{D-3}}
}
which is in agreement with the general results derived using   the crossing equation 
\refs{\AldayMF,\FitzpatrickYX,\KomargodskiEK}.

In the opposite limitof small impact parameters scattering,  $n \gg j$ and ${j \over n } > {1 \over \Delta_{gap}}$,    we have 
\eqn\limitlargespin{
\gamma(n,j) \sim - G n^2 \left({n \over j} \right)^{D-4} .
}
 
We have several comments to add to this story. First, these results should be universal and applicable to generic CFTs with large $N$ and large gap. To write the answer in an abstract form we have to use the relation of $G$ to the two-point function of stress tensors which is well-known and is roughly $G \sim {1 \over c_{T}}$.  It means that it should be possible to derive them using crossing equation which still be dominated by the stress tensor exchange. Probably the relevant limit is $z \to 0$, $\bar z \to 1$ with ${z \over 1 - \bar z}$ being fixed. It would be nice to reproduce the formulas above using crossing equations.

Second, we see that causality, or positivity of the time delay, implies the constraint $\gamma(n,s) < 0$,  which generalizes the ones that were previously known
\refs{\NachtmannMR,\FitzpatrickYX,\KomargodskiEK} for asymptotically large $s\gg n$. 
Again it would be very interesting to understand how to prove these constraints purely from the field theory point of view. 

Third, we see that considering double trace operators of the type $T_{\mu \nu} (\pa^2)^n \pa_{\mu_1} ... \pa_{\mu_j} T_{\rho \sigma}$ we get new structures due to the dependence on polarization which potentially lead to causality violations and bounds \bound, \boundac . Of course, the same is true about the double trace operators that involve the conserved current $J_{\mu}$. It will be very interesting to understand them 
from the purely CFT viewpoint.  Note that in the scattering process the polarizations of particles 3 and 4 in \kinematics\ 
can change relative to those of particles 2 and 4, so that
the phase shift is an operator that acts on this space of polarization tensors. 
 Both $t$-channel unitarity, as well as our considerations, constrain only some matrix elements of this general matrix. 
      While we  leave the general case for the future, we note that, 
 in some cases, we can  ensure  that the polarizations of 3 and 4 do not change by using conservation of angular momentum along the directions orthogonal to the
  impact parameter direction.    In these cases our considerations apply and we can say that the anomalous dimensions of the corresponding double trace operators
  should be negative. The positivity statement applies to the part of the phase shift that grows with the Mandelstam invariant $s$, which translate into the growth with 
  $n$ via \relationtoours . 
    
 However, in our case, this positivity requirement is not obvious from the CFT point of view. It would be nice to see whether this is a general requirement or is one
 that is present only in theories with a local bulk dual.

\newsec{Wormholes and Time Advances } 

General relativity has Lorentzian wormhole solutions that join far away points by short Einstein-Rosen bridges. 
The simplest configuration is the maximally extended Schwarzschild solution interpreted as an approximation to the metric of two 
distant black holes which share a single interior. As discussed in \FullerZZA , these solutions do not lead to a violation of causality in the ambient space 
because it is not possible to send signals through the wormhole \GallowayBR .

\ifig\wormholeadvance{ We consider a Lorentzian wormhole configuration that is described, near each wormhole, by the maximally extended 
Schwarzschild solution. We   send a (green) particle from the left very close to the past horizon. We  then send a (purple) particle from the right. (a) If this particle gets a 
time delay, it will fall into the singularity. (b) If the particle gets a time advance, then it can make it out of the other black hole and we would have a way 
of sending signals through the wormhole.  Here the blue lines represent the average position of the horizon of the black hole, defined by null lines that
are very far away from the two particles we send in. We assume that the impact parameter is much smaller than the Schwarzschild radius. }
 {\epsfxsize5.2in \epsfbox{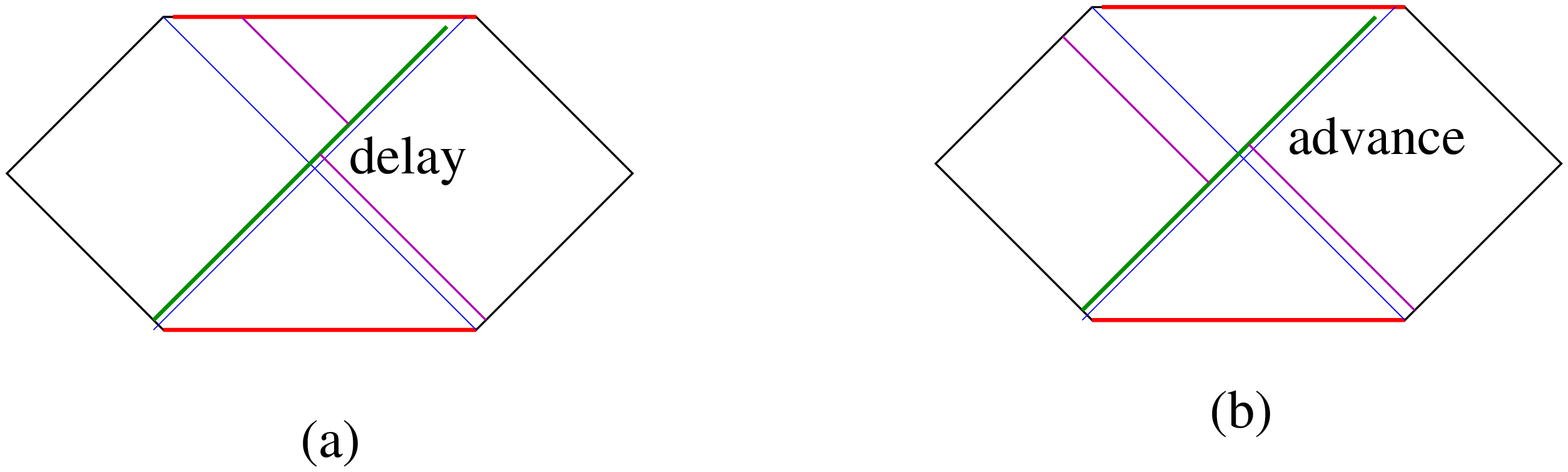} } 

The inability to send signals through the wormhole depends crucially on the fact that we have a Shapiro time {\it delay} as opposed to a time advance. 
For example, if one sends a fast moving particle from the left side, then a particle send from the right will suffer a time delay that will make it go into the singularity,
see e.g. \refs{\DrayHA,\ShenkerPQA} . 
However, if that particle were to suffer a time advance, as opposed to a time delay, then it would be able to go through the wormhole and we would have a 
violation of causality, see \wormholeadvance . 
Note that we can make the wormhole big enough that we can neglect the higher derivative corrections in the description of the background metric. It can also be 
big enough that we can neglect the backreaction of the two particles. So we are considering a situation where the impact parameter $b\ll $(Schwarzschild radius). In the $D=4$ case, the Schwarzschild radius acts at the IR cutoff of the logarithm. 

This impossibility of sending signals is crucial for interpreting the wormhole as an EPR state of two disconnected systems 
\refs{\IsraelUR,\MaldacenaKR,\MaldacenaXJA}.

\newsec{Cosmological Applications } 

The gravity wave non-gaussianities produced by inflation are a  direct measure of the graviton three-point vertex during inflation 
\refs{\MaldacenaVR,\MaldacenaNZ}. To leading order in the slow roll approximation we can do the computation in de Sitter space.
 The symmetries of de Sitter imply that only 
two different parity preserving structures are possible. These correspond to the two parity preserving   structures that we have in four-dimensional flat space. 
One is the one produced by the Einstein action and the other can be produced by a term in the action of the form $M_{pl}^2 \alpha_4 R^3$, where $R$ is a Riemann
tensor (not the Ricci tensor). 
 The relative size between the two types of gravity wave non-gaussianity is proportional to 
\eqn\ratiogar{  { \langle hhh \rangle_{R^3} 
\over \langle hh h \rangle_{Einstein} } \propto \alpha_4 H^4 
}
where $H$ is the 
Hubble scale during inflation. Of course, both are small compared to the two point function, 
 ${ \langle 3{\rm point } \rangle \over \langle 2{\rm point} \rangle^{3/2} } \sim { H \over M_{pl} }$. See \MaldacenaNZ\ for the explicit expressions. 

  Thus, if the gravity wave three-point function was measured { \it and it was found that this exotic new structure is present at a level comparable
to the Einstein one},   then  one concludes that $\alpha_4 $ is of the order of the Hubble scale. 
The considerations in this paper imply that there should also be new particles with spins $J> 2$ with masses comparable to the 
 Hubble scale during inflation. 
Thus, this would be an indirect evidence for  string theory during inflation. 

Note that $ \alpha_4 H^4 \sim 1 $ implies that supersymmetry had to be broken at the Planck scale and not at a lower scale, since the $+++$ and  $---$ structures are
 forbidden by supersymmetry.\foot{We thank Nima Arkani-Hamed for emphasizing this to us.} Let us be a bit more explicit about this point. If the short distance
 theory is supersymmetric, then the field theory Lagrangian does not contain the couplings giving rise to $\alpha_4$. Now, since supersymmetry is broken, this 
 three-point function could arise from integrating out massive particles. These are expected to contribute to $\alpha_4$ as 
 \eqn\constrib{ 
 \alpha_4 \sim { 1 \over M_{pl}^2}  \left( 
 { 1 \over m_B^2 } - { 1 \over m_F^2 }  \right) ~,~~~\longrightarrow ~~~~~~ \alpha_4 H^4 \sim  { H^2 \over M_{pl}^2 }  \ll 1 
 }
 which is very small. Where, to maximize the effect,  we assumed
  that the masses of the bosons and fermions, as well as their differences, are of order $H$. We then see from \ratiogar\ and \constrib\ that in this 
  supersymmetric scenario the contributions are very small. 
  Thus, in order for the right hand side of \ratiogar\ to be of order unity (or say a few percent) the supersymmetry should be broken at the Planck scale (during inflation) so that the three-point vertex is present in the original classical theory.  Notice that most of the string inflation models {\it do not } predict a large $\alpha_4$ since
  they are based on compactifications of the ten dimensional superstring.
  It should be a model where the string length  is comparable to the Hubble radius and with a very weak coupling to account for the small  experimental 
  upper bound for  $H/M_{pl}$ \AdeULN .

Note that if one imagines that inflation is given a dual description in the spirit of dS/CFT and the dual field theory is weakly coupled, then one expects that 
$\alpha_4 H^4 \sim 1   $. This is what happens in the Vasiliev theory \AnninosUI . Of course, this theory also contains massless higher spin particles. It is also
 not suitable for building an 
inflationary model because the scalar does not appear to obey the slow roll conditions.  

If the gravity waves produced by inflation are as large as to explain the signal seen by BICEP2 \AdeXNA , 
 then probing the gravity wave three-point functions might be possible (with a lot of optimism!).\foot{
At this time, there are alternative explanations for this signal \refs{\MortonsonBJA,\FlaugerQRA},   so we might have to wait till the dust settles. }

\newsec{Conclusions} 

In this paper, we   studied causality constraints on higher derivative corrections to the graviton three-point function. 
We considered a weakly coupled theory and studied higher derivative corrections which are important before the theory becomes strongly coupled. 
These are higher derivative corrections that arise in the classical regime of the theory. 

The constraints arise from  a thought experiment where we scatter two gravitons at relatively high energy and fixed impact parameter. 
The energy is high compared to the inverse of the impact parameter but low compared to the scale where the theory becomes strongly coupled. 
More explicitly, we have the very small overall coupling $G$ and we consider corrections to the three-point functions which scale as powers of
$\alpha p^2$ relative to the Einstein one. The three-point  amplitudes are very small because $G$ is very small. But $\alpha $ is a fixed quantity and
we look at impact parameters of the order of $b^2 \sim \alpha $. 
We found that when the impact parameter $b^2 \sim \alpha$, then we see a 
causality violation.
 In this impact parameter representation, and in the field theory regime (without higher spin particles), 
  the time delay comes from the  singularities in the $t$-channel, which is simply a 
pole at $t=0$ for the massless theory. More precisely it comes from the part that is quadratic in $s$ of its residue.
 In other words, terms going like 
$s^2/t$ at $t=0$. It is important that while $t=0$, the momentum transfer itself is non-zero.\foot{It is a null, non-zero momentum.}
  The overall sign of the time delay, then depends on 
  the contractions of the polarization tensors of the external particles with the momentum transfer in the $t$-channel.

We have argued that this type of tree level causality violation can only be fixed, at tree level, 
 by higher spin particles  at a mass scale  $m^2 \sim 1/\alpha $.  In string theory, this issue is fixed because the amplitude Reggeizes. Namely, it has a 
 behavior $ s^{2 + { \alpha' t \over 2} } $ for large $s$ and fixed $t$.  This is due to extended strings being exchanged in the $s$-channel. 
  If the amplitude Reggeizes, then corrections appear at a scale 
$b^2 \sim \alpha' \log (s \alpha')$. Due to the presence of the logarithm we did not find a sharp bound between the corrections to the graviton three
point amplitude, $\alpha$,  and the Regge slope $\alpha'$.

We should stress that in this discussion we have assumed that we have a weakly coupled gravitational theory. We have also assumed the notion 
of asymptotic causality which says that the causal structured determined by the far away regions of spacetime cannot be violated by its interior regions. 

The analysis in this paper was also motivated by trying to understand better the AdS/CFT correspondence. In particular, if we consider large $N$ gauge theories
we know that we have a weakly coupled theory in the bulk. However, we do not know under what conditions that weakly coupled theory is a local   in the bulk. 
It is clear that the absence of light massive  higher spin states is a requirement. Here we have tried to address the question of whether it is sufficient. 
We have only studied the simplest possible correction to gravity, its three-point function. We have argued that, as long as higher spin particles are very massive, 
there cannot be higher derivative corrections to the three-point functions.  Previous discussions argued against such corrections by saying that they would
make the theory strongly coupled at energies that are lower than the Planck scale \refs{\HeemskerkPN,\FitzpatrickZM}, but still parametrically larger than the scale of the corrections.\foot{In \refs{\HeemskerkPN,\FitzpatrickZM}, or in talks referring to those papers,  they impose the bound $\alpha \lesssim { 1 \over G (\Delta_{gap})^4 }$ (with $G \sim 1/N^2$). This bound comes 
from demanding that the theory remains perturbatively unitary at the scale $\Delta_{gap}$ where new particles appear. Here we argued for the stronger bound  $\alpha \lesssim { 1 \over 
\Delta_{gap}^2 }$.  }
Here we have
strengthened the bound by linking the scale of corrections to the appearance of new particles at the same scale.

As a more concrete statement, we are linking the values of the constants appearing in the stress tensor three-point functions 
 to the dimensions of the lightest particles with higher spins, $J> 2 $. In other words, 
 ${a - c \over c} \lesssim {1 \over \Delta_{gap}^2}$. Unfortunately, we could not determine the precise numerical constant in this inequality. 
  
Using the results in \refs{\CornalbaXK,\CornalbaXM,\CornalbaZB}, 
we can link the time delay for a high energy scattering process in the bulk 
      to the anomalous dimensions of certain double trace operators. These double trace operators have the rough form 
      $ T \partial_+^j (\partial^2)^{n} T $. They have both relative spin and  relative radial excitations. 
      The anomalous dimensions are $\gamma(n,j) \sim  - \delta(s,b)/\pi $, where the values of $b$ and $s$ are given in terms of $j,n$ in \relour \relationtoours . 
      The requirement that the time delay is positive leads to the statement that the anomalous dimensions for some of these operators should be negative. 
 
 For the de Sitter case, this analysis has {\it potentially} interesting phenomenological applications. 
 {\it If} the gravity wave three-point functions were measured, we expect to see the structure predicted by Einstein theory. 
 However a new structure  is also  possible. This new structure is the only one allowed by the approximate scale and conformal invariance of
 the inflationary phase. {\it If} such new structure was found with a strength comparable to the Einstein one, then it would be a direct signal of dramatically 
 new physics at the Hubble scale:  a tower of higher spin particles. Which is  a rather drastic departure from ordinary 
 field theory at the inflationary scales. It is not clear how likely this inflationary scenario is in the space of possible inflationary theories.

 \subsec{Open Problems } 
 
 It would be nice to derive these constraints in a more direct way. If one understood directly the constraints of unitarity and causality at the level of the four 
 point function, then one would not need to resort to the exponentiation argument we discussed in section 3.3. Furthermore, it might lead to sharper bounds
 that include numerical factors. 
 
 Our discussion of massive intermediate particles in 
 mixed representations was not complete.\foot{These are not present in $D=4$, so that the existence of an infinite tower of higher spin
 states is clear in $D=4$, or if $\alpha_4 \not =0$ in higher dimensions. In higher dimensional theories with only $\alpha_4=0$ but 
 non-zero $\alpha_2$,  in order to establish that the tower is really  infinite we need to rule out the possibility that the causality 
 problem is fixed with a finite number of mixed representations. We leave this to the future.  }
  These are representations that have maximal spin two in the $uv$-plane, but have additional 
 indices in the other directions (see appendix H). We suspect that  a finite number of these cannot solve the causality problem, but we did not prove it.  
 
In the AdS case, 
 it would be nice to derive the constraints   from the conformal bootstrap point of view. This is in the spirit of \HeemskerkPN , but it would involve 
 the stress tensor as an external state. One of the main messages from this paper is the importance of spin in order to derive constraints. The structure constants
 (or three-point functions) of operators with spin are very numerous but, as we have shown, there can be powerful constraints on them. 
 These constraints are not so easily seen when we scatter external operators with no spin. 
 Notice that, even the simplest bounds for $a$ and $c$ 
 discussed in \HofmanAR , which should be valid for arbitrary CFTs, 
 have not been derived from the conformal bootstrap approach. 

It would be nice to further constrain the interactions of all the higher spin particles and derive the general structure of the tree level theory. 
 This is the program that was pursued in the sixties and that led to string theory. However, we would like to know how unique string theory is.  
 Methods developed to tackle this problem might also be  useful  for analyzing large $N$ gauge theories such as large $N$ QCD. 

 It would also be nice to see whether in the de Sitter context there is a sharp bound for the gravity wave three-point correlators analogous to the one in 
 \HofmanAR\ for AdS correlators. Our de Sitter discussion assumed the local gravity description and a locally flat space discussion in the bulk, 
 so it applies most clearly when $\alpha_4 H^4$ is somewhat smaller than one, but still of order one compared with $H/M_{pl}$.

\newsec{Acknowledgments}

We would like to thank  N. Arkani-Hamed, S. Caron-Huot, A. Dymarsky, S. Giddings, S. He, D. Hofman, Y. Huang, Z. Komargodski, D. McGady, R. Myers, J. Penedones, L. Rodina, D. Simmons-Duffin, E. Skvortsov, A. Strominger and R. Wald for discussions.
X.C. and J.E.\ thank the Institute for Advanced Study for hospitality at the initial stages of this work.
J.E.\ is supported in part by MINECO and FEDER (grant FPA2011-22594), by Xunta de Galicia (GRC2013-024), and by the Spanish Consolider-Ingenio 2010 Programme CPAN (CSD2007-00042).
J.M.\ is supported in part by U.S.\ Department of Energy grant DE-SC0009988. 
The Centro de Estudios Cient\'\i ficos (CECs) is funded by the Chilean Government through the Centers of Excellence Base Financing Program of Conicyt.

\appendix{A}{Shapiro Time Delay}

The physical effect discussed in the paper is known in general relativity as the Shapiro time delay \ShapiroUW . The usual setup to discuss the Shapiro time delay is to consider propagation of light near a massive body (a star or a planet). Here we would like to consider a slight variation of it by considering propagation of light between two massive bodies. We consider the masses to be equal and consider the geodesic that is equally separated from each of them, such that the deflection is absent (each of the masses bends the trajectory in the opposite direction such that the net deflection is zero). The time delay on the other hand accumulates. Recall that the Schwarzschild metric takes the form
\eqn\metrci{ 
ds^2 = - \left( 1  - {r_s^{D-3} \over r^{D-3} }  \right) dt^2 f + { dr^2 \over 1  - {r_s^{D-3} \over r^{D-3} }  } + r^2 d\Omega^2 
}

\ifig\Shapiro{   We consider the metric produced by two massive sources at distance $2b$. We then send a particle between them  and measure the time delay. 
There is no deflection angle, but there is a non-zero time delay. We do the computation to leading order in the $r_S/b$ expansion, by simply consider a linear 
superposition of the two metrics.    }  {\epsfxsize1.2in \epsfbox{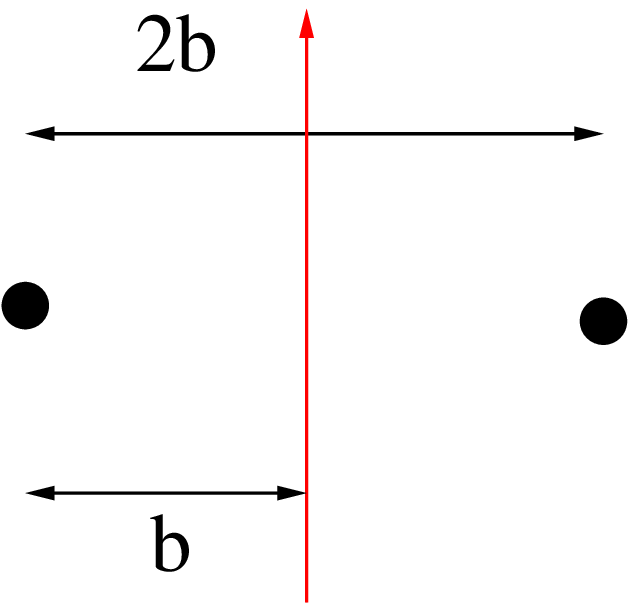} } 

We are interested in the metric produced by the superposition of two equal masses. In the Schwarzschild coordinates to first order in the mass, or $r_S$, we get 
\eqn\mfir{ 
ds^2  = ds^2_{Mink} + \sum_i { r_{s}^{D-3} \over r_i^{D-3} }(  dt^2 + dr_i^2 ) 
}
with $r_i = |\vec x - \vec x_i|$. Let us now consider two masses separated along the direction $x^1$, one at $x^1=-b$ and one at $x^1=b$. 
Then we consider a probe particle moving along the direction $x^{D-1}= z$. By symmetry, it will stay at $x^{1}=...=x^{D-2}=0$ if it starts there with velocity along the $z$ direction. We have that $r = \sqrt{ b^2 + z^2}$ for both masses. We also find that $dr = dz z/r $. We get 
\eqn\metrcg{ 
dt^2 (1 - 2  { r_{s}^{D-3} \over r^{D-3}}  ) = dz^2 ( 1 +  { r_{s}^{D-3} \over r^{D-3}} z^2/r^2 ) 
}
Then the time delay is 
\eqn\timedel{ 
\Delta t = \int_{-\infty}^{\infty} dz  r_s^{D-3} { 1 \over (b^2 + z^2)^{D-3\over 2 }} ( 1 + { z^2 \over z^2 + b^2 } ) = { r_s^{D-3} \over b^{D-4} } { (D-2) \sqrt{\pi} 
\Gamma( { d \over 2 } -2 ) \over 2 \Gamma({ D -1 \over 2 } ) } 
}

We now do the same for a particle moving at a velocity $v$. We find 
\eqn\metrcgns{ \eqalign{ 
& dt^2 (1 - 2  { r_{s}^{D-3} \over r^{D_3}}  ) = dz^2 \left( 1 +  { r_{s}^{D-3} \over r^{D_3}}  { z^2 \over z^2 + b^2 }   \right)  + d\tau^2 
\cr
& { dt \over d\tau}  (1 - 2  { r_{s}^{D-3} \over r^{D_3}}  )  = \gamma ~,~~~~~~~ \gamma^{-2} = 1 -v^2 
\cr  
& dt = { dz \over v } \left( 1 + (  2 +  { z^2 \over z^2 + b^2 }  - { 1 \over v^2 } )   { r_{s}^{D-3} \over r^{D_3} }\right) 
}}
The result of doing the integral over $z$ is the same as in \timedel\ multiplied by a factor of 
\eqn\newfac{ 
\Delta t = { 1 \over v } { ( 2 + { 1 \over D-3 } - { 1 \over v^2 } ) \over (1 + { 1 \over D-3 } ) } \Delta t_{\rm relativistic } 
}

We see that for slow velocities, we indeed find a negative time delay  (time advance) proportional to $1/v^3$. This is time advance relative to the particle 
moving with the same velocity $v$ in flat space (not relative to a particle moving at the speed of light!). 

We could have repeated the computation in different coordinate system. For example, we can consider the computation in the so-called isotropic coordinates given by
\eqn\isotrop{\eqalign{
d s^2 &=- \left( {1 - {r_s^{D-3} \over 4 r^{D-3}} \over 1 + {r_s^{D-3} \over 4 r^{D-3}} }\right)^2 d t^2 + \left( 1 + {r_s^{D-3} \over 4 r^{D-3}} \right)^{{4 \over D-3}} \sum_{i=1}^{D-1} dx_i^2, \cr
r^2 &= \sum_{i=1}^{D-1} x_i^2.
}}

One can easily check that the time delay for the geodesic in this coordinates coincides with the one computed in Schwarzschild coordinates to leading order in $r_S$.

\appendix{B}{Three-Point Amplitudes and Their Sums}

In this appendix we recollect different three-point amplitudes that involve a graviton and two other particles and 
present different polarization sums that appear in the time delay. In particular cases we reproduce the shock
wave computations but the results obtained using on-shell amplitudes are much more general and are valid in any 
theory with given on-shell three-point amplitudes. 

Let us be more explicit on the kinematics we are interested in. As defined in the bulk of the paper we consider the following kinematics
\eqn\momconfapp{ \eqalign{ 
p_{1 \,\mu} = & \left(p_u, { q^2 \over 16 p_u} , {\vec q \over 2 } \right)~,~~~~~p_{2 \, \mu} =  \left({ q^2 \over 16 p_v} ,p_v,- { \vec q \over 2} \right)~,~~~~~~~
\cr
p_{3 \, \mu} = & -  \left( p_u,  { q^2 \over 16 p_u} , - { \vec q \over 2}\right) ~,~~~~~
p_{4 \, \mu} =   -  \left({q^2 \over 16 p_v} , p_v , { \vec q \over 2 }\right) ~,~~~~~~~
\cr
s \simeq & 4 p_u p_v ~,~~~~~~t \simeq   - ( \vec q)^2   
}}

We define the polarization tensors as follows

\eqn\canrwiapp{\eqalign{ 
\eps^{ 1\, \mu } &= (-{ \vec q . \vec e_1 \over 2 p_u},0 , \vec e_1 )~,~~~~~\eps^{3\, \mu } =  ( { \vec q . \vec e_3 \over 2 p_u},0 , \vec e_3 ), \cr
\eps^{ 2\, \mu } &= (0,{ \vec q . \vec e_2 \over 2 p_v}, \vec e_2 ) ~,~~~~~\eps^{4\, \mu } =  (0, -{ \vec q . \vec e_4 \over 2 p_v} , \vec e_4 ).
}}

An important point is that all $\eps_i . p_j$ are of order $1$ and, thus, are sub-leading in the high energy limit compared to powers of $s$. Thus, in all our computations we can think of polarization tensors as being purely transverse, see \factfo .

We focus on the massless scalar, vector and graviton three-point amplitudes which are parameterized as follows
\eqn\threescalarandvector{\eqalign{
{\cal A}_{h h g} &=\sqrt{32 \pi G} \eps_{\mu \nu} p_1^{\mu} p_3^{\nu} \cr
{\cal A}_{\gamma \gamma g} &=  \sqrt{32 \pi G}   \eps_{\mu \nu} \left( \left[ p_1^{\mu} p_3^{\nu} (\eps_1 . \eps_3) -  \eps_1^{\mu} p_3^{\nu} (\eps_3 . p_1) - \eps_3^{\mu} p_1^{\nu} (\eps_1 . p_3)  \right] + \hat \alpha_2 p_1^{\mu} p_3^{\nu} (\eps_1 . p_3)  (\eps_3 . p_1)  \right), \cr
{\cal A}_{ggg} &= \sqrt{32 \pi G}  \left[ ( \epsilon_1 . \epsilon_2 \epsilon_3 . p_1 + \epsilon_1 . \epsilon_3 \epsilon_2 . p_3 + \epsilon_2 . \epsilon_3 \epsilon_1 . p_2)^2 \right. \cr
&\left. + \alpha_2 ( \epsilon_1 . \epsilon_2 \epsilon_3 . p_1 + \epsilon_1 . \epsilon_3 \epsilon_2 . p_3 + \epsilon_2 . \epsilon_3 \epsilon_1 . p_2) \epsilon_1.p_2 \epsilon_2 . p_3 \epsilon_3 . p_1 \right. \cr
&\left. + \alpha_4 (\epsilon_1.p_2 \epsilon_2 . p_3 \epsilon_3 . p_1)^2 \right].
}}
where we used the $\eps^{\mu \nu} \to \eps^{\mu} \eps^{\nu}$ form of the polarization tensor for the graviton . Using these three-point amplitudes we can compute the time delay using \phaseshift \ in the high energy limit. In the high energy limit the relevant part of the sum over graviton polarization tensors takes a very simple form
\eqn\simpleformcomplet{
\sum_{i} \eps^{i}_{\mu \nu}(q) (\eps^{i}_{\rho \sigma}(q))^{*} \sim {1 \over 2} \left( \eta_{\mu \rho} \eta_{\nu \sigma} +  \eta_{\mu \sigma} \eta_{\nu \rho} \right)
}
so that it leads to factors of $s^2$  when we contract with the $p_1$ or $p_3$ momenta from the left side and with $p_2$ or $p_4$ momenta on the right side, 
see \OnShell . Note that the large components of the external momenta are transverse to the momentum $\vec q$ of the intermediate particle, 
$p_1 + p_3 = ( 0,0,\vec q )$, see \momconfapp . 

The results are the following
\eqn\resultforamplitudesums{\eqalign{
\sum {\cal A}_{hhg} ( - i \pa_{\vec b}) {\cal A}_{hhg} ( - i \pa_{\vec b}) &=8 \pi G s^2 , \cr
\sum {\cal A}_{hhg} ( - i \pa_{\vec b}) {\cal A}_{\gamma \gamma g} ( - i \pa_{\vec b}) &=8 \pi G s^2 \left( e_2 . e_4 + \hat \alpha_2 e_2 . \pa_b e_4 . \pa_b \right),  \cr
\sum {\cal A}_{hhg} ( - i \pa_{\vec b}) {\cal A}_{g g g} ( - i \pa_{\vec b}) &= 8 \pi G s^2 \left( e_2^{i j} e_4^{i j} + \alpha_2 e_2^{i j} e_4^{i k} \pa_{b^j} \pa_{b^k} + \alpha_4 e_2^{i j} e_4^{k l} \pa_{b^i} \pa_{b^j} \pa_{b^k} \pa_{b^l} \right), \cr
\sum {\cal A}_{\gamma \gamma g} ( - i \pa_{\vec b}) {\cal A}_{\gamma \gamma g} ( - i \pa_{\vec b}) &=8 \pi G s^2 \left( e_1 . e_3 + \hat \alpha_2 e_1 . \pa_b  e_3 . \pa_b \right) \left( \eps_2 . e_4 + d_2 e_2 . \pa_b  \eps_4 . \pa_b \right), \cr
\sum {\cal A}_{\gamma \gamma g} ( - i \pa_{\vec b}) {\cal A}_{g g g} ( - i \pa_{\vec b}) &=8 \pi G s^2 \left( e_1 . e_3 + \hat \alpha_2 e_1 . \pa_b  e_3 . \pa_b \right) \cr
& (e_2^{i j} e_4^{i j} + \alpha_2 e_2^{i j} e_4^{i k} \pa_{b^j} \pa_{b^k} + \alpha_4 e_2^{i j} e_4^{k l} \pa_{b^i} \pa_{b^j} \pa_{b^k} \pa_{b^l}), \cr
\sum {\cal A}_{g g g} ( - i \pa_{\vec b}) {\cal A}_{g g g} ( - i \pa_{\vec b}) &=8 \pi G s^2 ( e_1^{i j} e_3^{i j} + \alpha_2 e_1^{i j} e_3^{i k} \pa_{b^j} \pa_{b^k} + \alpha_4 e_1^{i j} e_3^{k l} \pa_{b^i} \pa_{b^j} \pa_{b^k} \pa_{b^l}) \cr
& (e_2^{i j} e_4^{i j} + \alpha_2 e_2^{i j} e_4^{i k} \pa_{b^j} \pa_{b^k} + \alpha_4 e_2^{i j} e_4^{k l} \pa_{b^i} \pa_{b^j} \pa_{b^k} \pa_{b^l}).
}}
where all of the operators are acting on the propagator $ 1/b^{D-4}$. 
To reproduce the usual shock wave computations only the first three formulas are relevant. In the case of electrodynamics matching with \totaime\ is manifest. In the case of Gauss-Bonnet theory we have from \totaimeGB\  $\alpha_2 ={\lambda_{GB} \over 4} $, $\alpha_4 = 0$. 

Consider now the coupling of the graviton to a massive spin two particle which is a relevant amplitude for the discussion in the bulk of the paper. We get
\eqn\threescalarandvectorMassive{\eqalign{
{\cal A}_{gg \tilde g} &= \tilde \alpha_2 \eps_{\mu \nu}  \left[ \eps_1^{\mu} p_3^{\nu} (\eps_3 . p_1) + \eps_3^{\mu} p_1^{\nu} (\eps_1 . p_3) - p_1^{\mu} p_3^{\nu} (\eps_1 . \eps_3) - \eps_1^{\mu} \eps_3^{\nu} (p_1 . p_3)\right] \cr
& \left[(\eps_1 . \eps_3) (p_1 . p_3) - (\eps_3 . p_1) (\eps_1 . p_3) ] \right] \cr
&+ \tilde \alpha_4 \eps_{\mu \nu} p_1^{\mu} p_3^{\nu} \left[  (\eps_1 . p_3)  (\eps_3 . p_1) -  (\eps_1 . \eps_3)  (p_1 . p_3)  \right]^2.
}}
Notice that instead of three structures that are present in the massless amplitude we have only two. 

For the coupling to spin four particle or higher we can  have three structures \CostaMG . They take the same form as above with extra indices of particle of spin $J$ being contracted with momenta. An additional structure takes the form
\eqn\additional{\eqalign{
&\tilde c_{extra} \eps_{\mu \nu \rho \sigma} .p_1 ... p_1 \left[ \eps_1^{\mu} p_3^{\nu} (\eps_3 . p_1) + \eps_3^{\mu} p_1^{\nu} (\eps_1 . p_3) - p_1^{\mu} p_3^{\nu} (\eps_1 . \eps_3) - \eps_1^{\mu} \eps_3^{\nu} (p_1 . p_3)\right]\cr
&\left[ \eps_1^{\rho} p_3^{\sigma} (\eps_3 . p_1) + \eps_3^{\rho} p_1^{\sigma} (\eps_1 . p_3) - p_1^{\rho} p_3^{\sigma} (\eps_1 . \eps_3) - \eps_1^{\rho} \eps_3^{\sigma} (p_1 . p_3)\right] .
}}

We are interested in the contribution of new particles to the time delay. The computation is almost identical to the one we did for the graviton exchange. The first difference is that we have to compute the Fourier transform of the massive propagator in $D-2$ dimensions. The result is given by $({m \over b})^{{D-4 \over 2}} K_{{D-4 \over 2}} ( m b)$ which decays exponentially fast for $m b \gg 1$. 
The second difference comes from slightly different structure of the three-point functions.

Similarly, we can write down the contribution to the time delay of some higher spin particle the difference being that the sum of three-point amplitudes give $s^{J}$ and we have an extra structure in the three-point amplitude for external gravitons which is an analogous to the Einstein one.

The formulas above are valid in $D>4$, whereas in $D=4$ $\alpha_2$ and $\tilde \alpha_2$ structures above are absent\foot{They are identically zero.} but instead we have a parity odd structure. In $D \geq 5$ parity odd structures are absent in the class of amplitudes considered above.

\subsec{Scattering of a Scalar and a Graviton }

As an example let us reproduce the computation for the graviton that we did using the shock wave. 
It corresponds to scattering from the energetic scalar particle (we could have considered graviton-graviton scattering as well if we average of all polarizations of
gravitions 1 and 3).\foot{ More precisely we set polarizations one and three to be equal and then we sum over all of them. Physically we are considering a sequence
of coherent state with the various alternative polarizations. This makes  the $\alpha_2$ and $\alpha_4$ contributions vanish  in the ${\cal A}^{13I}$ part of the amplitude.}

As an example let us reproduce the computation for the graviton that we did using the shock wave. It corresponds to scattering from the energetic scalar particle (we could have considered graviton-graviton scattering as well).
\eqn\sumscalar{\eqalign{
\sum_{ {\rm states} } {\cal A}^{hhg} ( - i \pa_{\vec b}) {\cal A}^{g g g} ( - i \pa_{\vec b}) &\propto   G s^2 (  e_2^{i j} e_4^{i j} + \alpha_2 e_2^{i j} e_4^{i k} \pa_{b^j} \pa_{b^k} + \alpha_4 e_2^{i j} e_4^{k l} \pa_{b^i} \pa_{b^j} \pa_{b^k} \pa_{b^l}){ 1 \over b^{D-4}}
}}
Setting $\alpha_4 =0$,  $\alpha_2 = - {\lambda_{GB} \over 4}$   we reproduces the computation from the previous section \totaimeGB .

  Let us now study the constraints that follow from the positivity of the time delay.
Introducing new variables 
\eqn\energycorr{\eqalign{
t_2 &= {(D-2)(D-4) \alpha_2 \over  b^2} - {4 (D-4)(D-2) D \alpha_4 \over b^4} \cr
t_4 &= { (D-4)(D-2) D (D+2) \alpha_4 \over  b^4}
}}
we can write the phase shift in the form familiar from the study of the energy correlators
\eqn\phaseshift{
\delta(s,\vec b ) \sim 1 + t_2 \left({ (e.n)^2 \over e.e} - {1 \over D-2} \right) + t_4  \left({ (e.n)^4 \over (e.e)^2} - {2 \over D (D-2)} \right)
}
which coincides with the formula (3.6) in \refs{\CamanhoVW,\BuchelSK}.\foot{In \BuchelSK\ $d = D-1$ .} Here $\vec n = { \vec b \over |\vec b|}$ and 
$\vec e$ is the graviton polarization of particles two and four. 
 Thus, positivity constraints from causality are identical to the ones obtained in their analysis with identification of parameters as above \energycorr, namely
\eqn\positivity{\eqalign{
1- {t_2 \over D - 2 } - {2 t_4 \over D (D-2)} &\geq 0  \cr
\left( 1- {t_2 \over D-2} - {2 t_4 \over D (D-2)} \right) +{t_2 \over 2} &\geq 0 \cr
\left( 1- {t_2 \over D-2} - {2 t_4 \over D (D-2)} \right) +{D-3 \over D-2} (t_2 + t_4) &\geq 0
}}
So we get the bounds on $\alpha_2$ and $\alpha_4$ depending  for how small a $b$ we can trust the computation.
 If the computation is trustworthy for arbitrarily small $b$ we are forced to set $\alpha_2$ and $\alpha_4$ to zero.

\appendix{C}{The  QED  case }

 Note that  
 the  action  \conscte\ also arises in QED (quantum electrodynamics) 
 after we integrate out the massive electron  \DrummondPP . In that case $\hat \alpha_2 \propto { e^2 \over m^2 }$. 
  This is a one loop effect, suppressed by the coupling $e^2$.
  In this paper we have taken the coupling to be very small, so that we would have treated this $\hat \alpha$ as being essentially zero. 
  The discussion of this paper concentrated on the case that the higher curvature corrections were present at tree level, so that the causality problem 
  had to be solved by tree level physics. 
  In this appendix, we   consider this loop generated term in QED and we will show that the potential causality problem is solved by 
  one loop effects. 
    
      In QED, when we get to an impact parameter  of order $m^{-1}$ we cannot be satisfied with the low energy action \conscte . Fortunately
  the necessary diagrams were computed  in \BerendsAH . Using their results it is possible to go to the impact parameter representation (doing the Fourier 
  transform) and check that for 
  $b>1/m$ we get a result that agrees with the the simple Lagrangian \conscte , but for $b< 1/m $ we get a different result which displays no causality problem. 
  In other words, the potential causality problems arising from \conscte\ appear at $b \sim 1/\sqrt{|\alpha|} \sim e/m $, but at a larger distance, $b \sim 1/m$, 
  the computation should be already modified and we obtain results consistent with causality.\foot{ See \DubovskyAC\  and references therein 
  for a related discussion of this problem.} We can view this modification as arising from the propagation of electron 
  positron pairs along the $t$-channel. 

Notice, that, in addition,  
when the photon goes through the shock we can have electron positron pair creation.
 We can view this an another example of tidal excitations. Indeed in QED, the photon has a non-zero probability of becoming an electron positron pair. 

\appendix{D}{Causality and Unitarity for a Signal Model} 

In this appendix we review the constraints from causality and unitarity in the context of a simple signal model. 
We imagine a signal propagating along one dimension. We have an initial signal which is a function of time $f_{\rm in}(t)$
and an out-signal $f_{\rm out}(t)$ which, in Fourier space is given by $f_{\rm out}(\omega) = S(\omega) f_{\rm in}(\omega) $ or 
\eqn\outpf{ 
f_{\rm out}(t) =  \int dt' \int d \omega S(\omega ) e^{ - i \omega ( t - t' ) } f_{\rm in}(t') 
}
Causality implies that if  $f_{\rm in}(t') =0$ for $t'< 0$, then $f_{\rm out}(t) =0$ for $t< 0$. 
By unitarity we mean that the $L_2$ norm of the out-signal should be smaller than that of the in-signal $ \int d t |f_{\rm out} (t)|^2 
\leq \int dt |f_{\rm in}(t) |^2 $. 
Now it is well known that the Fourier transform of  a function which vanishes for $t<0$ is analytic in the upper half $\omega$ plane. 
This follows directly from the explicit integral expression for the Fourier transform. Then if $f_{\rm in}=0$ for $t< 0$ we find that 
  both $f_{\rm in}(\omega)$ and $f_{\rm out}(\omega )$ are analytic in the upper half plane. This also implies that 
$S(\omega)$, which is given  their ratio,  is also analytic. One might worry that $S(\omega)$ could have poles at zeros of $f_{\rm in}(\omega)$. However, 
we can change the location of the zeros of $f_{\rm in}(\omega)$ by choosing different functions.
 Therefore $S(\omega)$ is analytic in the upper half plane. 

We will now prove that unitarity implies that $|S(\omega)| \leq 1$ in the upper half plane.  
With some foresight, we pick a particular $f_{\rm in}(t)$ of the form 
\eqn\choosf{ 
f_{\rm in}(t) = e^{ - \gamma t } e^{ - i \omega_0 t }  { \theta(t)   \sqrt{ 2 \gamma}  }
}
with $\gamma>0$ and $\omega_0$ real. Note that $|| f_{\rm in } ||^2 = \int dt |f_{\rm in }(t)|^2 =1$. 
 For $Im(\omega)>0$ we can now write
 \eqn\boundfs{\eqalign{
|f_{\rm out}(\omega ) |^2 \leq &  \left| \int_0^\infty dt e^{ i \omega t } f_{\rm out}(t) \right| 
\leq \int_0^\infty dt  |e^{ i \omega t } |^2  \int_0^\infty dt |f_{\rm out}(t) |^2 ={ 1 \over 2 Im(\omega ) }  || f_{\rm out}  ||^2 
\cr
|f_{\rm out}(\omega ) |^2 \leq & { 1 \over 2 Im(\omega ) }  
\cr
|S(\omega )|^2 =& { |f_{\rm out}(\omega)|^2 \over | f_{\rm in }(\omega) |^2 } 
\leq \,    { 1 \over 2 Im(\omega) } { 1 \over | f_{in}(\omega )|^2}
}}
Here we used the Cauchy-Schwartz inequality. Note that $|e^{ i \omega t} |^2 = e^{ - 2 Im(\omega) t } $. We also used that $||f_{\rm out}||^2 \leq ||f_{\rm in}||^2 = 1$. 
We can now set $\omega = \omega_0 + i \gamma$ and find that  $f_{\rm in}(\omega_0 + i \gamma)  = 1/\sqrt{2 \gamma}$ for the specific function \choosf . 
Inserting this into \boundfs\ we then find that 
$|S(\omega_0 + i \gamma) |\leq 1$, which is what we wanted to prove, since $\omega_0$ and $\gamma$ are arbitrary. 

In conclusion,  we find that $S(\omega)$ should be analytic and bounded $|S(\omega ) | \leq 1$ 
 in the upper half plane. These are necessary and sufficient conditions.\foot{Note that some functions  which are analytic in the upper half plane, such as 
 $S(\omega) = e^{ i \omega^3 }$ are actually not causal.}

\ifig\ShockSignalModel{ 
We consider a $v$ independent perturbation that is localized in the $u$ direction, given here by the shaded region. 
We then consider signal propagating along the $u$ direction, which are 
$v$ dependent and demand causality. We can consider an $S$ matrix that connects the region before the perturbation to the region after the perturbation. 
}  {\epsfxsize2.2in \epsfbox{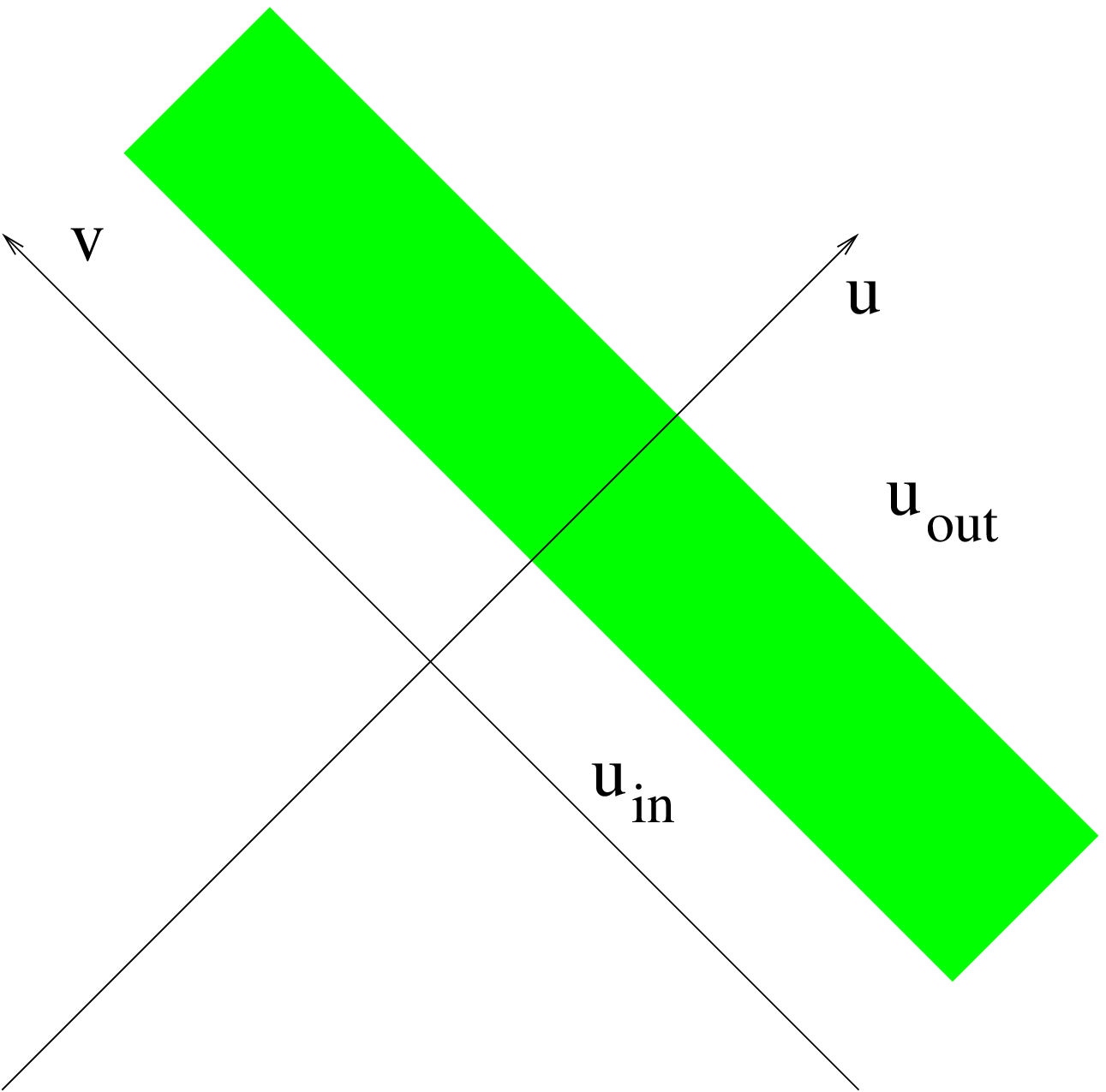} } 

Let us now briefly mention how this is connected to the field theory situation. We consider light cone coordinates $u$ and $v$. 
We consider a perturbation that is translation invariant in $v$ but is localized in the $u$ coordinate. We call this ``the shock''. 
We expand the fields in the 
$v$ coordinate at some $u_{\rm in}$ and then we expand them again at some $u_{\rm out}$ after the shock.
To make contact with the above discussion we call $t=v$ and  $p_v = -\omega$. We can expand the field as 
\eqn\fieldop{ 
\phi(t) = \int_0^\infty { d \omega \over \sqrt{\omega} } (  a_\omega e^{ - i \omega t } + a^\dagger_{\omega } e^{   i \omega t } ) 
}
We can do this for $\phi_{\rm in } $ and $\phi_{\rm out}$ in terms of $a_{\rm in}$ and $a_{\rm out}$. 
These oscillators then are related by 
\eqn\relosc{ 
a_{\omega , {\rm out} } = S(\omega ) a_{\omega , {\rm in } } ~,~~~~~~~a^\dagger_{\omega , {\rm out} } = S(\omega )^* a^\dagger_{\omega , {\rm in } }
}
This defines $S(\omega )$ for positive $\omega$. For negative $\omega$ 
 we can define $S(-\omega) = S(\omega)^*$. 
Alternatively, we can define 
\eqn\rrefs{ 
S(\omega ) = -\int_{-\infty}^\infty  dt e^{   i \omega t } [ \phi_{\rm out}(t) ,  i \partial_t \phi_{\rm in}(0)  ] =  -
\int_{0}^\infty  dt e^{   i \omega t } [ \phi_{\rm out}(t) ,  i \partial_t \phi_{\rm in}(0)  ] 
}
The commutation relations for the in and out oscillators require that $|S(\omega ) |^2 =1$. 

In a case where there is particle mixing, but no particle creation, 
then the fields have   indices $\phi^i_{\rm in }$ and  $\phi^j_{\rm out}$. Now $S$ is a matrix which 
obeys $S_{ij}(-\omega) = S_{ij}(\omega)^*$ and $S_{ij}(\omega) S_{kj}(\omega)^* = \delta_{ik}$ due to the commutation relations of the oscillators before and after the
shock.  In the preceding discussion we have neglected the transverse dimensions. We can now remedy that by including the momentum in the transverse dimensions as part of the indices we are discussing here. 

If we consider a signal that is made out of physical particles, one might correctly 
worry that the fact that $\omega > 0$ will preclude us from localizing the signal in time. 
In order to avoid this issue we 
 can consider a coherent state of the form 
\eqn\psidef{ 
 |\Psi \rangle = e^{ i \int dt f_{\rm in }(t ) \phi_{\rm in}(t) } | 0 \rangle
 }
 with a real function $f_{\rm in}$. This is a state that could be produced by adding a hermitian term to the Hamiltonian at some early time $u_{\rm in}$. 
 On this state we have the expectation values 
\eqn\expecval{ 
\langle \Psi | \partial_t \phi_{\rm in}(t) |\Psi \rangle =  f_{\rm in }(t) ~,~~~~~~~~~~~\langle \Psi | \partial_t \phi_{\rm out}(t) |\Psi \rangle = f_{\rm out }(t) ~
}
where the functions are related  as in the signal model.\foot{$f_{\rm out}$ is real if $f_{\rm in}$ is real when $S(-\omega) = S(\omega)^*$.}
Here we assumed a linear relation between the in- and out-signals.
Furthermore, we can also consider the expectation values of the normal ordered product 
$T_{vv} = T_{tt} = : \partial_t \phi(t)\partial_t \phi(t): $. When this is evaluated on the state \psidef , and integrated over $t$ we find that the answer is given by 
\eqn\momstr{ 
- P_v^{\rm in, out} = \int dv T_{vv} = \int dt (f_{\rm in, out}(t))^2 = || f_{\rm in , out} ||^2
}
 Thus, the condition that the total  light-cone momentum
 $P_v$ should not increase implies the norm condition $||f_{\rm out} ||^2 \leq || f_{\rm in} ||^2 $. 

More precisely, we can consider the signal $f_{\rm in}$ exciting a mode involving a graviton with a given polarization. 
The signal $f_{\rm out }$ is the same mode of the graviton. In addition, the initial  graviton could go into other massive particles.
Then the condition that the total $P_v$ in the out-graviton mode should be no bigger than the initial $P_v$, which was all contained in the graviton mode, leads
to the norm condition (or unitarity condition) for the signal model. 
In conclusion, the graviton-graviton matrix element $S_{gg}(\omega)$ obeys all the assumptions of the signal model. Therefore, it should be analytic 
and $|S_{gg}(\omega)| \leq 1$ in the upper
half plane. 

Note that we have assumed here a perfect $v$-translation symmetry for the perturbation that creates the shock. In our scattering problem, see \kinematics , 
particles 1 and 3 have small $p_v$ momentum. Thus, in this discussion, we have neglected this small momentum. This is reasonable for $s b^2 \gg 1$.

 \appendix{E}{Scattering in String Theory } 

String theory in flat space is the simplest example of a theory that follows into the category of   weakly coupled gravitational theories
with higher derivative corrections that are subject of our analysis. As explained in the introduction, a  motivation for this work was actually
to argue that string theory is inevitable, at least, under certain assumptions. 

It is well known that effective gravitational actions in string theory contain higher derivative corrections  at the string scale 
\refs{\GrossIV,\MetsaevZX}. 
In particular graviton three-point amplitudes can contain the  higher derivative terms that we constrained in this paper using causality. 
The potential problem is fixed by extra particles with string scale masses. Here we would like to understand how this is happening in detail.

Let us first recall the form of the three-point gravity amplitudes in bosonic, heterotic and type II string theories  \MetsaevYB
\eqn\threegravST{\eqalign{
{\cal A}^{ggg} &= \sqrt{32 \pi G }  \eps_1^{\mu_1 \mu_1 ' }  \eps_2^{\mu_2 \mu_2 ' }  \eps_3^{\mu_3 \mu_3 ' } N_{\mu_1 \mu_2 \mu_3} \bar N_{\mu_1 '  \mu_2 '  \mu_3 ' } \cr
N^{\mu_1 \mu_2 \mu_3}&= k_2^{\mu_1} \eta^{\mu_2 \mu_3} +  k_3^{\mu_2} \eta^{\mu_1 \mu_3} +  k_1^{\mu_3} \eta^{\mu_1 \mu_2} + {\alpha'  \over 2} \eps k_2^{\mu_1} k_3^{\mu_2} k_1^{\mu_3}
}}
and we have $\eps_{bos} = \bar \eps_{bos} = \eps_{het} = 1$ and $\eps_{II} = \bar \eps_{II} =\bar \eps_{het} = 0$ . Translating it to our notations we get
\eqn\ournot{\eqalign{
\alpha_2^{bos} &= 2 \alpha_2^{het} =  \alpha',~~~ ~~~~~~~~~~~~~\alpha_2^{II} = 0, \cr
\alpha_4^{bos} &=   {(\alpha')^2 \over 4},~~~~~~~~~~~~~~~~~~~~~  \alpha_4^{het} = \alpha_4^{II}=0.
}}
The vanishing of  some of these corrections can be understood from  supersymmetry, as explained in section 3.2. 
For the purposes of this paper, the type II case is not interesting since there are not corrections at all for the graviton three-point function. 

The high energy scattering problem in string theory was studied in a nice series of papers by  Amati, Ciafaloni and Veneziano \refs{\AmatiWQ,\AmatiUF,\AmatiTN,\AmatiZB,\AmatiTB}, see also e.g. \refs{\HorowitzBV ,\deVegaTS}.  Let us first review their picture. 
The scattering can be described in terms of a phase shift defined as 
\eqn\stringtheoryphaseshift{\eqalign{
\delta (\vec b, s) &= [{\rm POL}] \delta^{ACV} (\vec b, s), \cr
\delta^{ACV} (\vec b, s)&= \int {d^{D-2}\vec q \over (2 \pi)^{D-2}}  e^{i \vec q . \vec b}  C(s,t,u), \cr
C(s,t,u) &= {\Gamma (- {\alpha' s \over 4}) \Gamma (- {\alpha' t \over 4}) \Gamma (- {\alpha' u \over 4}) \over \Gamma ( 1 + {\alpha' s \over 4}) \Gamma ( 1 + {\alpha' t \over 4}) \Gamma (1 + {\alpha' u \over 4})  },
}}
where [POL] represents a factor that depends on the polarizations and is polynomial in the momenta. We will only need its form in a specific limit. 
  In the high energy limit $C(s,t,u)$ has the celebrated Regge behavior
\eqn\reggeofc{
C(s,t,u) \sim {  \Gamma(- {\alpha' t \over 4}) \over  \Gamma(1 + {\alpha' t \over 4})} \left(- i {s \alpha' \over 4} \right)^{-2+ {t \alpha' \over 2} } .
}

This Regge form is reflecting the creation of particles in the $s$-channel. The infinite sequence of $s$-channel poles is becoming a cut, 
the cut arising when $s \to s e^{ 2 \pi i }$. The creation of the massive $s$-channel states is also related to the 
  fact that we get an imaginary part from the $(-i)^{ t \alpha'\over 2}$ factor in \reggeofc\ is 
saying that the most likely process is to create a massive closed string, rather than scattering the gravitons.

For large $s$,  only small $q$ will contribute and we can approximate the prefactor in \reggeofc\ by $1/t$. Then the integral  becomes
\eqn\conclu{ 
\delta^{[ACV]} \propto  \int { d^{D-2} \vec q  \over 2(2 \pi)^{D-2} } { e^{ i \vec q . \vec b } \left(- i {s \alpha' \over 4} \right)^{ - { \vec q^{\, 2 } \alpha' \over 2} }\over \vec q^{\,2} } 
= { 1 \over ({\alpha'  Y \over 2 })^{D-4 \over 2 } }\int_0^1 d \rho \rho^{ D- 6 \over 2 } e^{ - { b^2 \over 2 \alpha'   Y } \rho } 
}
where $ Y = \log( -i s \alpha'/4) $. 
We have two characteristic behaviors, depending on whether $b^2$ is larger or smaller than $\alpha' \log(s \alpha')$. 
 For large $b$ we get the usual $1/b^{D-4}$ behavior. For small $b$ we get the result in brackets in \strinphs . 
Note that since the transition region occurs for a $b^2$ which is larger than $\alpha'$ by a $\log s $ factor, we can, a posteriori, justify the fact that we have 
approximated the prefactor in \reggeofc\ by $1/t$.

In our field theory discussion we had represented the phase shift as a sum over poles.  We can wonder how this applies to the string theory discussion. 
  Notice that from \reggeofc\ we  get   a Gaussian integrand factor 
of the form 
\eqn\gaussin{ 
e^{ i \vec q . \vec b } e^{ - (\vec q)^2 { \alpha' \over 2 } \log(-i  s \alpha'/4) }  
} 
 Let us 
assume that $\vec b = (b, 0,\cdots,0)$, so that it
 points along the first coordinates. When we do the integral over the first component of $\vec q$, call it $q_1$, 
 we get a saddle point for \gaussin\ at 
\eqn\saddl{ 
q_s = i { b \over \alpha' \log ( - i s\alpha'/4) } 
}
\ifig\SaddleContour{ We display the complex $q_1$ plane. We have displayed the poles in the $t$-channel by black crosses. The saddle point \saddl\ of the 
Gaussian integral has been denoted by a red cross. We have shifted the original integration contour for $q_1$, which was along the real axis, to the 
complex plane so that it passes through the saddle point \saddl . In the process we have picked up some of the poles in the $t$-channel.     }
 {\epsfxsize3.7in \epsfbox{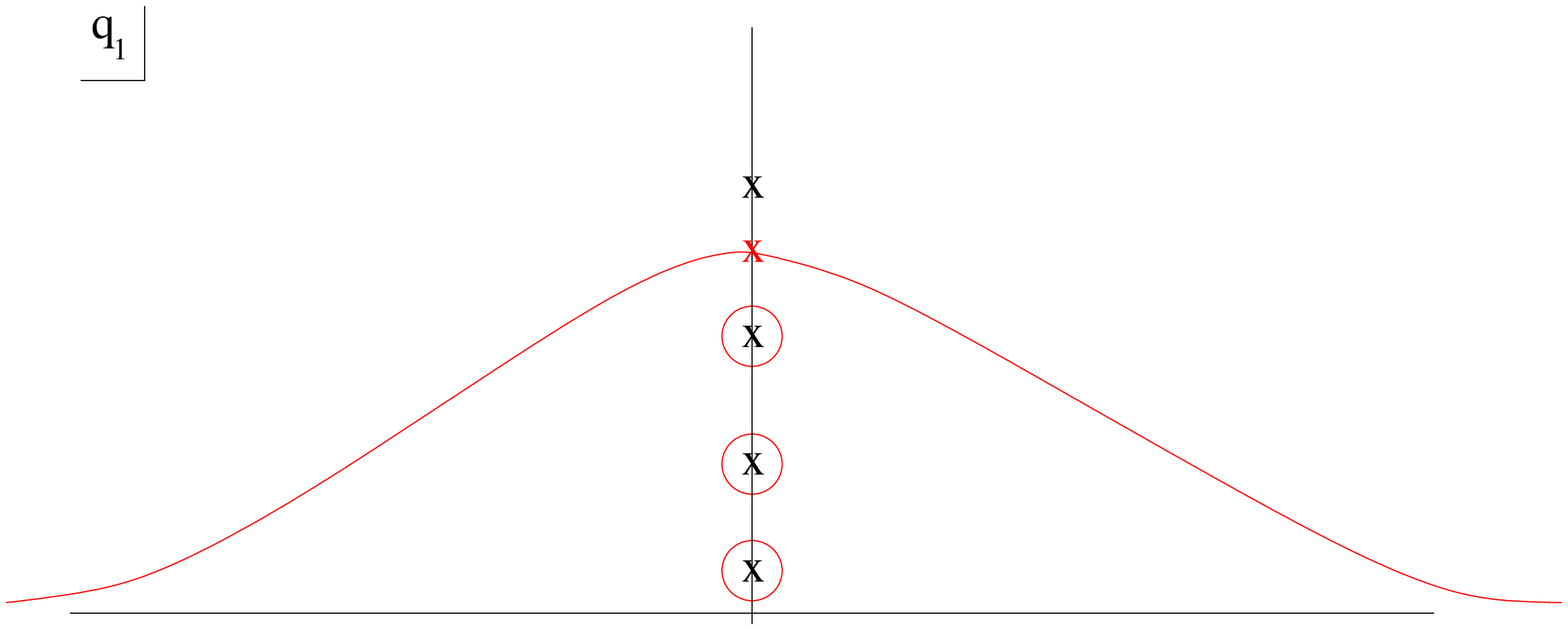} }
It is thus convenient to shift the contour to this location, \saddl ,  where this saddle point contribution gives something of the order of 
\eqn\saddlec{ 
 e^{ - { b^2 \over 2 \alpha' \log( -i s \alpha' /4) } }
 }
 This is not the whole answer, since by shifting the contour to this location,  we can pick up some poles from the prefactor \reggeofc , see \SaddleContour .
  We always pick up the 
 pole at $t=0$\foot{The location of this pole in the $q_1$ plane depends on $\vec q_{\rm rest}$ which we take to be real.},
  which was the center of our gravity discussion, but we can also pick up the poles at $t = { 4 \over \alpha' } n$ for $n <  -q_s^2 $, where 
 $q_s$ is given in \saddl . Notice that at these saddles the $t$ dependent part of the exponent gives us factors of $s^{ 2n}$ as we expect for the 
 corresponding spins\foot{ For a closed string with $N_L  = N_R=n$, the maximum spin is $J = 2 + 2 n$.}, once we take into account that [POL] contains a factor of $s^4$.  
 When $b$ is large
 $b^2 \gg \alpha' \log (s\alpha')  $, we pick up many poles, but when $b$ is small we only pick up the $t=0$ pole, but, even then, the integral is more 
 accurately computed using  \conclu  .  Note that the factor [POL] in string theory phase simplifies at $t=0$ and becomes the product of three-point amplitudes
 we discussed in the body of the paper and that we added as Pol in \strinphs . In other words [POL]$\to$Pol as $t\to 0$. 
 
 In particular,  note that the residues of  poles associated to the massive states go as $ { 1 \over (n !)^2} e^{ - b \sqrt{ 4 n \over \alpha'} }  s^{ 2 n}$. As a function of $n$, these 
 contributions decrease and then start increasing again with a transition at a value of $n$ corresponding to the saddle point \saddl \foot{Here we are assuming
 that both $b^2$ and $\log s \alpha' $ are large with a ratio larger than one but fixed, so that we can neglect the $1/(n!)^2$ in this discussion. This last factor 
 makes the sum converge for large $n$, but this convergent answer, as discussed below, is not the correct one. }. 
 
 We can ask: Why don't we include all poles in the $t$-channel?.  If we were to include all poles in the $t$-channel, we would obtain the wrong answer. The reason 
 it is wrong is that in string theory the argument that we can shift the contour is not correct because of contributions for large values of $q$. 
 Such large values of $q$ were never meant to be included in the integral, since the kinematics of the process we consider restricts the real values of $\vec q^{\, 2}$ 
 to be
 much smaller than $s$. In deriving the physical picture, we certainly assumed that $\vec q^{\, 2} \ll s $.\foot{{  Momentum conservation and on-shell conditions impose $q^2 < {s \over 4}$.}} Indeed, if we look at \reggeofc\ we get a very small 
 contribution from large real values of $\vec q^{ \, 2}$. On the other hand, if we were to keep $s$ fixed and we formally 
 look at large real values of $\vec q^{\, 2}$ in the
 original expression, \stringtheoryphaseshift , then we would encounter the $u$ channel poles. 
 The conclusion is that shifting the contour for the $q_1$ integral, while it can be done formally, it does not represent the real physical computation we want to do. 
 Approximating the integrand using \reggeofc , and then integrating gives the physically correct answer. 
 
 One can qualitatively say that, for large $b^2/( \alpha' \log s \alpha') $, we get a contribution of   some of the $t$-channel poles, as in \SaddleContour , and
 then the rest of the poles are completely resummed via the saddle point integral in \SaddleContour . Their contribution should be better thought of as coming 
 from the creation of extended objects in the $s$-channel. 
 
 Another remark we want to make is the following. It was shown in  \HorowitzBV\
  that the plane wave solution is a solution to all orders in the $\alpha'$ expansion. 
 This gravitational plane wave encodes the contribution from the $t=0$ pole. However, we have seen that sometimes we get a subleading contribution from the 
 other poles, due to massive states along the $t$-channel. These mean that the physical scattering process contains extra contributions not captured by the 
 plane wave\foot{This is not in contradiction with \HorowitzBV , since these extra terms can be viewed as a non-perturbative contribution in $\alpha'$.}.

 In string theory we can also take into account the tidal excitations.  In this case the phase shift can be viewed as an operator that maps the two initial gravitons to 
 two final generic string states.   
  \refs{\AmatiWQ,\AmatiUF,\AmatiTN,\AmatiZB,\AmatiTB} have shown that  this operator has the remarkably simple expression 
  $\hat \delta \propto \int d \sigma d\sigma'
 \delta_{\rm grav} ( \hat X_L(\sigma) - \hat X_R(\sigma') ) $ where $X_L$ and $X_R$ are the transverse space positions of the string on the worldsheet and 
 $\delta_{\rm grav}(b)$ is the ordinary gravity phase shift. This is valid for distances $b^2 \gg \alpha' \log(s \alpha')$. 
  The effects of these tidal excitations do not   help in resolving the causality issues discussed here and 
 are unrelated to the appearance of closed strings in the $s$-channel discussed above. 
 See \refs{\AmatiWQ,\AmatiUF,\AmatiTN,\AmatiZB,\AmatiTB} for further discussion.

\appendix{F}{Properties of the AdS Shock Wave}

Let us first examine the problem of higher derivative corrections for the AdS shock wave. As in the case of flat space 
the shock wave at hand continue to be an exact solution when arbitrary higher derivative corrections are included.

The argument for this is identical to the one in section 5 of \HorowitzGF . 
The vector $l_\mu = \partial_\mu u = \{ 1,0,0,0..\} $ in coordinates $u,v, y_i, z $. 
We can now compute the vector $V_\mu$ that the argument talks about. We find 
$V_\mu = \{0,\cdots,0, -2/z\}$. In other words $V_z = -2/z$ and the rest of the components are zero. 
This obeys that $V_\mu l^\mu =0$ are required in their argument. 

In addition, one can also compute the substracted Riemann tensor $$\check{R}_{\mu\nu\alpha\beta}=R_{\mu\nu\alpha\beta}+
2g_{\mu \lbrack \alpha}g_{\beta \rbrack \nu}$$ 
It is indeed of the form stated in \HorowitzGF\ with the symmetric tensor $K$ given by 
\eqn\tensk{\eqalign{
&K_{y^i,y^j} = { 1 \over 2} (  \delta_{ij} \partial_z h/z^3 - \partial_i \partial_j h/z^2 ) \,, \cr
&K_{z,y^i} = K_{y^i,z}= - {\partial_{y^i} \partial_z h \over z^2} \,, \cr
&K_{zz} = { 1 \over 2} (  \partial_z h/z^3 - \partial_z^2  h/z^2) \,,
}}
with the rest of the components  equal to zero. This $K$ obeys that $K_{\mu \nu} l^\nu =0$ as required by \HorowitzGF . Notice that this form of $K$
also leads to the equation of the form \laplac, when the Riemann tensor is used in Einstein's equations, namely $K^{\mu}_{\ \mu} = 0$. Once the former equation is imposed $\check R $ reduces to the Weyl tensor.  

Using this shock wave we can once again compute the time delay for different theories. It is convenient to evaluate Riemann tensor in the coordinates that make rotation symmetry manifest. The result is
\eqn\shockriemann{
\hat R_{u i j u}|_{\vec y = 0 ; z=1}=K_{i j}=-f(u)\left(1-\rho^2\right) {\varpi'(\rho)-\rho\varpi''(\rho)\over 8\rho}\left(n^i n^j- {1 \over D-2}\delta^{i j}\right)
}
where $i = 1, .., D-2$ so that we span $\vec y$ and $z$ components. In the bulk of the paper we consider propagation of different perturbations in the shock wave background described above. We first write the most general form of the second order equations of motion and then compute the time delay in the high 
energy limit.

Let us consider several limits of the shock wave to make contact with previous investigations of similar type. First, we expect to recover the flat space shock wave for probes that come close enough to the center of the shock or, equivalently, for $\rho \to 0$. Indeed, it is easy to check that this asymptotic 
is correctly recovered \limitsshock . 

Let us consider couple of other limits. We can fix $\vec y_0$ and send $z_0 \to 0$ in which case we have a source at the boundary
and the shock wave takes the form
\eqn\shockboundary{
h \sim { z^{D-3} \over (z^2 + |\vec y - \vec y_0|^2)^{D-2}},
}

which is exactly the shock wave considered in the energy correlator problem \HofmanAR. In the opposite limit $z_0 \to \infty$ we get
\eqn\shockhorizon{
h \sim z^{D-3},
}
and the time delay in this background was computed, for example, in \HofmanUG .
 
\appendix{G}{Time Advances and Time Machines }

In this appendix we would like to argue that a negative time delay enables us to build a time machine which 
leads to closed time-like curves. This is a standard argument \Tolman\  and the only thing we check is that the long range gravitational 
forces do not prevent us from setting it up. 
The setup we would like to consider is the following. We have two shock waves that correspond to energetic
particles with momenta $q_{1,u} = {\sqrt s \over 2}$ and $q_{2,v} = {\sqrt s \over 2}$ separated by distance $r$ 
in the transverse plane. The first shock is localized around $u=0$ and the second one around $v=0$. 
We would like the separation to be such that $r \gg r_{S}$, namely we are in the regime where the black hole formation does not occur
\eqn\conditionBH{
r^{D-3} \gg r_{S}^{D-3} = G \sqrt{s}.
}

Next we would like to consider a test particle that propagates through both shocks in such a way that it ends up at the same position where it started, thus, forming a closed time-like curve.

\ifig\CTC{a) We imagine a background that consists of two shock waves located at $u=0$ and $v=0$ widely separated in the transverse directions which is not presented on the figure.The arrows show  the motion of the probe massless particle projected on to the $u,v$ plane.  b) Same motion but projected on the transverse plane. The two background shocks are separated by $r$ and the probe passes at a short distance $b$ from each of them. The vertical region of the path in (a) corresponds to the horizontal motion in (b). 
 We can build a closed time-like curve as depicted on the picture by crossing this pair of shocks if time advances are allowed.
  We need mirrors to reverse the motion in the 
transverse plane as we pass through the shocks.  }  {\epsfxsize4.8in \epsfbox{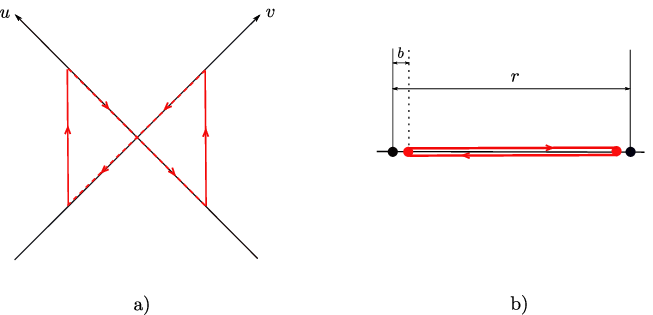} } 

The situation is depicted on \CTC . When the test particle crosses each of the shocks it gets shifted by $\Delta v = \Delta u \sim  {G \sqrt s \over b^{D-4}}$. Between the shock the particle travels the distance of order $r$. Thus, we want the time shift to be ${G \sqrt s \over b^{D-4}} = \left( {r_{S} \over b} \right)^{D-3} b \sim r \gg r_{S}$ which becomes
\eqn\conditionB{
\left( {r_{S} \over b} \right)^{D-4} \gg 1 .
}

We also want $b \gg \tilde r_{S}$ where $\tilde r_{S}$ is the Schwarzschild radius for the shock wave-test particle pair, $\tilde r_{S}^{D-3} \sim G \sqrt{ p \sqrt{s}}$, 
where $p$ is the energy of the probe particle.
 Together with \conditionB\ it implies $\sqrt s \gg p$ where $p$ is the energy of the probe particle. We also need that $p b\gg 1$. 
  The conclusion is that in $D>4$ we can construct closed time-like curve using negative time delays by choosing $b$,$r$ and $s$ appropriately.
For example, if we have a causality problem that appears at a scale $b^2 \sim \alpha_2$, then we take this value for $b$. Since we are at weak coupling, we know
that the Planck length $l_{p} \ll b$. We can then pick $\sqrt{s} l_{p} \sim X^{1 + a} ~, p l_p \sim X^{ 1 -a'} $, with $X = (b/l_p)^{D-3}$ and we can choose 
$ a >0 $, $a-a' < 0$, $ 1 - a' +1/(D-3) >0$ to ensure that $r_{S} \gg b$, $\tilde r_S \ll b$ and $ p b \gg 1$. We can achieve this with $a' =1/2$ and $a =1/4$, for example.

\appendix{H}{Representations That Couple to Two Gravitons}

Here we would like to understand better what are the representations of little group $SO(D-1)$ of massive particles that can couple to two gravitons. We are interested only in bosonic fields since a single  fermion does not couple to two gravitons. 
 
Let us consider the decay of a massive particle in its rest frame so that $p_2=(M,\vec 0)$. Gravitons produced have $\vec p_1 = - \vec p_3 = \vec p$.
We characterize the original particle by some polarization tensor $e_{i_1 ... i_ k}$ which has only spatial components and is traceless with respect to any pair of indices. We do not specify the symmetry properties of this tensor yet. 

We characterize gravitons by polarization tensors $e_1$ and $e_3$ such that $\vec e_1 . \vec p = \vec e_3 . \vec p = 0$. We have three type of contractions (see also \CostaMG)
\eqn\diffamplitudes{\eqalign{
{\cal A}_1 &= e_{i_1 ... i_k} p^{i_1} ... p^{i_k} (e_1 . e_3)^2, \cr
{\cal A}_2 &= e_{i_1 ... i_k} e_1^{i_1} e_3^{i_2} p^{i_3}... p^{i_k} (e_1 . e_3), \cr
{\cal A}_3 &= e_{i_1 ... i_k} e_1^{i_1} e_1^{i_2} e_3^{i_3} e_3^{i_4} p^{i_5}... p^{i_k} .
}}

The first amplitude ${\cal A}_1$ exists only for particles in the symmetric traceless representations (Young tableau that consists of  single horizontal row with $k$  boxes). Actually all three amplitudes are are allowed for symmetric representation and we discussed them in the bulk of the paper in detail. 

For the second and third amplitude we can add more rows to the Young diagram. Properties of these, so-called, mixed-symmetry tensors are nicely reviewed, for example, in appendix E of \DidenkoDWA. By thinking about the representation in terms of tensors which are manifestly anti-symmetric with respect to indices in a given column we can read off possible representations. In particular the fact that we have only three different vectors means that we can have at most three rows.

Let us write all the amplitudes in a covariant manner. The general prescription is the following 
\eqn\prescription{
  e^i_{1} e^j_3 \to E^{\mu \nu}_{13} \equiv \eps_1^{\mu} p_3^{\nu} (\eps_3 . p_1) + \eps_3^{\mu} p_1^{\nu}  (\eps_1 . p_3)  - p_1^{\mu} p_3^{\nu} (\eps_1 . \eps_3) - \eps_1^{\mu} \eps_3^{\nu} (p_1 . p_3).
}
 We then have for the amplitudes
\eqn\diffamplitudesB{\eqalign{
{\cal A}_1 &=  \eps_{\mu_1 ... \mu_k} p_1^{\mu_1} ... p_1^{\mu_k} \left[ (\eps_1 . \eps_3) (p_1. p_3) - (\eps_1 . p_3) (\eps_3 . p_1) \right]^2 , \cr
{\cal A}_2 &=  \eps_{\mu_1 ... \mu_k}   E_{13}^{\mu_1 \mu_2} p_1^{\mu_3}... p_1^{\mu_k}  \left[ (\eps_1 . \eps_3) (p_1. p_3) - (\eps_1 . p_3) (\eps_3 . p_1) \right], \cr
{\cal A}_3 &= \eps_{\mu_1 ... \mu_k} E_{13}^{\mu_1 \mu_3} E_{13}^{\mu_2 \mu_4}p_1^{\mu_5}... p_1^{\mu_k} .
}}

To compute the time delay we need to use the completeness relation
\eqn\completeness{
\sum_{i} \eps_{\nu_1 ... \nu_k}^* \eps_{\mu_1 ... \mu_k} = \Pi_{ \nu_1 ... \nu_k | \mu_1 ... \mu_k }
}
and contract both sides of the amplitude, where $\Pi$ is a projector on to the space orthogonal to the intermediate momentum.

As the next step we would like to focus on those diagrams that produce $s^a$ with $a \geq 2$ for the amplitude. Everything that is less is irrelevant for causality violation. In the language of Young tableaux it corresponds to having $a \geq 2$ boxes in the first row. It will be curious to understand 
if mixed symmetry fields with $a=2$ can resolve the causality problem we observed in the bulk of the paper. As an example of such field would be $(2,2)$ (a square with four boxes) or $(2,2,2)$ fields (a vertical rectangle of $2 \times 3$ boxes). There is a short list of representations
of this kind.   

One could wonder whether these particles alone (without the infinite tower of higher spin particles) 
could solve the causality problem. We think that the answer is no. 
We leave a full 
 exploration of this question for the future.

\listrefs

\bye